\documentclass[prd,aps,floats,onecolumn,twoside,preprintnumbers,
superscriptaddress,floatfix,nofootinbib,showpacs]{revtex4}
\usepackage{graphicx,colordvi,pstricks,rotating,amsmath}
\usepackage{amssymb,epstopdf,bbm}
\usepackage{slashed} 
\usepackage[T1]{fontenc}
\usepackage{epsfig}

\begin{document}

\title{Seiberg-Witten map invariant scatterings}

\author{Du\v sko Latas}
\affiliation{University of Belgrade, Faculty of Physics, P.O Box 44, Belgrade Serbia}
\email{latas@ipb.ac.rs}
\author{Josip Trampeti\'{c}}
\affiliation{Ru\dj er Bo\v{s}kovi\'{c} Institute, Division of Experimental Physics, Bijeni\v{c}ka 54, 10000 Zagreb, Croatia}
\email{josip.trampetic@irb.hr}
\affiliation{Max-Planck-Institut f\"ur Physik, (Werner-Heisenberg-Institut), F\"ohringer Ring 6, 
D-80805 M\"unchen, Germany}
\email{trampeti@mppmu.mpg.de}
\author{Jiangyang You}
\affiliation{Ru\dj er Bo\v{s}kovi\'{c} Institute, Division of Physical Chemistry, Bijeni\v{c}ka 54, 10000 Zagreb, Croatia}
\email{jiangyang.you@irb.hr}
 
\newcommand{\tr}{\hbox{tr}}
\def\BOX{\mathord{\vbox{\hrule\hbox{\vrule\hskip 3pt\vbox{\vskip
3pt\vskip 3pt}\hskip 3pt\vrule}\hrule}\hskip 1pt}}
 
\date{\today}  

\begin{abstract}
We investigate scattering amplitudes of the reversible $\theta$-exact Seiberg-Witten 
map-based noncommutative (NC) quantum electrodynamics and show explicitly the SW map invariance for all types of tree-level NCQED two-by-two processes, including M\o ller, Bhabha, Compton, pair annihilation, pair production and light-by-light (LbyL: $\gamma\gamma\to\gamma\gamma)$ scatterings. We apply our NCQED results to the $\gamma\gamma\to\gamma\gamma$ and $\gamma\gamma\to\ell^+\ell^-$ exclusive channels, convoluted to the ultraperipheral lead $^{208}$Pb ion-ion collisions, where LbyL scattering  being recently measured by the ATLAS Collaboration at the LHC. We have demonstrated that $\gamma\gamma\to\gamma\gamma$ is the more appropriate channel to probe scale $\Lambda_{\rm NC}$. We also find that the transverse($\varphi$) dependence of the NC $\rm PbPb(\gamma\gamma)\to Pb^{\ast}Pb^{\ast}\gamma\gamma$ differential cross section shows large anisotropy near the peak NC contribution for a pure spacelike NC parameter. So the $\varphi$ variation of the differential cross section is likely an appropriate signature for the NCQED contributions with the pure space-space noncommutative parameter, given that enough events could be available for determining the anisotropy. 
\end{abstract}

 \pacs{02.40.Gh,11.10.Nx, 11.15.-q, 11.30.Pb}

\maketitle

\section{Introduction}

Lorentz symmetry, as a cornerstone of modern physics, appears with the Einstein formulation of the special 
theory of relativity in 1905. A classical example maintaining that Lorentz symmetry is the Dirac-Born-Infeld model \cite{Dirac:1928hu,Dirac:1931kp,Born:1934gh}, which introduces an upper bound for the electric field near the electron eliminating electron self-energy divergences, while another is famous 
Euler-Heisenberg classical Maxwell Lagrangian density \cite{Heisenberg:1935qt}. 
Since than, Lorentz symmetry is one of the symmetries whose violation has been challenged many 
times by many experimental attempts and bounds obtained.   

In the modern string theory framework regarding the effect of spontaneous breaking of 
Lorentz symmetry, one of the most striking observations is that via tensor field vacuum 
expectation values the low-energy effective theory can be expressed 
by usual gauge fields or gauge fields deformed by space-time noncommutativity (NC). 
These two versions of effective theory can be subsequently connected by highly nonlocal  expressions, 
called the Seiberg-Witten (SW) map \cite{Seiberg:1999vs}. 

While the simplest quantized coordinates argument would set the noncommutative scale to be at Planck scale, string theory does not specify directly the vacuum expectation value for the $B^{\mu \nu}$ field, therefore,  
the corresponding antisymmetric tensor $\theta^{\mu \nu}$ governing NC space-time deformations could 
bear an unknown value which may be assessed by experiments 
\cite{Douglas:2001ba,Szabo:2001kg,Szabo:2009tn,Bigatti:1999iz}.  NC gauge field theories are therefore studied intensively as perturbative quantum field theories from both theoretical (mathematical physics and field theory) 
\cite{Kontsevich:1997vb, Madore:2000en,Jurco:2000fb,Jurco:2001my,Jurco:2001kp,arXiv0711.2965B,arXiv0909.4259B,Mehen:2000vs,Zanon:2000nq,Jurco:2001rq,Jackiw:2001jb,Connes,Madore} and phenomenological viewpoints \cite{Bichl:2001nf,Bichl:2001cq,Calmet:2001na,Aschieri:2002mc,Hinchliffe:2002km,Martin:2013gma,Banerjee:2001un,Martin:2002nr,Brandt:2003fx,Banerjee:2003vc,Buric:2006wm,Latas:2007eu,Buric:2007ix,Martin:2009sg,Martin:2009vg,Buric:2010wd,Badelek:2001xb,Hewett:2000zp,Rizzo:2002yr,Behr:2002wx,Godfrey:2001yy,Altschul:2006pv,Garg:2011aa,Mathews:2000we,Baek:2001ty,
Schupp:2002up,Minkowski:2003jg,Ohl:2004tn,Melic:2005su,Tamarit:2008vy,Alboteanu:2006hh,
Alboteanu:2007bp,Alboteanu:2007by,He:2006yy,Buric:2007qx,Das:2007dn,Selvaganapathy:2016jrl,
Horvat:2010sr,Horvat:2012vn,Horvat:2017gfm,Horvat:2017aqf,Alfaro:2020njh}.
 
The SW map turns out to be not only of theoretical importance, but also highly instrumental 
in defining quasirealistic phenomenological quantum gauge field theory models 
\cite{Liu:2000ad,Liu:2000mja,Okawa:2001mv,Banerjee:2004rs,
Brace:2001rd,Brace:2001fj,Cerchiai:2002ss,Barnich:2002pb,Barnich:2003wq,Martin:2008xa,
Martin:2015nna,arXiv0711.2965B,arXiv0909.4259B,Horvat:2011qn,Martin:2012aw}. 
In recent years some crucial unitarity properties 
\cite{Gomis:2000zz,Aharony:2000gz,Carroll:2001ws,Kostelecky:2000mm,Cohen:2006ky,SheikhJabbari:2008nc} of the perturbatively quantized NC gauge theories  \cite{Filk:1996dm,Martin:1999aq}, with and without the SW map, have also been successfully linked up via the so-called $\theta$-exact-SW-map{\footnote{J.T. thanks L. Alvarez-Gaume for pointing out, during our discussions in the early years of SW, 
the importance of $\theta$-exact SW map, i.e. on the necessity to resum of all originally 
$\theta$-expanded SW map solutions at the end of a day.}} 
expansion technique leading to the quantum effect of UV-IR mixing 
\cite{Bigatti:1999iz,Minwalla:1999px,VanRaamsdonk:2000rr,Hayakawa:1999yt,Hayakawa:1999zf,Hayakawa:2000zi,Matusis:2000jf,Ruiz:2000hu,Khoze:2000sy,Armoni:2003va,Ferrari:2004ex,Zeiner:2007,Schupp:2008fs,Horvat:2011iv,Horvat:2011bs,Horvat:2011qg,Horvat:2013rga,Trampetic:2014dea,Trampetic:2015zma,Horvat:2015aca,Martin:2020ddo}.  Additional to the Moyal manifold, it was shown that UV/IR mixing manifests itself on the $\kappa$-Minkowski \cite{Grosse:2005iz,Meljanac:2011cs} and Snyder \cite{Snyder:1946qz,Snyder:1947nq,Meljanac:2017grw,Meljanac:2017jyk} spaces,  as well. We note that there are also connections between UV-IR mixing and other ideas like: 
vacuum birefringence \cite{Abel:2006wj}, holography \cite{Cohen:1998zx,Horvat:2010km}, 
weak gravity conjecture, naturalness and the hierarchy problem 
\cite{Huang:2006tz,Li:2006jja,Palti:2017elp,Lust:2017wrl,Craig:2019zbn,Koren:2020biu}, and noncommutative 
Aharony-Bergman-Jafferis-Maldacena (NCABJM) theory \cite{Martin:2017nhg}, all known as a possible windows 
to quantum gravity \cite{AmelinoCamelia:1997gz,ArkaniHamed:1998rs,ArkaniHamed:2006dz,Abel:2006hk}.

In recent years tremendous progress has been achieved in the field of the on-shell 
scattering amplitudes of quantum gauge theories 
\cite{Arkani-Hamed:2017mur,Mizera:2019blq}. 
The NC (super) Yang-Mills (NC(S)YM) without the SW map have been found to share a good amount of recently discovered abstract scattering amplitude structures with their commutative counterparts at both tree and one-loop levels \cite{Raju:2009yx,Huang:2010fc}. 
On the other hand, scattering amplitudes of SW mapped NC gauge theories 
are mainly studied on an explicit process-by-process basis. 
An expansion over the NC parameter $\theta$ was also used in various early studies which covered 
a fairly large number of processes. Furthermore, while it has been shown formally that  
SW map-based equivalence exists between on-shell background field effective actions 
\cite{Martin:2016zon,Martin:2016hji,Martin:2016saw}, there is not yet a work that studies explicitly 
the scattering amplitudes in an NC gauge theory defined via a reversible SW map 
and compares with its counterpart without the SW map. 

In this work, we fill this vacancy by demonstrating explicitly that all tree-level two-by-two scattering 
amplitudes in an NCQED model defined with a reversible $\theta$-exact SW map are identical 
to their unmapped counterparts which is compatible with the formal equivalence proven before \cite{Martin:2016zon,Martin:2016hji,Martin:2016saw}. 

As applications of this newly found identity, we revisit the NCQED two-by-two processes and, in particular, light-by-light  scatterings and lepton pair production processes in the context of  ATLAS ultraperipheral lead 
$^{208}$Pb ion-ion collision measurements
\cite{Aaboud:2017bwk,ATLAS-CONF-2019-002,Aad:2019ock,ATLAS-CONF-2020-010} 
and show that LbyL scattering is better probe to the NC scale of these two. ATLAS results  
were already compared with the standard model (SM) predictions \cite{dEnterria:2013zqi,Klusek-Gawenda:2016euz,dEnterria} and used in constraining various models beyond the SM \cite{Knapen:2016moh,Ellis:2017edi,Akmansoy:2018xvd,Kostelecky:2018yfa,J.:2019bws,Horvat:2020ycy,Bekli:2020unl}. 
 
The paper is structured as follows:\\
In Sec. II. we describe the Moyal NC deformation and the SW map-induced actions by means of 
the $\theta$-exact SW maps for a photon and charged fermion.  
In Sec. III. we prove the SW map invariance of scatterings amplitudes in the matter and 
the gauge sectors of NCQED. In Secs. IV-VI. we revisit and compute the cross section of exclusive processes in NCQED and SM: M\o ller, Bhabha, Compton (first computed by Klein and Nishina 1929 \cite{Klein-Nishina}), dilepton pair annihilations and productions, and $\gamma\gamma\to\gamma\gamma$ scatterings
(first published by Karplus and Neuman in 1951 \cite{Karplus:1950zz}) showing complementarity between NCQED and QCD  \cite{Hewett:2000zp,Baek:2001ty,Badelek:2001xb,Rizzo:2002yr,Mathews:2000we,Godfrey:2001yy}, 
and present them in the form of 3D figures of ratios of cross sections 
${\sigma_{\rm NC}}/{\sigma_{\rm SM}}$ as functions of the incoming  energy and
the NC scale. Section VII is devoted to the SM and NCQED 
computations of recent ATLAS Collaboration measured  $\rm PbPb(\gamma\gamma)\to Pb^{\ast}Pb^{\ast}\gamma\gamma$ and $\rm PbPb(\gamma\gamma)\to Pb^{\ast}Pb^{\ast}\ell^+\ell^-$ collision cross sections.  Section VIII contains discussions and conclusions, respectively. In Appendixes we use main conventions and notations, like equations of motions (EOM) and on shell conditions from Peskin and Schroeder \cite{Peskin:1995ev}, and give the NCQED Feynman rules (FRs), the NC phase factors including $\varphi$ integrals, and details of vanishing of the SW map-induced amplitude for 
$\gamma\gamma\to\gamma\gamma$ scattering. For relevant integrals and other mathematical issues among Wolfram {\it Mathematica} \cite{mathematica}, we are using mainly Gradshteyn and Ryzhik  \cite{Gradshteyn}. 

\section{SEIBERG-WITTEN MAP-INDUCED NCQED ACTION}

\subsection{General SW map considerations}

We consider the minimal NCQED model in terms of the minimal NC gauge and charge fermion fields, 
$A_\mu$ and $\Psi$ respectively, which lives in the adjoint representation of the NC gauge group U(N):
\begin{eqnarray}
S^{\rm min}&=&\int-\frac{1}{4}F^{\mu\nu}\star F_{\mu\nu}+\bar\Psi \star(i\slashed{D}-m)\Psi,
\label{NCminAction}\\
F_{\mu\nu}&=&\partial_\mu A_\nu - \partial_\nu A_\mu -i[A_\mu\stackrel{\star}{,}A_\nu], \;D_\mu\Psi=\partial_\mu\Psi - i[A_\mu\stackrel{\star}{,}{\Psi}],
\nonumber
\end{eqnarray}
where ($\star$) denote the Moyal-Weyl star-product. The relation between the NC and commutative quantities or fields is given by the $\theta$-exact Seiberg-Witten map. In the above action (\ref{NCminAction}) we do apply the generalized SW maps of the NC gauge parameter $\Lambda$, NC gauge and fermion fields $A_\mu$ and $\Psi$, in terms of commutative gauge parameter $\lambda$ and commutative--physical--gauge and fermion fields $a_\mu$ and $\psi$, respectively.

Extending results obtained in Refs. \cite{Martin:2012aw,Martin:2015nna}, by applying a method based on the following consistency condition transformations \cite{Jurco:2001my,Jurco:2001rq} 
\begin{gather}
\delta_\Lambda A_\mu \equiv \partial_\mu\Lambda+ i[\Lambda \stackrel{\star}{,}A_\mu] 
=\delta_\lambda A_\mu[a_\mu]\,,
\label{deltaA}\\
\delta_\Lambda F_{\mu\nu} \equiv i[\Lambda \stackrel{\star}{,}F_{\mu\nu}]=\delta_\lambda F_{\mu\nu}[a_\mu]\,,
\label{deltaF}\\
\Lambda[[\lambda_1,\lambda_2],a_\mu]
=[\Lambda[\lambda_1,a_\mu]\stackrel{\star}{,}
\Lambda[\lambda_2,a_\mu]]
+i\delta_{\lambda_1}\Lambda[\lambda_2,a_\mu]-i\delta_{\lambda_2}
\Lambda[\lambda_1,a_\mu],
\label{SWrecurs}
\end{gather}
we have been working out in detail the $\theta$-exact SW map for the noncommutative gauge field strength 
$F_{\mu\nu}$  up to the third order of ordinary U(N) gauge field $a_\mu$ \cite{Trampetic:2015zma}.  
So in terms of commutative U(1) gauge field strength $f_{\mu\nu}$, we have found the following solution:
\begin{equation}
F_{\mu\nu}\left(e\cdot a_\mu,\theta^{\mu\nu}\right)=e f_{\mu\nu}+F^{e^2}_{\mu\nu}
+F^{e^3}_{\mu\nu}+\mathcal O\left(e^4\right), \;f_{\mu\nu}=\partial_\mu a_\nu-\partial_\nu a_\mu, 
\label{2.2}
\end{equation}
where the term for the SW map up to the $e^2$ order $F^{e^2}_{\mu\nu}$ 
is fairly universal \cite{Horvat:2013rga,Trampetic:2014dea,Trampetic:2015zma}
\begin{equation}
F^{e^2}_{\mu\nu}(x)=e^2\theta^{ij}\Big( f_{\mu i}\star_{2}f_{\nu j}-a_i\star_2\partial_jf_{\mu\nu}\Big).
\label{2.3}
\end{equation}
Here, we use well-known Moyal-Weyl $\star$- and the $\star_2$-product definitions: 
\begin{equation}
(f\star g)(x)=f(x)e^{i\frac{\partial_x\theta\partial_y}{2}}g(y)\Big|_{x=y}\;\;
\Longrightarrow\;\;
f(x)\star_2 g(x)=f(x)\frac{\sin\frac{{\partial_x}\theta{\partial_y}}{2}}{\frac{{\partial_x}\theta{\partial_y}}{2}}g(y)\Bigg|_{x=y},
\label{star2}
 \end{equation}
where, in fact, $\star_2$-product is nonassociative; however, it is commutative.

Now we give the $e^3$ power of the $\theta$-exact field strength 
$F^{e^3}_{\mu\nu}$ [SW map (I)-induced solution in Eq. (65) from Ref. \cite{Trampetic:2015zma}], 
needed for all further computations: 
\begin{equation}
\begin{split}
F^{e^3}_{\mu\nu}(x)=
\frac{e^3}{2}\theta^{ij}\theta^{kl}&\bigg(\left[f_{\mu k}f_{\nu i} f_{l j}\right]_{\star_{3'}}
+\left[f_{\nu l}f_{\mu i}f_{kj}\right]_{\star_{3'}}-\left[f_{\nu l}a_i\partial_j f_{\mu k}\right]_{\star_{3'}}
-\left[f_{\mu k}a_i\partial_j f_{\nu l}\right]_{\star_{3'}}
\\&-\left[a_k\partial_l\left(f_{\mu i}f_{\nu j}\right)\right]_{\star_{3'}}
+\left[a_i\partial_j a_k \partial_l f_{\mu\nu}\right]_{\star_{3'}}
+\left[\partial_l f_{\mu\nu}a_i\partial_j a_k\right]_{\star_{3'}}
\\&+\left[a_k a_i \partial_l\partial_j f_{\mu\nu}\right]_{\star_{3'}}
-\frac{1}{2}\Big(\left[a_i\partial_k a_j\partial_l f_{\mu\nu}\right]_{\star_{3'}}
+\left[\partial_l f_{\mu\nu}a_i\partial_k a_j\right]_{\star_{3'}}\Big)\bigg),
\label{2.4}
\end{split}
\end{equation}
with new star product being defined in Refs. \cite{Trampetic:2015zma,Horvat:2015aca}.

A general SW map expansion of the noncommutative U(N) charged fermion field $\Psi(x)$, 
in terms of commutative gauge and fermion fields ($a,\psi$) respectively, is given as
\begin{eqnarray}
\Psi(x)=\Psi^{(0)}(x)+\Psi^{(1)}(x)+\Psi^{(2)}(x)+\cdot\cdot\cdot
=\psi-\frac{1}{2}\theta^{ij}a_i\bullet\partial_j\psi+\mathcal O (a^2)\psi,
\label{Psi}
\end{eqnarray}
where the right-hand side of Eq. (\ref{Psi}) represents the leading-order charged fermion SW map, derived long ago \cite{Seiberg:1999vs}.
\noindent
The bullet product ($\bullet$) in Eq. (\ref{Psi}) is defined as
\begin{equation}
(f\bullet g)(x)=f(x)\frac{e^{i\frac{\partial_x\theta\partial_y}{2}}-1}{i\frac{\partial_x\theta\partial_y}{2}}g(y)\bigg|_{x=y}
\;\;\Longrightarrow\;\;\int\,(f\bullet g)h=\int\, f(g\bullet h).
\label{bullet}
\end{equation}
Above the $\bullet$ product is not associative, but it satisfies the integral identity, 
which is easy to prove by expanding, integrating by part, and resumming the resulting series. 

At the end of this section, note that, to obtain the complete two-electron--two-photon vertex $\Gamma^{\mu\nu}(\bar e e\gamma\gamma)$ (necessary for any quantum-loop computation)  
in momentum space, we also need a second order in gauge field $\theta$-exact SW map for 
the electron field, $\Psi^{(2)}(x)$. Using the SW differential equation method \cite{Martin:2012aw} 
in the same way as we did in Ref. \cite{Trampetic:2015zma}, we have also obtained the $\Psi^{(2)}(x)$ 
term of the SW map expansion (\ref{Psi}) explicitly:
\begin{equation}
\begin{split}
\Psi^{(2)}(x)&=i\int dp dq dk e^{-i(p+q+k)x}\,\:\frac{1}{4}\theta^{ij}\bigg\{\tilde a_i(p)\tilde a_j(q)\tilde\psi(k)\frac{e^{-\frac{i}{2}(p\theta q+p\theta k+q\theta k)}-1}{-\frac{i}{2}(p\theta q+p\theta k+q\theta k)}
\\&-\theta^{mn}\bigg[a_i(p)\tilde a_m(q)\tilde\psi(k)(q+k)_j k_n
\frac{1}{\frac{1}{2}{q\theta k}}\bigg( \frac{e^{-\frac{i}{2}(p\theta q+p\theta k+q\theta k)}-1}{-\frac{i}{2}(p\theta q+p\theta k+q\theta k)} -  \frac{e^{-\frac{i}{2}(p\theta q+p\theta k)}-1}{-\frac{i}{2}(p\theta q+p\theta k)} \bigg)
\\&-\frac{1}{2}\,k_j \Big(2q_n\tilde a_m(p)\tilde a_i(q)-q_i\tilde a_m(p)\tilde a_n(q)\Big)\tilde \psi(k)
\frac{1}{\frac{1}{2}p\theta q}\bigg(
\frac{e^{-\frac{i}{2}(p\theta q+p\theta k+q\theta k)}-1}{-\frac{i}{2}(p\theta q+p\theta k+q\theta k)}
-\frac{e^{-\frac{i}{2}(p\theta k+q\theta k-p\theta q)}-1}{-\frac{i}{2}(p\theta k+q\theta k-p\theta q)}\bigg)
\bigg]\bigg\}.
\end{split}
\label{SWPsi^2}
\end{equation}

\subsection{ Minimal $\theta$-exact SW mapped NCQED action} 

From minimal noncommutative U(N) action (\ref{NCminAction}) we perform 
the SW map expansion in terms of commutative --physical-- fields ($a_\mu,\psi$)  
and write the generalized manifestly gauge invariant actions with a minimal number of fields: 
\begin{equation}
S^{\rm min}\,\stackrel{\rm SW}{=}\,
S_{\rm U(1)}+S_{a^3}+S_{a^4}+S_{\bar\psi a\psi}+S_{\bar\psi a^2\psi}+\cdot\cdot\cdot,
\label{S}
\end{equation}
The solution of the $\theta$-exact SW map in terms of commutative gauge field strengths $f^{\mu\nu}$ was resolved 
from the SW differential equation \cite{Trampetic:2015zma} giving the following minimal 
action\footnote{See discussion regarding gauge 
freedom parameters in \cite{Trampetic:2015zma,Horvat:2015aca}, at the beginning of Appendix A.b.}: 
\begin{gather}
S_{\rm U(1)}=\int-\frac{1}{4}f_{\mu\nu}f^{\mu\nu}+i\bar\psi\slashed{\partial}\psi\,,
\label{U1}\\
S_{a^3}=-\frac{e}{2}\int\theta^{ij}f^{\mu\nu}\Big(f_{\mu i}\star_2 f_{\nu j}-\frac{1}{4}f_{ij}\star_2f_{\mu\nu}\Big),
\label{f3}\\
S_{a^4}=-\frac{e^2}{4}\theta^{ij}\theta^{kl}\int\,(f_{\mu i}\star_2 f_{\nu j})(f^\mu_{\;\,\; k}\star_2 f^\nu_{\;\,\; l})-(f_{ij}\star_2 f_{\mu\nu})(f^\mu_{\;\,\; k}\star_2 f^\nu_{\;\,\; l})
\nonumber\\
+2 f^{\mu\nu}[f_{\mu i}f_{\nu k}f_{jl}]_{\star_{3'}}+2 f^{\mu\nu}\Big(a_i\star_2\partial_j(f_{\mu k}\star_2 f_{\nu l})-[f_{\mu k} a_i\partial_j f_{\nu l}]_{\star_{3'}}-[a_i f_{\mu k}\partial_j f_{\nu l}]_{\star_{3'}}\Big)
\nonumber\\
-\frac{1}{4}f^{\mu\nu}\left[f_{\mu\nu}f_{ik}f_{jl}\right]_{\star_{3'}}+\frac{1}{8}\left(f^{\mu\nu}\star_2 f_{ij}\right)\left(f_{kl}\star_2 f_{\mu\nu}\right)+\frac{1}{2}\theta^{pq}f^{\mu\nu}\left[\partial_i f_{jk} f_{lp}\partial_q f_{\mu\nu}\right]_{\mathcal M_{\rm (I)}},
\label{f4}
\end{gather}
where additional star products could be found in Refs. \cite{Trampetic:2015zma,Horvat:2015aca}.

\subsubsection{Minimal matter sector: Relevant parts up to the $e^2$ order}

Next we give $\theta$-exact electron-photon action up to the $e^1$ order $S_{\bar\psi a\psi}$ as
\begin{equation}
\begin{split}
S_{\bar\psi a\psi}=\int\,&\bar\psi\slashed{a}\star\psi-\frac{i}{2}\theta^{ij}\Big[(\partial_j\bar\psi\bullet a_i)(\slashed{\partial}+im)\psi+\bar\psi(\slashed{\partial}+im)(a_i\bullet\partial_j\psi)\Big]
\\=\int\,&\bar\psi \slashed{a}\psi+\frac{i}{4}\theta^{ij}\bar\psi\Big[2 (\partial_i \slashed{a}-\slashed{\partial}a_i)\bullet\partial_j\psi-f_{ij}\bullet(\slashed{\partial}+im)\psi\Big].
\end{split}
\label{Spee}
\end{equation}

There are also two-electron--two-photon action relevant terms $S_{\bar\psi a^2\psi}$ 
with additional parts which are proportional to free field equation for either $\psi$ or $\bar\psi$. 
Therefore, since they vanish due to the EOM they are irrelevant to the tree-level 
scattering processes we are heading for. Bearing this in mind we collect now 
the relevant terms at the $e^2$ order which does not vanish on shell $S_{\bar\psi a^2\psi}\big|_{\rm relevant}$:
\begin{equation}
\begin{split}
S_{\bar\psi a^2\psi}\big|_{\rm relevant}=-\frac{1}{2}\int\,&\theta^{ij}\Big[(\partial_j\bar\psi\bullet a_i)\slashed{a}\star\psi+\bar\psi\slashed{a}\star(a_i\bullet\partial_j\psi)
+\bar\psi\big(a_i\star_2(2\partial_j \slashed{a}-\slashed{\partial}a_j)\big)\star\psi
+\frac{i}{2}\theta^{kl}(\partial_j\bar\psi\bullet a_i)\big(\slashed{\partial} a_k\bullet \partial_l\psi\big)\Big].
\end{split}
\label{Sppee}
\end{equation}

Extracting the relevant matter part of the two-electron--two-photon coupling terms we 
integrate (\ref{Sppee}) by part in such way that all partial derivatives hit $\psi$, thus changing the form of $S_{\bar\psi a^2\psi}$ to
\begin{equation}
\begin{split}
S_{\bar\psi a^2\psi}\big|_{\rm relevant}=\frac{1}{2}\int&\theta^{ij}\Big[\bar\psi \partial_j\big(a_i\bullet(\slashed{a}\star\psi)\big)-\bar\psi\slashed{a}\star(a_i\bullet\partial_j\psi)
-\bar\psi\big(a_i\star_2(2\partial_j \slashed{a}-\slashed{\partial}a_j)\big)\star\psi
-\frac{i}{2}\theta^{kl}\bar\psi \partial_ja_i\bullet\big(\slashed{\partial} a_k\bullet \partial_l\psi\big)\Big].
\end{split}
\label{Sppeer}
\end{equation}
From the above actions (\ref{f3},\ref{f4},\ref{Spee}, and \ref{Sppeer}), we obtain the Feynman rules which are given in  Appendix A, respectively.

\subsubsection{Minimal gauge sector: Relevant parts of three- and four-photon couplings}

Since the triple-photon interaction, based on a fairly universal second-order order SW map for the gauge field, is
already given as Eq. (\ref{f3}), here we continue with higher-order contribution. 

Expanding (\ref{NCminAction}) up to the order $a_\mu^4$ by using (\ref{S}) 
we get the following general form for the four-photon interaction:
\begin{equation}
S^{e^2}\equiv S_{a^4}=-\frac{1}{4e^2}\int\,\Big(F^{e^2}_{\mu\nu}F^{ e^2\mu\nu}+2ef^{\mu\nu}F_{\mu\nu}^{e^3}\Big),
\label{Se2}
\end{equation}
where the two distinct solutions for the $e^3$ order gauge field strength have been found and given explicitly in Ref. \cite{Trampetic:2015zma}.  After using solution (I) [Eq. (65) \cite{Trampetic:2015zma}) from (\ref{2.3})$-$(\ref{2.4}) and (\ref{Se2}), we obtain 
\begin{equation}
\begin{split}
-\frac{1}{4e^2}\int\,F^{e^2}_{\mu\nu}F^{ e^2\mu\nu}&=
-\frac{e^2}{4}\int\,\theta^{ij}\big(f_{\mu i}\star_2f_{\nu j}-a_i\star_2\partial_j f_{\mu\nu}\big)\theta^{kl}\big(f^\mu_{\;k}\star_2f^\nu_{\;l}-a_k\star_2\partial_l f^{\mu\nu}\big).
\label{R4p-1}
\end{split}
\end{equation}
In the second term of Eq. (\ref{Se2}),  we are not using the complete (\ref{2.4}) for $F_{\mu\nu}^{e^3}$, 
but only the relevant part, which then for the universal gauge field SW map $A_\mu^{e^2}=-\frac{e^2}{2}\theta^{ij}a_i\star_2(\partial_j a_\mu+f_{j\mu})$ gives 
\begin{equation}
\begin{split}
-\frac{1}{2e}\int\,f^{\mu\nu}\Big[F_{\mu\nu}^{e^3}\Big]_{\rm relevant}&=-\frac{1}{e}\int\,f^{\mu\nu}[iea_\mu\stackrel{\star}{,}A^{e^2}_\nu]=-\frac{i}{e}\int\,[f^{\mu\nu}\stackrel{\star}{,}ea_\mu]A^{e^2}_\nu
\label{R4p-2}
\end{split}
\end{equation}
producing
\begin{equation}
-\frac{1}{2e}\int\,f^{\mu\nu}\Big[F_{\mu\nu}^{e^3}\Big]_{\rm relevant}=
\frac{e^2}{2}\int\,\theta^{ij}\big(\partial_i f^{\mu\nu}\star_2\partial_j a_\mu\big)\theta^{kl}\big(a_k\star_2(\partial_l a_\nu+f_{l\nu})\big).
\label{R4p-3}
\end{equation}
Altogether, from Eq. (\ref{Se2}) we then have 
\begin{equation}
\begin{split}
S^{e^2}\big|_{\rm relevant}&=-\frac{e^2}{4}\int\,\theta^{ij}\theta^{kl}\Big[\big(f_{\mu i}\star_2f_{\nu j}\big)\big(f^\mu_{\;k}\star_2 f^\nu_{\;l}\big)
-2\big(a_i\star_2\partial_j f_{\mu\nu}\big)\big(f^\mu_{\;k}\star_2 
f^\nu_{\; j}\big)
\\
&\phantom{XXXXXXX}+
\big(a_i\star_2\partial_j f_{\mu\nu}\big)\big(a_k\star_2\partial_l f^{\mu\nu}\big)
-2\big(\partial_i f^{\mu\nu}\star_2\partial_j a_\mu\big)\big(a_k\star_2(\partial_l a_\nu+f_{l\nu})\big)
\Big].
\label{R4p-4}
\end{split}
\end{equation}
From the above action (\ref{R4p-4}), we obtain the relevant FRs and give them in Appendix A.

\subsection{Pametrizations of the NC $\theta$ matrix and unitarity of the NCQED theory
}

We parametrize the $4\times4$ antisymmetric $\theta$ matrix as follows:
\begin{equation}
\theta_{\mu\nu}=\frac{c_{\mu\nu}}{\Lambda_{\rm NC}^2}=\frac{1}{\Lambda_{\rm NC}^2}
\begin{pmatrix}
0&c_{01}&c_{02}&c_{03}\\
-c_{01}&0&c_{12}&c_{13}\\
-c_{02}&-c_{12}&0&c_{23}\\
-c_{03}&-c_{13}&-c_{23}&0
\end{pmatrix},
\label{Cmatrix}
\end{equation}
where the $c_{\mu\nu}$-matrix elements are identical in all reference frames resulting in Lorentz symmetry breaking and run within $[0,1]$ values. The $c_{0i}$ elements are related to the time-space (timelike) noncommutativity defined by the direction of the background 
$\bf E_{\theta}=\frac{1}{\Lambda_{\rm NC}^2}(  {\rm c_{01}},  {\rm c_{02}},  {\rm c_{03}})$ field,
while the  $c_{ij}$ coefficients are related to the space-space (spacelike) NC and are defined by the direction of the background 
$\bf B_{\theta}=\frac{1}{\Lambda_{\rm NC}^2} ({\rm c_{23}},  -{\rm c_{13}}, {\rm c_{12}})$ field.

It is important is to stress that, for the case $\theta^{0i}\not= 0, \forall \;i$, the NC theory is not unitary, i.e. acausal  behavior {\cite{Gomis:2000zz,Aharony:2000gz}}. Only the condition $\theta^{0i}= 0$ keeps the theory unitary and causal. However, to keep the possibility of nonzero space and time NC contributions to the novel particle physics effects, i.e., the physical noncommutative effects, we invoke a covariant generalization of $\theta^{0i}=0$ requirements through the  so-called unitarity condition, known as the quantum or perturbative unitarity condition \cite{Carroll:2001ws}. Namely, the theory with $\theta^{0i}\not= 0$ can be converted into one with only $\theta^{ij}\not= 0$ by a suitable observer Lorentz transformation, since the presence of observer Lorentz invariance implies that there are no difficulties with perturbative unitarity provided by $\theta_{\mu\nu}\theta^{\mu\nu}>0$. Analogous methods do apply for noncommutative theories with $\theta^{0i}\not= 0$, which lead to the so-called lightlike noncommutativity defined in Ref. \cite{Aharony:2000gz}. One may combine both spacelike and lightlike into the following set of requirements for unitarity:
\begin{equation}
{\bf B}^2_{\theta} \geq {\bf E}^2_{\theta}\;\:;\:\;{\bf B}_{\theta} \cdot {\bf E}_{\theta}=0.
\label{likelightunitar}
\end{equation} 
The lightlike noncommutativity takes place when the equality holds in the first part of Eq. \eqref{likelightunitar} otherwise, the conditions allow a boost to $B_\theta\neq E_\theta=0$, i.e. the spacelike noncommutativity.


\section{SEIBERG-WITTEN MAP INVARIANT SCATTERING AMPLITUDES}

\subsection{Electron and positron processes}

In accord with Figs.\ref{fig:FD4}-\ref{fig:FD6} here we use the two-by-two kinematics 
${\bf1}(k_1)+{\bf2}(k_2)\to {\bf3}(k_3)+{\bf4}(k_4)$, and 4-momentum conservation $k_1+k_2= k_3+k_4$ which defines Mandelstam variables for arbitrary $2\to2$ scattering:
\begin{equation}
s=(k_1+k_2)^2=(k_3+k_4)^2,\; t=(k_1-k_4)^2=(k_3-k_2)^2, \;u=(k_2-k_4)^2=(k_1-k_3)^2.
\label{Mandelstam}
\end{equation}

\subsubsection{M\o ller scattering: $e^-e^-\to e^-e^-$}

\begin{figure*}[t]
\begin{center}
\includegraphics[width=14cm,angle=0]{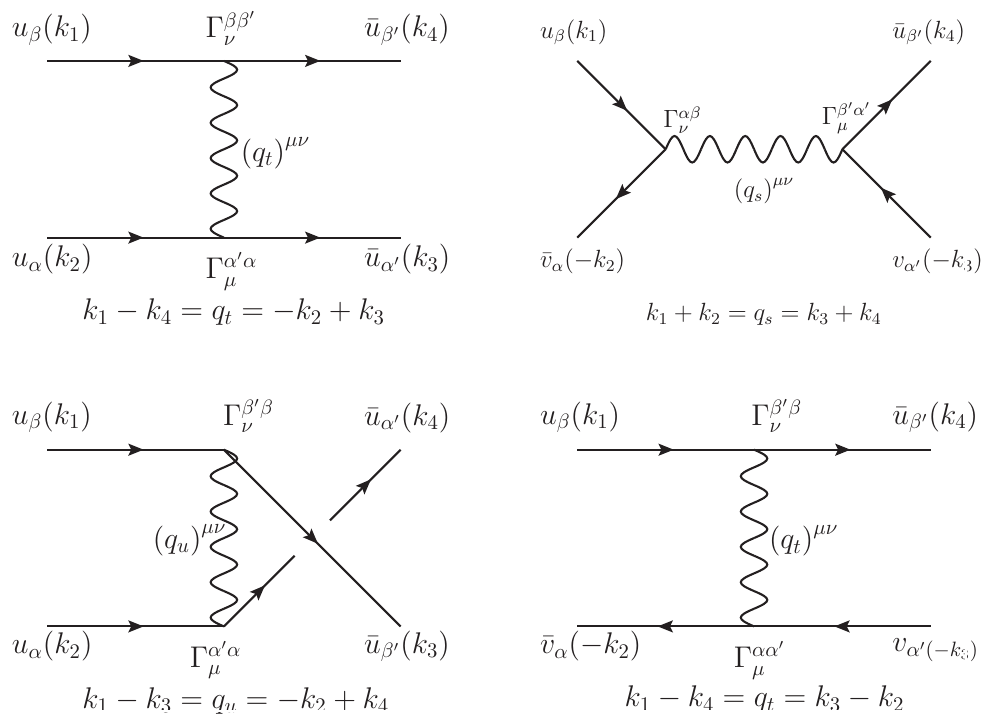}
\end{center}
\caption{Left panel: Feynman diagrams contributing to the elastic M\o ller scatterings 
$e^-(k_1)e^-(k_2) \to e^-(k_3)e^-(k_4)$. Right panel: Feynman diagrams contributing to the elastic Bhabha scatterings $e^-(k_1)e^+(k_2) \to e^+(k_3)e^-(k_4)$. Both are presented with EOM and on-shell conditions from Peskin and Schroeder \cite{Peskin:1995ev}. Momenta $k_i,i=1,2,3,4,$ placed in the counterclockwise way start in the left upper corner, while definitions of respecting Mandelstam variable and channels $(s,t,u)$ are given in Eq. (\ref{Mandelstam}).}
\label{fig:FD4}
\end{figure*}
First, we examine the NC effects arising from the contributions relevant to the M\o ller amplitude ${\cal M}^{\rm M}$. Summing  diagrams in Fig.\ref{fig:FD4} by using Feynman rules (\ref{FRpee1}) and/or (\ref{FRpee2}) we read out the following $(S-1)\equiv S^M$ matrix element:
\begin{equation}
S^{\rm M}\big(e^-(k_1)\;e^-(k_2) \to e^-(k_3)\;e^-(k_4)\big)=(2\pi)^4\;\delta^{(4)}(k_1+k_2-k_3-k_4) \;i
{\cal M}^{\rm M}_{I+II}
\label{Moller}
\end{equation}
Here ${\cal M}^{\rm M}_{I+II}$ represents sum of the non-SW($I$) map-induced 
and the SW$(II)$ map-induced contributions to the amplitude, respectively. 
Performing trivial integrations over the $\delta$ functions and splitting amplitudes 
into non-SW mapped NCQED $({\cal M}_{It}, {\cal M}_{Iu}$) and 
the SW mapped parts of NCQED $({\cal M}_{IIt}, {\cal M}_{IIu})$, respectively, 
we have the following M\o ller scattering amplitude ${\cal M}^{\rm M}_{I+II}$:
\begin{eqnarray}
{\cal M}^{\rm M}_{I+II}&=&{\cal M}_t+{\cal M}_u=
({\cal M}_{I}+{\cal M}_{II})_t+({\cal M}_{I}+{\cal M}_{II})_u,
\label{calM1+2}\\
{\cal M}_t
&=&
{\bar u}(k_4)\,\Gamma_{I+II}^\mu(-q_t,k_1)\,u(k_1)
\frac{g_{\mu\nu}}{q_t^2}\;
{\bar u}(k_3)\,\Gamma_{I+II}^\nu(q_t,k_2)\,u(k_2),
\label{calM1}\\
{\cal M}_u
&=&-{\bar u}(k_3)\,\Gamma_{I+II}^\mu(-q_u,k_1)\,u(k_1)
\frac{g_{\mu\nu}}{q_u^2}\;
{\bar u}(k_4)\,\Gamma_{I+II}^\nu(q_u,k_2)\,u(k_2),
\label{calM2}
\end{eqnarray}
with vertices $\Gamma_I$ and $\Gamma_{II}$ being given in Appendix A as FRs (\ref{FRpee1}( and (\ref{FRpee2}), 
while propagators we are taken from Ref. \cite{Peskin:1995ev}, respectively. 
So the above amplitude ${\cal M}_{It}$ represents the $t$-channel contribution 
from the NCQED without SW map, while ${\cal M}_{IIt}$ stands for 
the $t$-channel amplitude arising in the NCQED theory with the SW map. 
The same notations are going to be used for $s$- and $u$-channel amplitudes, further on.

Using detailed vertices from (\ref{FRpee1}) and (\ref{FRpee2}) and applying them in amplitudes (\ref{calM1}) and (\ref{calM2}), we find that all the SW map-induced NCQED parts of amplitudes and/or diagrams in the left part of Fig.\ref{fig:FD4}, 
including both the pure NC $\Gamma_{II\mu}\times \Gamma^\mu_{II}$ and the cross terms of the type 
$\Gamma_{I\mu}\times \Gamma^\mu_{II}$ respectively, due to the free field equations vanish on shell trivially, 
\begin{equation}
\big({\cal M}_{IIt}\big)_{(\ref{calM1})}=\big({\cal M}_{IIu}\big)_{(\ref{calM2})}=0,
\label{callMM}
\end{equation}
giving the SM result which is shifted only by the noncommutative phases:
\begin{eqnarray}
{\cal M}_{It}&=&e^2
\;e^{-\frac{i}{2}(k_4\theta k_1+k_3\theta k_2)}\frac{1}{t}\;{\bar u}(k_4)\gamma^\mu u(k_1)\;{\bar u}(k_3)\gamma_\mu u(k_2),
\label{calM1M}\\
{\cal M}_{Iu}&=&-e^2
\;e^{\frac{i}{2}(k_4\theta k_2+k_3\theta k_1)}\frac{1}{u}\;{\bar u}(k_4)\gamma_\nu u(k_2)\;{\bar u}(k_3)\gamma^\nu u(k_1).
\label{calM2M}
\end{eqnarray}
This property was first observed for fermions in the adjoint representation of the NC $\rm U(1)$ some years ago~\cite{Horvat:2011iv}.

For the in-out momentum conservation $\delta^{(4)}(k_1+k_2-k_3-k_4)$ one can show the following relations, in general:
\begin{equation}
({\it 1}.)\;\;k_1\theta k_3-k_4\theta k_2=k_3\theta k_2+k_4\theta k_1\;\; {\rm and}\;\;
({\it 2}.)\;\;k_1\theta k_2-k_4\theta k_3=k_3\theta k_2-k_4\theta k_1.
\label{formsI;II}
\end{equation}
Using the relation $({\it 1}.)$ above one can show that NCQED M\o ller scattering amplitude shares the same permutation symmetry as its QED counterpart and respects the fermion statistics accordingly.

\subsubsection{Bhabha scattering: $e^-e^+\to e^-e^+$}

Second, we examine the NC effects arising from the contributions of the two relevant diagrams to the Bhabha amplitude ${\cal M}^{\rm B}$. From the sum of right diagrams in Fig.\ref{fig:FD4} we read out the following $(S-1)$ matrix element:
\begin{equation}
S^{\rm B}\big(e^+(k_2)e^-(k_1)\to e^+(k_3)e^-(k_4)\big)=(2\pi)^4\;\delta^{(4)}(k_1+k_2-k_3-k_4) \;i{\cal M}^{\rm B}_{I+II}
\label{Bhabha}
\end{equation}
Trivial integrations over the $\delta$ functions and splitting amplitudes into parts coming 
from non-SW mapped NCQED $({\cal B}_{It}, {\cal B}_{Iu}$) 
from the SW mapped parts of NCQED $({\cal B}_{IIt}, {\cal B}_{IIs})$, respectively, 
we have the following Bhabha scattering amplitude ${\cal M}_{I+II}^{\rm B}$:
\begin{eqnarray}
&&{\hspace{-1cm}}{\cal M}_{I+II}^{\rm B}={\cal B}_t+{\cal B}_s=
({\cal B}_{I}+{\cal B}_{II})_t+({\cal B}_{I}+{\cal B}_{II})_s,
\label{calB1+2}\\
&&{\hspace{-0.2cm}}{\cal B}_t=-{\bar v}(-k_2)\,\Gamma_{I+II}^\mu(q_t,-k_3)\,v(-k_3)
\frac{g_{\mu\nu}}{q_t^2}\;{\bar u}(k_4)\,\Gamma_{I+II}^\nu(-q_t,k_1)\,u(k_1),
\label{calB1}\\
&&{\hspace{-0.2cm}}{\cal B}_s=
{\bar v}(-k_2)\,\Gamma_{I+II}^\mu(-q_s,k_1)\,u(k_1)
\frac{g_{\mu\nu}}{q_s^2}\;{\bar u}(k_4)\,\Gamma_{I+II}^\nu(q_s,-k_3)\,v(-k_3).
\label{calB2}
\end{eqnarray}

Using vertices from (\ref{FRpee1}) and (\ref{FRpee2}) and applying in the amplitudes (\ref{calB1}) and (\ref{calB2}) we show, due to the free field equations, the trivial on-shell vanishing of the SW map induced amplitudes 
\begin{equation}
\big({\cal B}_{IIt}\big)_{(\ref{calB1})}=\big({\cal B}_{IIs}\big)_{(\ref{calB2})}=0,
\label{callMB}
\end{equation}
and producing the SM result shifted by the noncommutative phases:
\begin{eqnarray}
{\cal B}_{It}&=&-e^2
\;e^{-\frac{i}{2}(k_4\theta k_1-k_3\theta k_2)}\frac{1}{t}\;{\bar u}(k_4)\gamma_\mu u(k_1)\;{\bar v}(-k_2)\gamma^\mu v(-k_3),
\label{calM1B}\\
{\cal B}_{Is}&=&e^2
\;e^{\frac{i}{2}(k_4\theta k_3+k_2\theta k_1)}\frac{1}{s}\;{\bar u}(k_4)\gamma_\nu v(-k_3)\;{\bar v}(-k_2)\,\gamma^\nu\,u(k_1).
\label{calM2B}
\end{eqnarray}

\subsection{Electron and photon processes}

\subsubsection{Compton scattering: $\gamma \,e \to \gamma\,e$}

\begin{figure}[t]
\begin{center}
\includegraphics[width=11cm,angle=0]{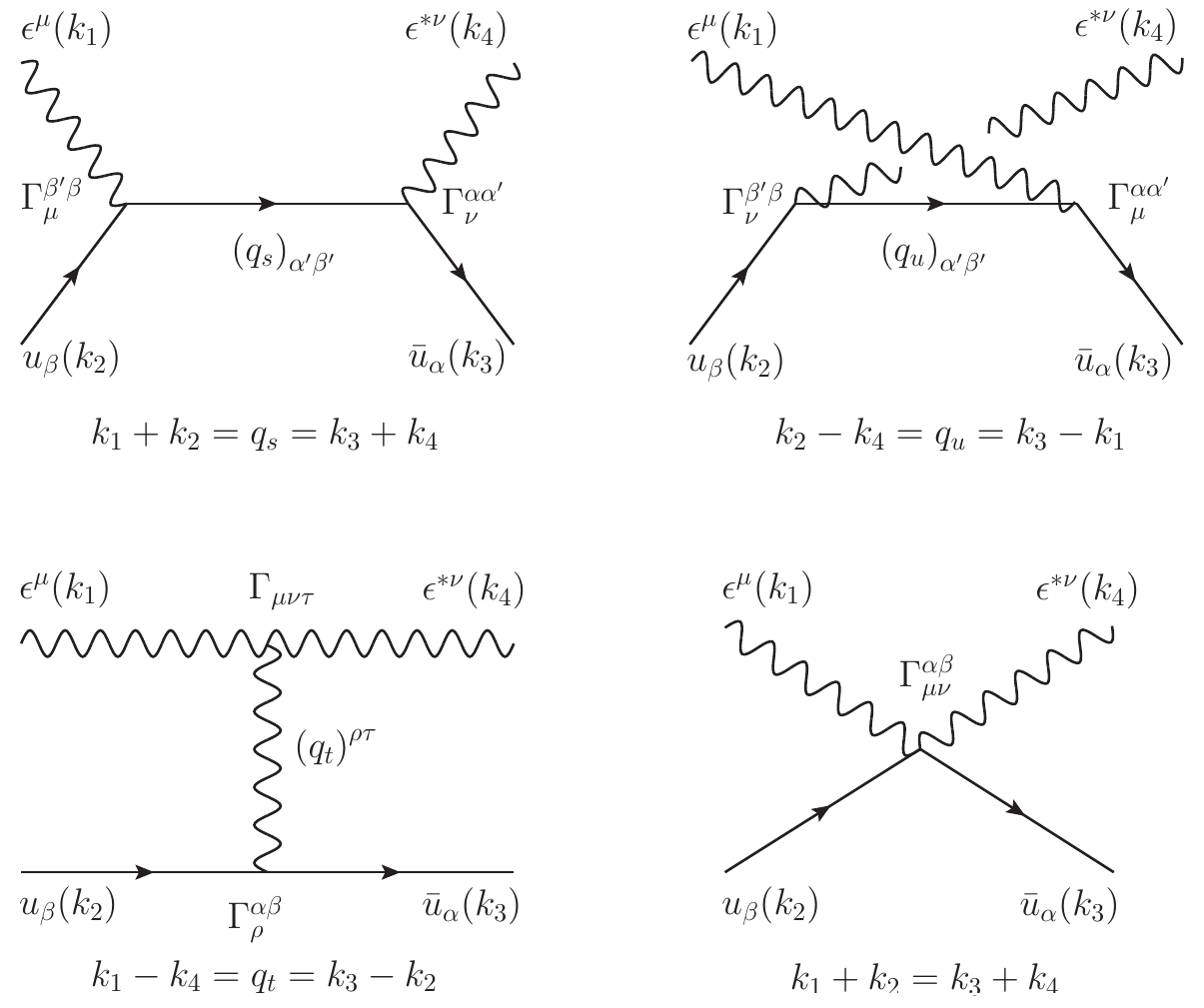}
\end{center}
\caption{Feynman diagrams contributing to the Compton scatterings $e^-\gamma \to e^-\gamma$, 
with free field conditions as usual from Peskin and Schroeder \cite{Peskin:1995ev}. Momenta $k_i,\;i=1,2,3,4,$ 
placed  in the counterclockwise way start in the left upper corner, with respecting 
Mandelstam variable or channels $(s,t,u)$ being given in Eq. (\ref{Mandelstam}).}
\label{fig:FD1}
\end{figure}
From the sum of diagrams in Fig.\ref{fig:FD1} we read out the following Compton scattering 
$(S-1)\equiv S^{\rm C}$ matrix element:
\begin{equation}
S^{\rm C}\big(\gamma(k_1)\,e^-(k_2) \to e^-(k_3)\,\gamma(k_4)\big)=(2\pi)^4\;\delta^{(4)}(k_1+k_2-k_3-k_4)
 \;i{\cal M}^{\rm C}_{I+II}
\label{ComptS}
\end{equation}
with the total amplitude ${\cal M}^{\rm C}_{I+II}$ being split into the non-SW$(I)$ and the SW($II$) map-induced contributions to the electron and photon interactions and/or Compton scattering amplitude, respectively.  
Feynman rules for vertices $(\Gamma_{I+II}^\mu$)'s being given in Appendix A, [Eqs. (\ref{FRpee1})-(\ref{FReepp})], while propagators are defined in Ref. \cite{Peskin:1995ev}, respectively. 
After performing trivial integrations over $\delta$ functions we obtain the following sum of amplitudes:
\begin{equation}
\begin{split}
{\cal M}^{\rm C}_{I+II}&=\epsilon_{\mu}(k_1)\epsilon^{\ast}_{\nu}(k_4){\bar u}(k_3){\rm C}^{\mu\nu}_{I+II}u(k_2),\;\;\bar{{\rm C}}=\gamma_0{{\rm C}^\dagger}\gamma_0,
\\
{\rm C}^{\mu\nu}_{I+II}&=-i\,\Bigg[\Gamma_{I+II}^\mu \frac{\slashed{k_1}+\slashed{k_2}+m_e}{q_s^2-m_e^2}\;
\Gamma_{I+II}^{\nu}+\Gamma_{I+II}^{\nu} \frac{\slashed{k_2}-\slashed{k_4}+m_e}{q_u^2-m_e^2}\;\Gamma_{I+II}^\mu
-\frac{1}{q_t^2}\Gamma_{I+II\,\rho}^{{}}\,\Gamma_{I+II}^{\mu\nu\rho}+\Gamma_{I+II}^{\mu\nu}\Bigg],
\\
{\cal M}^{\rm C}_{I+II}&=
{\cal M}^{\rm C}_{Is}+{\cal M}^{\rm C}_{It}+{\cal M}^{\rm C}_{Iu}+{\cal M}^{\rm C}_{I4}+{\cal M}^{\rm C}_{IIs}
+{\cal M}^{\rm  C}_{IIt}+{\cal M}^{\rm C}_{IIu}+{\cal M}^{\rm C}_{II4},
\end{split}
\label{AmpCtot}
\end{equation}
where $\gamma_0{\Gamma_\mu^\dagger}\gamma_0=\Gamma_\mu^{\ast}$,
$\gamma_0{\Gamma_{\mu\nu}^\dagger}\gamma_0=\Gamma^{\ast}_{\mu\nu}$, and
$\gamma_0{\Gamma_{\mu\nu\rho}^\dagger}\gamma_0=\Gamma^{\ast}_{\mu\nu\rho}$.
Above, amplitude ${\cal M}^{\rm C}_{Is}$ represents the $s$-channel contribution 
from the NCQED without SW map, while ${\cal M}^{\rm C}_{IIs}$ 
stands for the $s$-channel amplitude induced into the NCQED by the SW($II$) map. 
The same is valid for the $u$- and $t$-channels, respectively. For instance, only the first term 
$\Gamma^{\mu}_{I}$ in (\ref{FRpee1}) represents the Feynman rule for the NCQED with non-SW($I$) 
map-induced terms, while the additional two terms in Eq. (\ref{FRpee1}) arise due to the SW($II$) map. 
Second, only the first line $V^{\mu_1\mu_2\mu_3}_{I}$ of Eq. (\ref{FgA}), 
represents the FR for the NCQED theory with non-SW($I$) map-induced contributions. 
Other notations of the diagrams in Fig.\ref{fig:FD1} and corresponding contributions  given 
in Eqs. (\ref{AmpCtot}) are self-evident, and Ward identities in matter sector being satisfied. 

Note that the four-field-vertex contact term in Fig.\ref{fig:FD1} does not exists without the SW$(I)$ map, 
i.e., only the nonvanishing part is induced by the SW$(II)$ map. In Appendix A, this is anticipated by writing explicitly in Eq. (\ref{FReepp}) vanishing of that contact term part on shell due to the free field equations: 
$(\Gamma_{I}^{\mu\nu})_{\alpha\beta}=0$. Thus, we called it as irrelevant. However, the SW($II$) map-induced term $(\Gamma_{II}^{\mu\nu})_{\alpha\beta}\not= 0$ in Eq. (\ref{FReepp}) is necessary to cancel 
the SW($II$) map-induced peaces in the other three diagrams in Fig.\ref{fig:FD1}. 
Next, we show that by explicit computations of $s$-, $u$-, and $t$-channel 
three-field-vertex diagrams from Fig.\ref{fig:FD1} and Eqs. (\ref{AmpCtot}) for the on-shell case.
\begin{eqnarray}
{\cal M}^{\rm C}_{Is}&=&-ie^2\epsilon_{\mu}(k_1)\epsilon_\nu^{\ast}(k_4)
e^{-\frac{i}{2}(k_1\theta k_2-k_4\theta k_3)}{\bar u}(k_3)\gamma^{\mu}
\frac{\slashed{k_1}+\slashed{k_2}+m_e}{(k_1+k_2)^2-m_e^2}\gamma^{\nu} u(k_2),
\label{MsI}\\
{\cal M}^{\rm C}_{IIs}&=&-ie^2\epsilon_{\mu}(k_1)\epsilon_\nu^{\ast}(k_4)
{\bar u}(k_3)\Big[\frac{i}{2}e^{\frac{i}{2}(k_4\theta k_3)}F_\bullet(k_1,k_2) \gamma^{\mu}(\theta k_2)^{\nu}
\label{MsII}\\
&-&\frac{i}{2}e^{-\frac{i}{2}(k_1\theta k_2)}F_\bullet(-k_4,k_3) \gamma^{\nu}
(\theta k_3)^{\mu}
+\frac{1}{4}F_\bullet(-k_4,k_3)F_\bullet(k_1,k_2)\slashed{k_1}(\theta k_3)^{\mu}(\theta k_2)^{\nu}\Big] u(k_2),
\nonumber\\
{\cal M}^{\rm C}_{It}&=&\frac{4e^2}{(k_3-k_2)^2}\epsilon_{\mu}(k_1)\epsilon_\nu^{\ast}(k_4)e^{-\frac{i}{2}(k_3\theta k_2)}\sin{\frac{k_1\theta k_4}{2}}{\bar u}(k_3)
\Big[g^{\nu\mu}\slashed{k_1}-\gamma^{\nu} k_1^{\mu}-{k_4}^{\nu}\gamma^{\mu}\Big]u(k_2),
\nonumber\\
\label{MtI}\\
{\cal M}^{\rm C}_{IIt}&=&e^2\epsilon_{\mu}(k_1)\epsilon_\nu^{\ast}(k_4)
e^{-\frac{i}{2}(k_3\theta k_2)}F_{\star_2}(k_1,-k_4){\bar u}(k_3)
\Big[\theta^{\nu\mu}\slashed{k_1}+\gamma^{\nu}(\theta k_1)^{\mu}
-(\theta k_4)^{\nu}\gamma^{\mu}  \Big]u(k_2),
\nonumber\\
\label{MtII}\\
{\cal M}^{\rm C}_{Iu}&=&-ie^2\epsilon_{\mu}(k_1)\epsilon_\nu^{\ast}(k_4)
e^{-\frac{i}{2}(k_1\theta k_3-k_4\theta k_2)}{\bar u}(k_3)\gamma^{\mu}
\frac{\slashed{k_2}-\slashed{k_4}+m_e}{(k_2-k_4)^2-m_e^2}\gamma^{\nu} u(k_2),
\label{MuI}\\
{\cal M}^{\rm C}_{IIu}&=&-ie^2\epsilon_{\mu}(k_1)\epsilon_\nu^{\ast}(k_4)
{\bar u}(k_3)\Big[-\frac{i}{2}e^{\frac{i}{2}(k_4\theta k_2)}F_\bullet(k_1,k_3)
(\theta k_3)^{\nu}\gamma^{\mu}
\label{MuII}\\
&+&\frac{i}{2}e^{-\frac{i}{2}(k_1\theta k_3)}F_\bullet(-k_4,k_2)
(\theta k_2)^{\mu}\gamma^{\nu}
-\frac{1}{4}F_\bullet(k_1,k_3)F_\bullet(-k_4,k_2)\slashed{k_4}(\theta k_3)^{\nu}(\theta k_2)^{\mu}\Big] u(k_2).
\label{MtB}
\nonumber
\end{eqnarray}
Finally from Appendix A using FR (\ref{FReepp}), in the notation of the fourth diagram in Fig.\ref{fig:FD1}, we have
\begin{gather}
{\cal M}^{\rm C}_{I4}=0,
\label{4fvI}
\\
{\cal M}^{\rm C}_{II4}={\cal M}^{\rm C}_{II4s}+{\cal M}^{\rm C}_{II4t}+{\cal M}^{\rm C}_{II4u},
\label{4fvII}
\end{gather}
where
\begin{gather}
\begin{split}
{\cal M}^{\rm C}_{II4s}=&\frac{e^2}{2}\epsilon_{\mu}(k_1)\epsilon_\nu^{\ast}(k_4){\bar u}(k_3)\Big[\gamma^{\nu}(\theta k_3)^{\mu}e^{-\frac{i}{2}(k_1\theta k_2)}F_\bullet(-k_4,k_3)-(\theta k_2)^{\nu}\gamma^{\mu}e^{\frac{i}{2}(k_4\theta k_3)}F_\bullet(k_1,k_2)
\\&+\frac{i}{2}F_\bullet(-k_4,k_3)F_\bullet(k_1,k_2)(\theta k_2)^{\nu}(\theta k_3)^{\mu} \slashed{k_1}\Big]u(k_2),
\\
{\cal M}^{\rm C}_{II4t}=&\frac{e^2}{2}\epsilon_{\mu}(k_1)\epsilon_\nu^{\ast}(k_4){\bar u}(k_3)\Big[2\big((\theta k_4)^{\nu}\gamma^{\mu}-\gamma^{\nu}
(\theta k_1)^{\mu}\big)-\theta^{\nu\mu}\big(\slashed{k_1}+\slashed{k_4}\big)\Big]e^{-\frac{i}{2}(k_3\theta k_2)}F_{\star_2}(k_1,-k_4)
u(k_2),
\\
{\cal M}^{\rm C}_{II4u}=&\frac{e^2}{2}\epsilon_{\mu}(k_1)\epsilon_\nu^{\ast}(k_4){\bar u}(k_3)\Big[(\theta k_3)^{\nu}\gamma^{\mu}e^{\frac{i}{2}(k_4\theta k_2)}F_\bullet(k_1,k_3)-\gamma^{\nu}(\theta k_2)^{\mu}e^{-\frac{i}{2}(k_1\theta k_3)}F_\bullet(-k_4,k_2)
\\&-\frac{i}{2}F_\bullet(k_1,k_3)F_\bullet(-k_4,k_2)(\theta k_3)^{\nu}(\theta k_2)^{\mu} \slashed{k_4}\Big]u(k_2).
\end{split}
\end{gather}
It can then be shown that
\begin{equation}
{\cal M}^{\rm C}_{IIs}+{\cal M}^{\rm C}_{II4s}={\cal M}^{\rm C}_{IIt}+{\cal M}^{\rm C}_{II4t}={\cal M}^{\rm C}_{IIu}+{\cal M}^{\rm C}_{II4u}=0.
\label{ComptSW(II)}
\end{equation}
Thus, the sum of the SW map-induced contributions to the on-shell Compton scattering amplitude vanishes.

\subsubsection{Dilepton pair annihilation$\;|\;$production: $\ell^+ \ell^-\to\gamma\gamma\;|\;\gamma\gamma\to\ell^+ \ell^-$}

\begin{figure}[t]
\begin{center}
\includegraphics[width=8cm,angle=0]{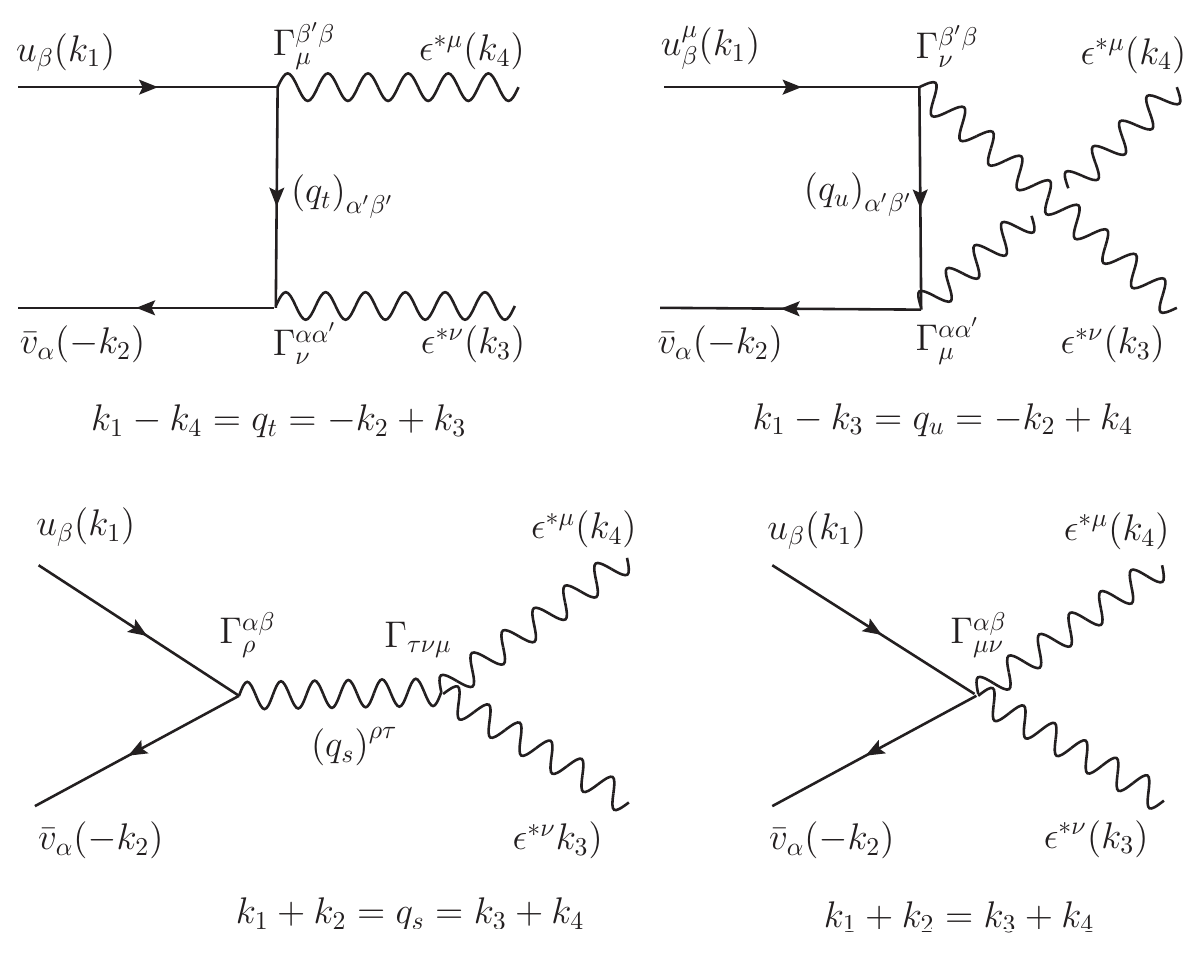}\;\;\;\;\;\;\;\;\;\;\;\:
\includegraphics[width=8cm,angle=0]{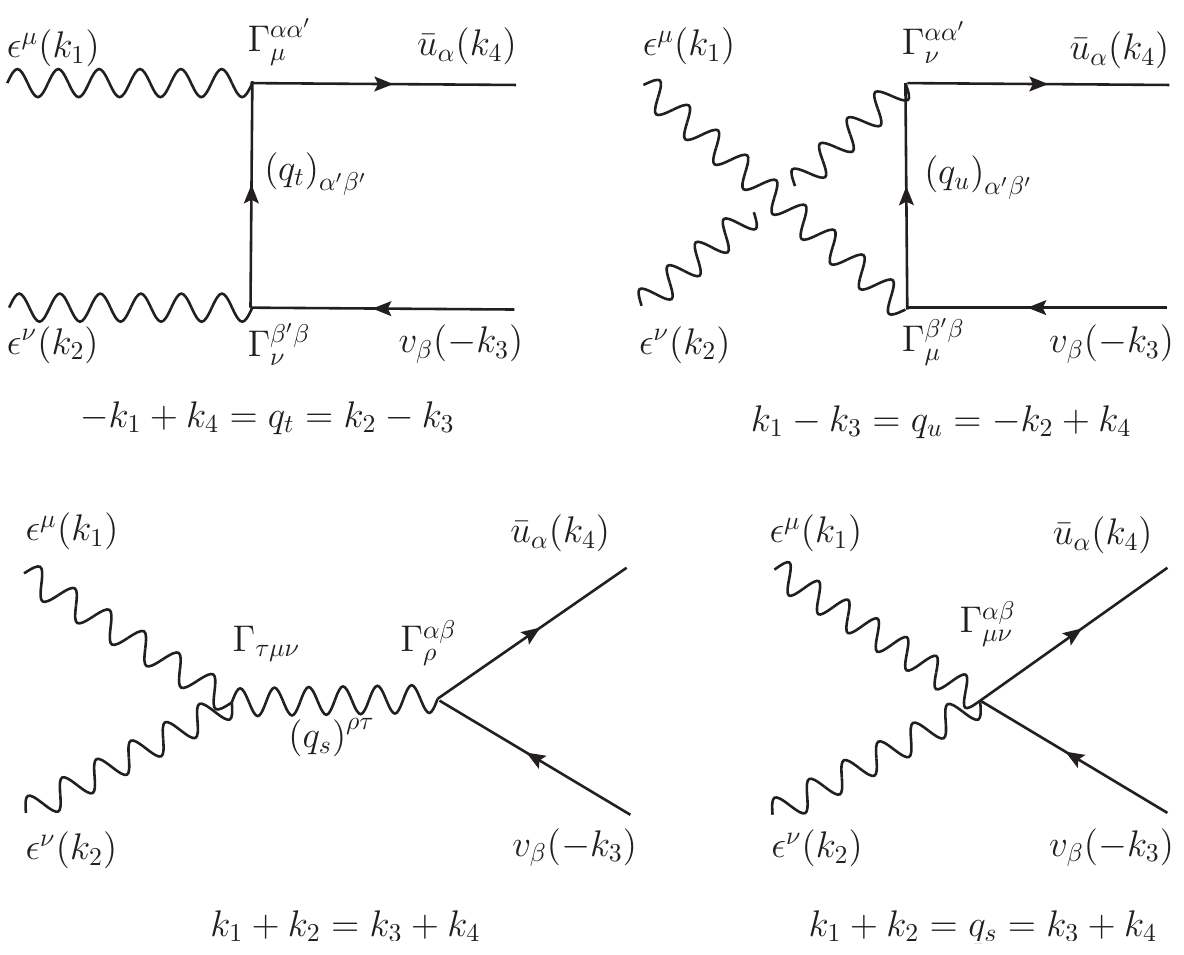}
\end{center}
\caption{Left panel: Feynman diagrams contributing to the $e^-(k_1)e^+(k_2) \to \gamma(k_3)\gamma(k_4)$ 
pair annihilation. Right panel: diagrams contributing to the $\gamma(k_1)\gamma(k_2) \to e^-(k_3)e^+(k_4)$ pair production. Free field conditions are used from Peskin and Schroeder \cite{Peskin:1995ev}.}
\label{fig:FD2}
\end{figure}
From the sum of diagrams, for annihilation (left panel in Fig.\ref{fig:FD2}), and for production (right panel in Fig.\ref{fig:FD2}), we read out the following $(S-1)$ matrix elements, i.e. the ${\cal M}^{\rm A}$ and 
${\cal M}^{\rm P}$ amplitudes, respectively:
\begin{equation}
S^{\rm A}\big(e^-(k_1)\;e^+(k_2) \to \gamma(k_3)\;\gamma(k_4)\big)
=(2\pi)^4  \;\delta^{(4)}(k_1+k_2-k_3-k_4) \;i{\cal M}^{\rm A}_{I+II}\\
\label{ComptSe+e-}
\end{equation}
\begin{equation}
S^{\rm P}\big(\gamma(k_1)\;\gamma(k_2) \to e^-(k_3)\;e^+(k_4)\big)=(2\pi)^4 \;
\delta^{(4)}(k_1+k_2-k_3-k_4) \;i{\cal M}^{\rm P}_{I+II}
\label{ProdSgg}
\end{equation}

Because of the same topological and Lorentz structures of contributing diagrams in left and right panels in Fig.\ref{fig:FD2} as in the Compton scattering (Fig.\ref{fig:FD1}), the same conclusion of vanishing the SW$(II)$ map-induced contributions is valid for the pair annihilation and pair production processes, respectively. 
So those, together with (\ref{callMM}), (\ref{callMB}), and (\ref{ComptSW(II)}), 
represent the extraordinary property of SW maps in the matter sector of NCQED. 
The same property we shall prove for the pure NCQED gauge sector.

\subsection{Photon-photon process}
\begin{figure}[t]
\begin{center}
\includegraphics[width=14cm,angle=0]{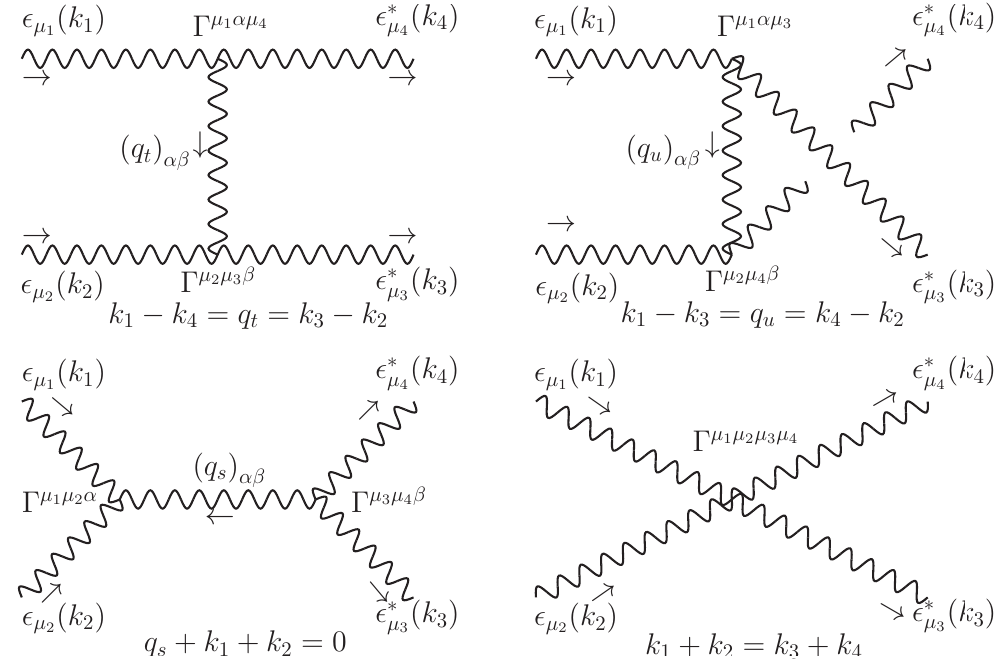}
\end{center}
\caption{Feynman diagrams contributing to the exclusive LbyL scatterings $\gamma(k_1)\gamma(k_2)\to \gamma(k_3)\gamma(k_4)$, with free field conditions from Peskin and Schroeder \cite{Peskin:1995ev}. 
Separated arrows indicate the flow of momenta. Outer momenta $k_i,i=1,2,3,4,$ are placed in the counterclockwise way, with respecting definitions of Mandelstam variable/channels $(s,t,u)$ being given in Eq. (\ref{Mandelstam}).
}
\label{fig:FD6}
\end{figure}
In this section we present noncommutative effects arising from the contributions of 
the four relevant diagrams to the $\gamma\gamma\to\gamma\gamma$ 
scattering amplitude ${\cal M}^\gamma$ from Fig.\ref{fig:FD6}. 
We shall further on consider two different basis to express our total amplitude after summing up diagrams. 
Those are, first the amplitudes in terms of the Lorentz decomposition basis. 
Second are the photon amplitudes expressed in the helicity basis, 
along the analysis of  \cite{Raju:2009yx,Huang:2010fc,Arkani-Hamed:2017mur}, i.e., on the line of the 
scattering amplitude ideas in recent trends of theoretical physics progress. 

Lorentz decomposition is important for proof the SW$(II)$ map invariance of 
scattering amplitudes on shell, respectively. 
Now note that after or when proving of vanishing of the SW$(II)$ map-induced contributions, 
the Ward identities have to be shown only for the non-SW($I$) map induced parts of the scattering amplitudes, which we shall actually perform first in the next subsections.
\begin{equation}
S^\gamma\big(\gamma(k_1)\gamma(k_2)\to\gamma(k_3)\gamma(k_4)\big)
=(2\pi)^4\delta^{(4)}(k_1+k_2-k_3-k_4) \;i{\cal M}^\gamma_{I+II}
\label{4g}
\end{equation}
Here each vertex $\Gamma$ in the diagrams is the sum of the non-SW$(I)$ map, and SW$(II)$ map-induced relevant parts, from FRs (\ref{Fg})--(\ref{F.W3}). By splitting (\ref{4g}) into non-SW $(\Gamma_I)$ and SW $(\Gamma_{II})$ parts, we have the following amplitude for the sum of $s-, t-$, and $u$-channel three-photon diagrams, plus the four-photon one as:
\begin{equation}
\begin{split}
&{\cal M}^\gamma_{I+II}
= \epsilon_{\mu_1}(k_1)\epsilon_{\mu_2}(k_2)
\epsilon^{\ast}_{\mu_3}(k_3){\epsilon}^{\ast}_{\mu_4}(k_4) 
\\
&
\cdot\bigg\{ \Gamma_{I+II}^{\mu_1\mu_2\alpha}(k_1,k_2,q_s)\frac{-ig_{\alpha\beta}}{q^2_s}
\Gamma_{I+II}^{\mu_3\mu_4\beta}(-k_3,-k_4,-q_s)
+\Gamma_{I+II}^{\mu_1\alpha\mu_4}(k_1,-q_t,-k_4)\;\frac{-ig_{\alpha\beta}}{q^2_t}
\Gamma_{I+II}^{\mu_2\mu_3\beta}(k_2,-k_3,q_t)
\\
&\phantom{,}
+\Gamma_{I+II}^{\mu_1\alpha\mu_3}(k_1,-q_u,-k_3)\frac{-ig_{\alpha\beta}}{q^2_u}
\Gamma_{I+II}^{\mu_2\mu_4\beta}(k_2,-k_4,q_u)
+\Gamma_{I+II}^{\mu_1\mu_2\mu_3\mu_4}(k_1,k_2,-k_3,-k_4)\bigg\} 
=i{\cal M}^{3\gamma}_{I+II}+i{\cal M}^{4\gamma}_{I+II}.
\end{split}
\label{4gstu}
\end{equation} 

\subsubsection{Non-SW($I$) map-induced $\gamma\gamma\to\gamma\gamma$ amplitudes and Ward identities}

\noindent
Total non-SW($I$) map-induced contributions are:
\begin{equation}
\begin{split}
i{\cal M}^\gamma_I&=i{\cal M}^{3\gamma}_{I}+i{\cal M}^{4\gamma}_{I}=
 \epsilon_{\mu_1}(k_1)\epsilon_{\mu_2}(k_2)
\epsilon^{\ast}_{\mu_3}(k_3){\epsilon}^{\ast}_{\mu_4}(k_4) 
\\ 
&\phantom{X}
\cdot\bigg\{\frac{1}{s}\;\Gamma_{I\;\;\;\;\;\;\alpha }^{\mu_1\mu_2}(k_1,k_2,q_s)\;\Gamma_I^{\mu_3\mu_4\alpha}(-k_3,-k_4,-q_s)
+\frac{1}{t}\Gamma_{I\;\;\;\;\;\;\alpha}^{\mu_1\mu_4}(k_1,-q_t,-k_4)\;\Gamma_I^{\mu_2\mu_3\alpha}(k_2,-k_3,q_t)
\\
&\phantom{xx}
+\frac{1}{u}\;\Gamma_{I\;\;\;\;\;\;\alpha}^{\mu_1\mu_3}(k_1,-q_u,-k_3)\;\Gamma_I^{\mu_2\mu_4\alpha}(k_2,-k_4,q_u)
+i\Gamma^{\mu_1\mu_2\mu_3\mu_4}_I(k_1,k_2,-k_3,-k_4)\bigg\}. 
\end{split}
\label{4gIstu}
\end{equation} 
Taking into account FRs and the decompositions of vertices, the total on-shell non-SW map-induced amplitude (\ref{4gIstu}) shall have the following form:
\begin{equation}
i {\cal M}^\gamma_I= a_1 \sin\frac{k_1\theta k_4}{2} \sin\frac{k_2\theta k_3}{2} + a_2 \sin\frac{k_1\theta k_3}{2} \sin\frac{k_2\theta k_4}{2} + a_3 \sin\frac{k_1\theta k_2}{2} \sin\frac{k_3\theta k_4}{2},
\label{MIstu}
\end{equation}
where the coefficients $a_i, i=1,2,3$ shall be determined later by explicit computations. One could simplify $i {{\cal M}^\gamma_I}$ by employing the following Moyal-Weyl $\star$-product Jacobi identity in momentum space:
\begin{equation}
\sin\frac{k_1\theta k_2}{2} \sin\frac{k_3\theta k_4}{2}-\sin\frac{k_1\theta k_3}{2} \sin\frac{k_2\theta k_4}{2}
+\sin\frac{k_2\theta k_3}{2}\sin\frac{k_1\theta k_4}{2} =0.
\label{JIstu}
\end{equation}
One of the momenta, $k_4$, has a fixed place, and the remaining three are cyclically permuted. 
The left side vanishes due to the momentum conservation and antisymmetry among two indices of 
$\theta^{\mu_1\mu_2}$. 

Note that the structure of our Jacobi identity in momentum space (\ref{JIstu}), generated by 
the Moyal-Weyl $\star$ product applied to the $s$-, $t$-, $u$-channel scattering amplitudes given in Fig.\ref{fig:FD6},  arises from color-kinematic correspondence between 
the U(1) and the $\rm U_\star(1)$ gauge theories  \cite{Raju:2009yx,Huang:2010fc,Arkani-Hamed:2017mur}. 
Taking into account the SW map, that is, after the $\rm {U_\star(1)}\stackrel{SW}{\longrightarrow}{U(1)}$, 
one has only usual U(1) gauge symmetry. The same is valid for the inverse SW map, thus, one could say 
that the structure (\ref{JIstu}) is invariant under the U(1) SW map. Finding concrete frameworks of 
the alternative derivation which provide a geometric interpretation of such structure, 
in color-kinematic correspondence or duality between gauge theory and perturbative gravity 
at the tree-level scattering amplitudes, was recently performed and published in \cite{Mizera:2019blq}.

Using Eqs. (\ref{4gIstu}) and (\ref{JIstu}) and FRs from Appendix A, we get usual Lorentz decomposition 
of the on-shell non-SW map amplitude $({\cal M}^\gamma_I\big|_{Lorentz})$ in the following form:
\begin{equation}
\begin{split}
&{\cal M}^\gamma_I\big|_{Lorentz}=\Big[{\cal M}^{3\gamma}_{I}+{\cal M}^{4\gamma}_{I}\Big]_{Lorentz}=-i4e^2 \epsilon_{\mu_1}(k_1)\epsilon_{\mu_2}(k_2)
\epsilon^{\ast}_{\mu_3}(k_3){\epsilon}^{\ast}_{\mu_4}(k_4) 
\\&
\cdot \bigg\{ \sin \frac{k_1\theta k_2}{2} \sin \frac{k_3 \theta k_4 }{2} \bigg[ g^{\mu_1\mu_2} g^{\mu_3\mu_4} + g^{\mu_1\mu_3} g^{\mu_2\mu_4} - 2 g^{\mu_1\mu_4} g^{\mu_2\mu_3}
\\ &
+ \frac{1}{s}\Big(2 k_2^{\mu_1} g_{\alpha}^{\mu_2}-2 k_1^{\mu_2} g_{\alpha}^{\mu_1} + (k_1-k_2)_\alpha g^{\mu_1\mu_2} \Big) \Big(2 k_3^{\mu_4} g^{\alpha\mu_3}-2 k_4^{\mu_3} g^{\alpha\mu_4} + (k_4-k_3)^\alpha g^{\mu_3\mu_4} \Big) 
\\ &
+ \frac{1}{u}\Big(2 k_3^{\mu_1} g_{\alpha}^{\mu_3}+2 k_1^{\mu_3} g_{\alpha}^{\mu_1} - (k_1+k_3)_\alpha g^{\mu_1\mu_3} \Big) \Big(2 k_2^{\mu_4} g^{\alpha\mu_2}+2 k_4^{\mu_2} g^{\alpha\mu_4} - (k_4 + k_2)^\alpha g^{\mu_2\mu_4} \Big) \bigg]
\\ & \phantom{.} -\sin \frac{k_1\theta k_4}{2} \sin \frac{k_2 \theta k_3}{2} \bigg[ 
g^{\mu_1\mu_3} g^{\mu_2\mu_4} + g^{\mu_1\mu_4} g^{\mu_2\mu_3}-2 g^{\mu_1\mu_2} g^{\mu_3\mu_4} 
\\ & 
- \frac{1}{t}\Big(2 k_1^{\mu_4} g_{\alpha}^{\mu_1} + 2 k_4^{\mu_1} g_{\alpha}^{\mu_4} - (k_1 + k_4)_\alpha g^{\mu_1\mu_4} \Big) \Big(2 k_2^{\mu_3} g^{\alpha\mu_2} + 2 k_3^{\mu_2} g^{\alpha\mu_3}- (k_2+k_3)^\alpha g^{\mu_2\mu_3}\Big) 
\\ &  
- \frac{1}{u}\Big(2 k_3^{\mu_1} g_{\alpha}^{\mu_3}+2 k_1^{\mu_3} g_{\alpha}^{\mu_1} - (k_1+k_3)_\alpha g^{\mu_1\mu_3} \Big) \Big(2 k_2^{\mu_4} g^{\alpha\mu_2}+2 k_4^{\mu_2} g^{\alpha\mu_4} - (k_2+k_4)^\alpha g^{\mu_2\mu_4} \Big)\bigg] \bigg\}. 
\label{ForGammaAmplitude2}
\end{split}
\end{equation}
From the above it is easy to show that Ward identities in the pure gauge sector are fulfilled too. Namely, with the replacement $\epsilon^{\mu_i}(k_i)\to k_i^{\mu_i}$ in Eq. (\ref{ForGammaAmplitude2}), we have found that  
$k_i^{\mu_i}\cdot\big({\cal M}^\gamma_I\big|_{Lorentz}\big)_{\mu_i} =0, \forall i=1,...,4$; 
i.e., it always vanishes,  as it should.\\
\phantom{from where it is easy to see that the Ward identities in the pure gauge sector are toXXXXXXXXXXXXXXX}Q.E.D.

\subsubsection{SW($II$) map-induced $\gamma\gamma\to\gamma\gamma$ amplitude}

Because of our first result for Compton scattering in momentum space (\ref{ComptSW(II)}), i.e. that 
the $\theta$-exact SW($II$) map-induced contributions from the sum of diagrams in Fig.\ref{fig:FD1} 
vanish nontrivilay, it was clear that we have to prove that the same property holds in the pure gauge sector of NCQED, i.e. that from Fig.\ref{fig:FD6} the sum of s-, t-, and u-channels 3-photon diagrams plus 4-photon term from (\ref{F.9}), vanish too. After extracting SW$(II)$ map induced amplitude ${\cal M}^\gamma_{II}\big|_{Lorentz}$ 
from (\ref{4gstu}) we have to show that: 

\begin{equation}
\begin{split}&
{\cal M}^\gamma_{II}\Big|_{Lorentz}=\Big[{\cal M}^{3\gamma}_{II}+{\cal M}^{4\gamma}_{II}\Big]_{Lorentz}
=-i \epsilon_{\mu_1}(k_1)\epsilon_{\mu_2}(k_2)\epsilon^{\ast}_{\mu_3}(k_3){\epsilon}^{\ast}_{\mu_4}(k_4)
\\&\phantom{.}
\cdot\bigg\{\frac{1}{s}\Big[\Gamma_{I\;\;\;\;\;\;\alpha}^{\mu_1\mu_2}(k_1,k_2,q_s)\;\Gamma_{II}^{\mu_3\mu_4\alpha}(-k_3,-k_4,-q_s)
+\Gamma_{II\;\;\;\;\alpha}^{\mu_1\mu_2}(k_1,k_2,q_s)\;\Gamma_{I}^{\mu_3\mu_4\alpha}(-k_3,-k_4,-q_s)
\\&\phantom{XX}
+\Gamma_{II\;\;\;\;\alpha}^{\mu_1\mu_2}(k_1,k_2,q_s)\;\Gamma_{II}^{\mu_4\mu_3\alpha}(-k_3,-k_4,-q_s)\Big]
\\&\phantom{x}
+\frac{1}{t}\Big[\Gamma_{I\;\;\;\;\;\;\alpha}^{\mu_1\mu_4}(k_1,-q_t,-k_4)\;\Gamma_{II}^{\mu_2\mu_3\alpha}(k_2,-k_3,q_t)
+\Gamma_{II\;\;\;\;\alpha}^{\mu_1\mu_4}(k_1,-q_t,-k_4)\;\Gamma_{I}^{\mu_2\mu_3\alpha}(k_2,-k_3,q_t)
\\&\phantom{XX}
+\Gamma_{II\;\;\;\;\alpha}^{\mu_1\mu_4}(k_1,-q_t,-k_4)\;\Gamma_{II}^{\mu_2\mu_3\alpha}(k_2,-k_3,q_t)\Big]
\\&\phantom{x}
+\frac{1}{u}\Big[\Gamma_{I\;\;\;\;\;\;\alpha}^{\mu_1\mu_3}(k_1,-q_u,-k_3)\;\Gamma_{II}^{\mu_2\mu_4\alpha}(k_2,-k_4,q_u)
+\Gamma_{II\;\;\;\;\alpha}^{\mu_1\mu_3}(k_1,-q_u,-k_3)\;\Gamma_{I}^{\mu_2\mu_4\alpha}(k_2,-k_4,q_u)
\\&\phantom{XX}
+\Gamma_{II\;\;\;\;\alpha}^{\mu_1\mu_3}(k_1,-q_u,-k_3)\;\Gamma_{II}^{\mu_2\mu_4\alpha}(k_2,-k_4,q_u)\Big]
+i\Gamma^{\mu_1\mu_2\mu_3\mu_4}_{II}(k_1,k_2,-k_3,-k_4) \bigg\}=0. 
\end{split}
\label{4gIIstu}
\end{equation} 

The underlying idea for the SW map invariance of scattering amplitudes is that for each vertex term in 
Feynman rules generated by the invertible SW map there exists a collection of diagrams to cancel it on shell. The general proof for such cancellation is already  sketched in our equivalence and duality papers \cite{Martin:2016hji,Martin:2016saw}. Proof of Eq. (\ref{4gIIstu}) is given in Appendix C.

\section{CROSS SECTIONS FOR ELECTRON AND POSITRON SCATTERINGS}

Completing SW maps for the gauge field (photon) and charged fermion field (electron), 
and constructing the minimal $\theta$-exact U(1) photon-electron action up to the second order 
in coupling constant ($e^2$), we apply  NCQED to the following collision processes 
\cite{Hewett:2000zp,Godfrey:2001yy,Altschul:2006pv,Garg:2011aa,Mathews:2000we,Baek:2001ty}: 
M\o ller $(e^-e^- \to e^-e^-)$, and Bhabha $(e^-e^+ \to e^-e^+)$ scatterings, 
Compton scattering $(e\gamma\to e\gamma)$ \cite{Godfrey:2001yy,Mathews:2000we}, 
dilepton pair annihilation ($ \ell^+ \ell^-\to\gamma\gamma$)\cite{Hewett:2000zp}, dilepton pair production 
($\gamma\gamma\to \ell^+ \ell^-)$ \cite{Baek:2001ty}, and photon-photon 
$(\gamma\gamma\to\gamma\gamma)$ scatterings \cite{Hewett:2000zp}, respectively.
We find that an oscillatory dependence is induced in these processes due to the NC matrix $\theta^{\mu\nu}$. In this article,  we shall compute unpolarized total cross sections for head-to-head collision processes within the same general frame --we call it the noncentral mass (NCM) frame-- and also in the central mass (CM) frame obtained form the NCM by taking appropriate limits, respectively. 

\subsection{M\o ller scattering cross section}

From the sum of matrix elements (\ref{calM1M}) and( \ref{calM2M}), with the square of total amplitude and after summing and averaging over spins for the massless unpolarized case, we find 
\begin{equation}
\frac{1}{4}{\sum\limits_{\rm spins}}\big|{\cal M}^{\rm M}\big|^2
=\frac{1}{4}{\sum\limits_{\rm spins}}\left|{\cal M}^{}_{It}+{\cal M}^{}_{Iu}\right|^2
=2e^4\bigg[\frac{s^2+u^2}{t^2}+\frac{s^2+t^2}{u^2}+2\frac{s^2}{tu}
\cos(k_3\theta k_2+k_4\theta k_1)\bigg].
\label{AmplitaM}
\end{equation}

Phases in the pure $t$- and $u$-channel terms of (\ref{AmplitaM}) cancel out, 
while total phases in the interference cross terms of Eq. 
(\ref{AmplitaM}) turn into the $\mp$ of the form $(\it 1.)$ in Eq. (\ref{formsI;II}), 
i.e., $\mp 2(k_3\theta k_2+k_4\theta k_1)$, respectively. 
From the above, one obtains the differential cross section in the head-to-head massless M\o ller scattering. 
Using (\ref{B3}) and (\ref{E4}), and after rearranging  Eq. (\ref{AmplitaM}) we obtain three terms representing QED contributions in agreement with Ref. \cite{Peskin:1995ev}, while the fourth one is coming from the NCQED phase factors, confirming the expression earlier given in Ref. \cite{Hewett:2000zp}.

To obtain the total cross section with the NC correction included we shall compute 
the phase space integral in a general --NCM-- frame of head-to-head collision. In the NCM frame after integration over $\vec k_3$, the delta function over energy $\omega_3$ then yields a differential cross section 
formula for an arbitrary unpolarised two-by-two process in the polar coordinate system: 
\begin{equation}
\frac{d\sigma}{d\Omega}\Bigg|_{\rm NCM}=
\Big(\frac{\omega_4}{4\pi s}\Big)^2\frac{1}{4}{\sum\limits_{\rm spins}}\big|{\cal M}\big|^2, \;\; 
d\Omega=\,\sin\vartheta d\vartheta d\varphi.
\label{TCrossSectC}
\end{equation}

Since we shall work out all processes for the massless case the cross section (\ref{TCrossSectC}) 
represents quite general form in the NCM frame. For relevant processes we just have to compute all terms in ${\sum\limits_{\rm spins}}\big|{\cal M}\big|^2$, and than integrate over the final state angles.

We next compute the M\o ller cross section of head-to-head collision for the massless case in the NCM frame defined in Eqs. (\ref{SphCoordN}). Using  Eq. (\ref{SphCoordphaseGener}), the relevant phase factor for M\o ller, Eqs (\ref{B3}) and (\ref{E4}), for the general case of space-time noncommutativity we obtain contributions to the M\o ller scattering, for all types of noncommutativity. 

Taking into account two identical particles in the final state by integrating over $d\Omega$ we obtain the  NC part of the 
M\o ller cross section in the NCM frame as a function of only incoming energies $\omega_{1,2}$ and the NC parameters $M$ and $\Lambda_{\rm NC}$, respectively: 
\begin{equation}
\begin{split}
&\sigma^{\rm M}_{\rm NC}\Big|_{\rm NCM}
=-\frac{4\pi\alpha^2}{s}\int^1_{-1}\frac{dx}{1-x^2}
\bigg[1-J_0\bigg(\frac{\omega_1\omega_4}{\Lambda^2_{\rm NC}}M\sqrt{1-x^2}\bigg)\bigg],\;x=\cos\vartheta,
\end{split}
\label{NCTCrossSectMnCM}
\end{equation}
with $M$ given in Eq. (\ref{E4}). Expanding the above cross section in the regime of small energies and large 
$\rm\Lambda_{NC}$, one obtains
\begin{equation}
\sigma^{\rm M}_{\rm NC}\Big|_{\rm NCM}
=-\frac{\pi\alpha^2M^2}{\Lambda^4_{\rm NC}}\int^{1}_{-1}dx\frac{\omega_1^3\omega_2}{\big(\omega_1(1-x)+\omega_2(1+x)\big)^2},
\label{NCTCrossSectMnMCSmall}
\end{equation}
which is regular at both $x\to \pm 1$ limits. Therefore the full destructive cross section~\eqref{NCTCrossSectMnCM} is regular at these limits too.

In the CM frame $\omega_1=\omega_2=\omega_4$; thus, $M\rightarrow2\tilde M$. For small argument expansion \eqref{NCTCrossSectMnCM} simplifies into an asymptotic cross section at low energies,  suppressed by the NC scale $\Lambda_{\rm NC}$ to the fourth power:
 \begin{equation}
\begin{split}
&\sigma^{\rm M}_{\rm NC}\Big|_{\rm CM}
\simeq-\frac{\pi\alpha^2s}{2\Lambda^4_{\rm NC}}\tilde M^2,\;\;\tilde M= \sqrt{{c_{13}^2+c_{23}^2}}.
\end{split}
\label{NCTCrossSectMCM} 
\end{equation}
We notice that $\tilde M$ is proportional to the length of the projection of the vector ${\bf\hat B}_\theta=\Lambda^2_{\rm NC}{\bf B}_\theta$ onto the plane transverse to the incoming particle axis. So for the same length of ${\bf B}_\theta$ the noncommutative contribution would be maximized when it is on the transverse plane, as expected. 

\subsection{Bhabha scattering cross section}

From the sum of matrix elements (\ref{calM1B},\ref{calM2B}), 
and after summing and averaging  over spins for the massless unpolarized case, we have
\begin{equation}
\frac{1}{4}{\sum\limits_{\rm spins}}\big|{\cal M}^{\rm B}\big|^2
=\frac{1}{4}{\sum\limits_{\rm spins}}\left|{\cal B}^{}_{It}+{\cal B}^{}_{Iu}\right|^2
=2e^4\bigg[\frac{s^2+u^2}{t^2}+\frac{u^2+t^2}{s^2}+2\frac{u^2}{ts}\cos(k_3\theta k_2-k_4\theta k_1)\bigg],
\label{AmplitaB}
\end{equation}
which gives the differential cross section for the head-to-head massless unpolarized Bhabha collision.
Again the phases in the pure $t$- and $s$-channel terms of (\ref{AmplitaB}) cancel out, while total phases in the interference cross terms of (\ref{AmplitaB}) turn into the $\mp$ of the form $(\it 2.)$ in Eq. (\ref{formsI;II}), i.e. $\mp 2(k_3\theta k_2-k_4\theta k_1)$, respectively. 
To determine phases in the NCM frame (\ref{SphCoordN}) we are using relevant phase factor from (\ref{SphCoordphaseGener}). The phase factors for Bhabha from (\ref{B3}) and (\ref{E4}) for NCM  shows that the Bhabha scattering receives contributions from all NC types.

The same procedure as for M\o ller, gives the noncommutative part of the Bhabha differential cross section in the NCM frame, and as a function of only incoming energies $\omega_1$ and $\omega_2$ and the NC parameters $c_{03}$, $A$, and $\Lambda_{\rm NC}$, respectively: 
\begin{equation}
\begin{split}
&\frac{d\sigma^{\rm B}_{\rm NC}}{dx}\Bigg|_{\rm NCM}
=\frac{\pi\alpha^2}{s}\Big(\frac{\omega_4}{\omega_1}\Big)^3\frac{(1+x)^2}{1-x}
\bigg[1-\cos\Big(\frac{2\omega_1\omega_4}{\Lambda^2_{\rm NC}}c_{03}\;(1-x)\Big)
J_0\Big(\frac{\omega_1\omega_4}{\Lambda^2_{\rm NC}}A\sqrt{1-x^2}\Big)\bigg].
\end{split}
\label{NCTCrossSectBnCM}
\end{equation}

\begin{figure}[t]
\begin{center}
\includegraphics[width=8cm,angle=0]{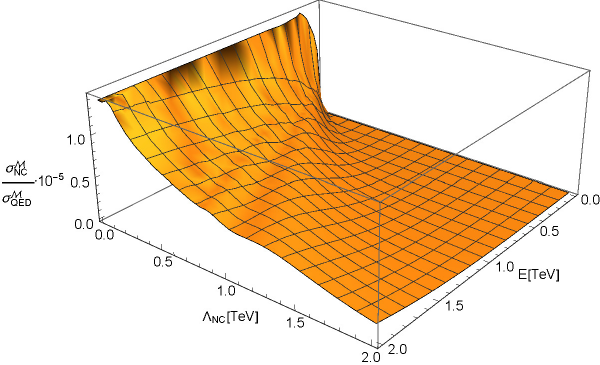}\;\;\;\;
\includegraphics[width=8cm,angle=0]{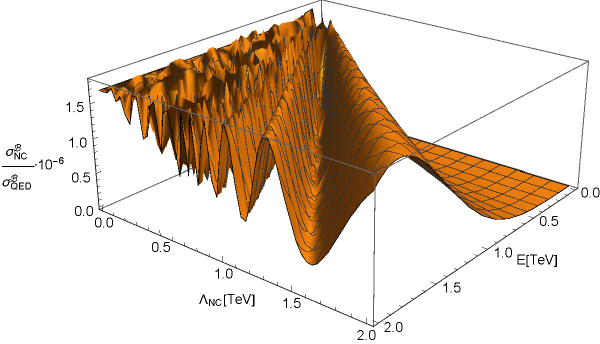}
\end{center}
\caption{Left: M\o ller cross section ratio with 
the NC correction (\ref{NCTCrossSectMnCM}) included, with the pure QED cross section, 
given as a function of incoming energy E and scale $\Lambda_{\rm NC}$ for the spacelike NC $|\tilde M|=\sqrt{c_{13}^2+c_{23}^2}\lesssim1$. Right: Bhabha cross section ratio with the NC correction (\ref{NCTCrossSectBnCM}) included, with pure QED cross section, as a function of energy E and scale $\Lambda_{\rm NC}$ for the lightlike NC: $|c_{13}|\lesssim1$. Both are computed numerically for the cutoff $\epsilon=10^{-7}$.} 
\label{fig:MollerBhabha}
\end{figure}
The small argument expansions of cosine and Bessel functions in 
(\ref{NCTCrossSectBnCM}) gives
\begin{equation}
\begin{split}
&\sigma^{\rm B}_{\rm NC}\Big|_{\rm NCM}
=\frac{2\pi\alpha^2\omega_1^3\omega^4_2}{\Lambda^4_{\rm NC}}
\int^{1}_{-1}dx(1+x)^2\frac{A^2 (1+x)+8c_{03}^2(1-x)}{\big(\omega_1(1-x)+\omega_2(1+x)\big)^5},
\end{split}
\label{NCTCrossSectBnCMSmall}
\end{equation}
which is regular at $x\to\pm 1$ limits.

It is interesting that Bhabha cross section in the CM frame of head-to-head collision 
for the massless case, i.e. where $\omega_1=\omega_2=\omega_4$, depends only on ${\bf E}_\theta$ as $A\rightarrow2\sqrt{{c_{01}^2+c_{02}^2}}$ [see Eq. (\ref{E4})]. 
This means that in the CM frame and for the pure spacelike noncommutativity where ${\bf E}_\theta=0$, the  NC contribution to Bhabha cross section vanishes.  However the general condition (\ref{likelightunitar}) still allows the NC contribution to be nonzero. In this case, one can express ${\bf E}_\theta$ as projection along the z axis ${\bf E}_\theta^{\parallel}$ as well as perpendicular to the z axis ${\bf E}_\theta^{\bot}$, so that $A=2\Lambda^{2}_{\rm NC}|{\bf E}_\theta^{\bot}|$ and $|c_{03}|=\Lambda^{2}_{\rm NC}|{\bf E}_\theta^{\parallel}|$. The cross section (\ref{NCTCrossSectBnCMSmall}) for energies below the scale $\Lambda_{\rm NC}$ in the CM frame then simplifies to 
\begin{equation}
\sigma^{\rm B}_{\rm NC}\Big|_{\rm CM}
\simeq
\frac{\pi\alpha^2 s}{12}\Big(2{\bf E}_\theta^2+{\bf E}_\theta^{\bot\;2}\Big).
\label{NCTCrossSectBCM}
\end{equation}
Thus for fixed $|{\bf E}_\theta|$ (which in the lightlike noncommutativity $|{\bf B}_\theta|=|{\bf E}_\theta|\propto \Lambda^{-2}_{\rm NC}$ case fixes the NC scale too), the NC contribution would maximize when ${\bf E}_\theta={\bf E}_\theta^\bot$.

Finally, we illustrate oscillatory behavior of M\o ller and Bhabha scattering cross sections in 3D plots from Fig.\ref{fig:MollerBhabha}, which show almost no oscillation for M\o ller due to the absence of cosine, while there is wild oscillation for the Bhabha case. 

\section{CROSS SECTIONS FOR ELECTRON AND PHOTON PROCESSES} 

In this section, we analyze Comptons $\ell^\pm\gamma\to \ell^\pm\gamma$, as well as dilepton pair $\ell^+ \ell^- \to \gamma\gamma$ annihilation, and pair $\gamma\gamma\to\ell^+ \ell^-$ productions ($\ell=e,\mu,\tau$) and  
$\gamma\gamma\to\gamma\gamma$, respectively. The above processes in NCQED are important by themselves, however,   diphoton and dilepton productions together with Bhabha scatterings also serve as a source of the so-called, long-distance effect contributions to the exclusive $\gamma\gamma\to\gamma\gamma$ scatterings and shall be, in particular, considered later. 

\subsection{Compton scattering cross section}

After rearranging phases (\ref{formsI;II}), we obtain the sum of the remaining non-SW map-induced amplitudes as:
\begin{equation}
{\cal M}^{\rm C}={\cal M}^{\rm C}_{Is}+{\cal M}^{\rm C}_{It}+{\cal M}^{\rm C}_{Iu}=
e^{i\frac{k_2\theta k_3}{2}}\Big(e^{-i\frac{k_1\theta k_4}{2}}
\big({\cal M}^{\rm C}_{s}-{\cal M}^{\rm C}_{t}\big)+e^{i\frac{k_1\theta k_4}{2}}
\big({\cal M}^{\rm C}_{u}+{\cal M}^{\rm C}_{t}\big)\Big),
\label{Totamp}
\end{equation}
\begin{eqnarray}
{\cal M}^{\rm C}_{s}&=&-i e^2\epsilon_{\mu}(k_1)\epsilon_\nu^{\ast}(k_4){\bar u}(k_3)\gamma^{\mu}\frac{\slashed{k_1}+\slashed{k_2}+m_e}{(k_1+k_2)^2-m_e^2}\gamma^{\nu} u(k_2),
\label{MCs}\\
{\cal M}^{\rm C}_{t}&=&-i e^2\epsilon_{\mu}(k_1)\epsilon^{\ast}_\nu(k_4)
\frac{2}{(k_1-k_4)^2}{\bar u}(k_3)
\Big(g^{\mu\nu}\slashed{k_1}-{k_1}^{\mu}\gamma^{\nu}-{k_4}^{\nu}\gamma^{\mu}
 \Big)u(k_2),
\label{MCt}
\\
{\cal M}^{\rm C}_{u}&=&-i e^2\epsilon_{\mu}(k_1)\epsilon^{\ast}_\nu(k_4)
{\bar u}(k_3)\gamma^{\nu}\frac{\slashed{k_2}-\slashed{k_4}+m_e}{(k_1-k_3)^2-m_e^2}\gamma^{\mu} u(k_2),
\label{MCu}
\end{eqnarray}
where two terms (\ref{MCs}) and (\ref{MCu}) without NC phases arising from the first two 
diagrams in Fig.\ref{fig:FD1}, are the same as in Klein-Nishina \cite{Klein-Nishina} expressions, 
while (\ref{MCt}) with extracted phases from (\ref{Totamp}) represents 
NC correction coming from the three-photon vertex contribution to the $t$-channel fully described by the non-SW map term (\ref{MtI}). 

After average over initial spins and summations over final spins with contractions over photon polarizations, 
we compute the absolute square of the ${\cal M}^{\rm C}$ for the unpolarized case as:
\begin{eqnarray}
{\sum\limits_{\rm Spins}}\big|{\cal M}^{\rm C}\big|^2&=&\big({{\cal M}^{\rm C}_s}^\dagger-{{\cal M}^{\rm C}_t}^\dagger\big)\big({{\cal M}^{\rm  C}_s}
-{{\cal M}^{\rm C}_t}\big)+\big({{\cal M}^{\rm C}_u}^\dagger+{{\cal M}^{\rm C}_t}^\dagger\big)\big({{\cal M}^{\rm C}_u}+{{\cal M}^{\rm C}_t}\big)
\nonumber\\
&+&e^{ik_1\theta k_4}\big({{\cal M}^{\rm C}_s}^\dagger-{{\cal M}^{\rm C}_t}^\dagger\big)
\big({\cal M}^{\rm C}_u+{\cal M}^{\rm C}_t\big)
+e^{-ik_1\theta k_4}\big({{\cal M}^{\rm C}_u}^\dagger+{{\cal M}^{\rm C}_t}^\dagger\big)
\big({\cal M}^{\rm C}_s-{\cal M}^{\rm C}_t\big).
\label{Totampsquer} 
\end{eqnarray}
From the above equation, it is clear that $\big|{\cal M}^{\rm C}_s\big|^2+\big|{\cal M}^{\rm C}_u\big|^2$ builds Klein-Nishina formula. Also in the massless average we have found that ${{\cal M}^{\rm C}_s}^\dagger{\cal M}^{\rm C}_u={{\cal M}^{\rm C}_u}^\dagger{\cal M}^{\rm C}_s=0$,  ${{\cal M}^{\rm C}_s}^\dagger{\cal M}^{\rm C}_t
={{\cal M}^{\rm C}_t}^\dagger{\cal M}^{\rm C}_s$, and ${{\cal M}^{\rm C}_u}^\dagger{\cal M}^{\rm C}_t={{\cal M}^{\rm C}_t}^\dagger{\cal M}^{\rm C}_u$.

By extracting out the ordinary QED contributions, following Peskin  and  Schroeder \cite{Peskin:1995ev}, in terms of Mandelstam variables and for the massless unpolarized case, from Eq. (\ref{Totampsquer}), we have obtained the noncommutative correction to the absolute square of the full Compton amplitude 
${\cal M}^{\rm C}$:
\begin{equation}
\begin{split}
\frac{1}{4}{\sum\limits_{\rm spins}}\big|{\cal M}_{\rm NC}^{\rm C}\big|^2
&=\frac{1}{4}{\sum\limits_{\rm spins}}\big|{\cal M}^{\rm C}\big|^2-\frac{1}{4}{\sum\limits_{\rm Spins}}\Big(\big|{\cal M}^{\rm C}_s\big|^2+\big|{\cal M}^{\rm C}_u\big|^2\Big)
= \frac{1}{2}(1-\cos k_1\theta k_4)
\big({{\cal M}^{\rm C}_t}^\dagger+{{\cal M}^{\rm C}_u}^\dagger-{{\cal M}^{\rm C}_s}^\dagger\big){\cal M}^{\rm C}_t
\\&={4e^4}
(1-\cos k_1\theta k_4) \bigg[\frac{(k_1k_2)^2+(k_1k_3)^2}{(k_1k_4)^2}
-2\frac{m_e^2}{k_1k_4}+\frac{m_e^4}{(k_1k_2)(k_2k_4)}\bigg]_{m_e\to0}=8e^4 \frac{s^2+u^2}{t^2}\sin^2{\frac{k_1\theta k_4}{2}}.
\label{NCampsquer}
\end{split}
\end{equation}

Jumping shortly to Eqs. (\ref{Totamp})-(\ref{NCampsquer}), we found a 
differential cross section for massless Compton scattering confirming results obtained before in  \cite{Godfrey:2001yy,Mathews:2000we,Baek:2001ty}, where first two terms represents the Klein-Nishina formula for standard QED. However, the third term (\ref{NCampsquer}) containing the phase factor arises from the $t$-channel  diagram  in Fig.\ref{fig:FD1} containing a triple-photon coupling vertex, existing in all Moyal versions of the NCQFT \cite{Godfrey:2001yy}, and all of its interference with the $s$- and $u$-channel diagrams and with the interference between the $s$- and $u$-channel diagrams, as presented in Eq. (\ref{Totampsquer}). 

Continuing with NCM frame, having definitions (\ref{SphCoordN}), and (\ref{Mandelstam2}), and after variable change ($\cos\vartheta\to x$) and integrating over $\varphi$, we obtain the following massless NC part of Compton differential cross section as function of incoming energies $\omega_{1,2}$, the NC parameters $c_{03}$ and $C$, and the NC scale 
$\Lambda_{\rm NC}$, [see Eq. (\ref{E4})], respectively: 
\begin{equation}
\begin{split}
\frac{d\sigma^{\rm C}_{\rm NC}}{dx}\Bigg|_{\rm NCM}&
=8\pi\alpha^2\frac{\omega_4^2}{s^2}
\;\bigg[1+2\frac{\omega_2}{\omega_1}\frac{{1+x}}{1-x}+2\frac{\omega_2^2}{\omega_1^2}\frac{(1+x)^2}{(1-x)^2}\bigg]
\bigg[1-\cos\bigg(\frac{\omega_1\omega_4}{\Lambda^2_{\rm NC}}c_{03}(x-1)\bigg)
J_0\bigg(\frac{\omega_1\omega_4}{\Lambda^2_{\rm NC}}\;C\sqrt{1-x^2}\bigg)\bigg],
\end{split}
\label{NCTCrossSectCnCM}
\end{equation}
while in the CM frame, where $\omega_1=\omega_2=\omega_4$, of head-to-head collision and, 
from Eq. (\ref{NCTCrossSectCnCM}) 
we obtain the following NC correction to the cross section:
\begin{equation}
\sigma^{\rm C}_{\rm NC}\Big|_{\rm CM}=\frac{2\pi\alpha^2}{s}\int^{1}_{-1}dx
\frac{4+(1+x)^2}{(1-x)^2}
\bigg[1-\cos\Big(\frac{sc_{03}}{4\Lambda^2_{\rm NC}}(x-1)\Big)
J_0\Big(\frac{sC}{4\Lambda^2_{\rm NC}}\sqrt{1-x^2}\Big)\bigg],
\label{NCTCrossSectCCM}
\end{equation}
which is valid for various types of noncommutativity. 

\subsubsection{Collinear singularity in NCQED Compton  scattering}

\begin{figure}
\begin{center}
\includegraphics[width=14cm,angle=0]{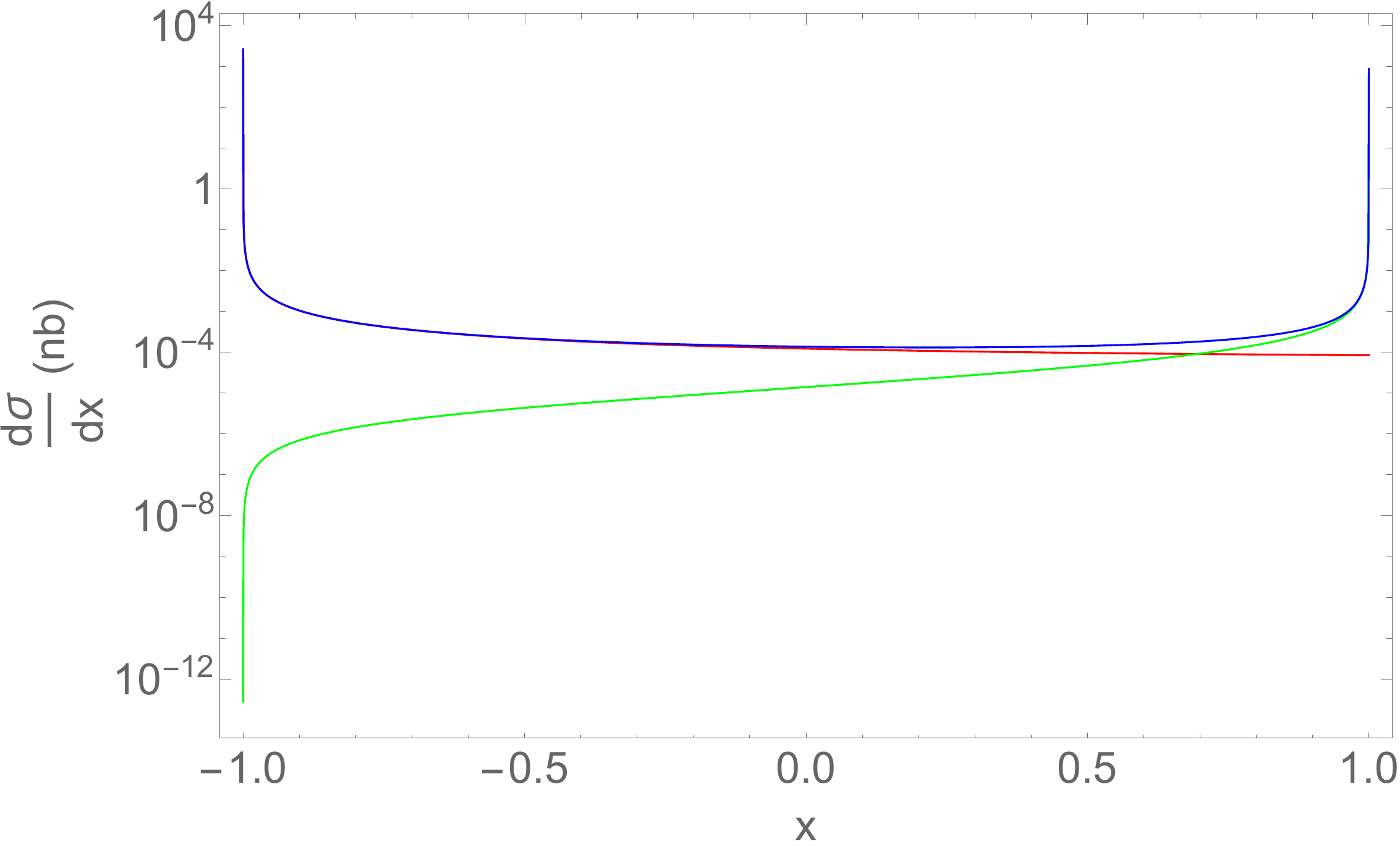}
\end{center}
\caption{Massless NCQED Compton scattering differential cross section--in log scale--versus $x=\cos\vartheta$, with $\vartheta$ being the polar angle. The plots are evaluated from QED and Eq. (\ref{NCTCrossSectCnCM}) for head-to-head collision of a photon with energy $\omega_1=200$ GeV and an electron with energy $\omega_2=250$~GeV. The red line presents the QED, green the NCQED (\ref{NCTCrossSectCnCM}) and blue the total (QED+NCQED) differential cross section, respectively. The collinear divergences of QED and NCQED contributions can be easily noticed.
}
\label{Comptoncolinear}
\end{figure}

It is easy to notice the similarity between the NCQED amplitude square, and the quark-gluon scattering amplitude $qg\to qg$ in QCD, which has the $t$-channel term (\ref{NCampsquer}) and (\ref{NCTCrossSectCnCM}) added to the amplitude square besides the term that is identical to the Compton scattering in the commutative QED, which makes the complete (QED+NCQED) differential cross section for a massless electron divergent at both forward and backward limits respectively, as shown in Fig.~\ref{Comptoncolinear}.
Next we can also notice that the NCQED $\gamma e\to\gamma e$ process possesses a $t$-channel collinear singularity similar to $qg\to qg$. On the other hand, these two singularities are not exactly the same. The NCQED $\gamma e\to\gamma e$ amplitude square diverges by $t^{-1}$ because of the NC factor as shown in Eqs. (\ref{NCampsquer})-(\ref{NCTCrossSectCCM}) \footnote{Inspecting more carefully Eqs. (\ref{NCTCrossSectCnCM}) and (\ref{NCTCrossSectCCM}), we notice that at $\left(\frac{1}{\Lambda_{\rm NC}}\right)^8$ order and higher, the collinear singularity of Compton scattering in NCQED gets eliminated by the arbitrary NC factor. However for timelike and/or lightlike noncommutativity, due to the cosine, collinear singularity gets eliminated starting at $\left(\frac{1}{\Lambda_{\rm NC}}\right)^4$ order.}, while $qg\to qg$ diverges by $t^{-2}$ \cite{Peskin:1995ev}. 

Since to our knowledge the collinear singularity in the NCQED Compton process has not attracted many studies, we invest here some efforts to study its properties. One question one may ask is whether this collinear singularity can be suppressed by certain reasonable choices of NC parameter $\theta^{\mu\nu}$. Limiting to the head-to-head scattering scenario~\eqref{NCTCrossSectCnCM} and~\eqref{NCTCrossSectCCM}, we notice that the collinear singularity is suppressed if $C=\sqrt{(c_{01}-c_{13})^2+(c_{02}-c_{23})^2}=0$. Now a nonvanishing NC contribution to Compton cross section exists only if $c_{03}\neq 0$ when $C=0$. Once we inspect these constraints more carefully we notice that $\theta^{\mu\nu}$ would have to be neither spacelike nor lightlike NC, to satisfy both conditions.
We conclude that it is not possible to suppress the collinear singularity by an appropriate choice of $\theta^{\mu\nu}$.

Next, we attempt to introduce a regulator for the collinear singularity in the NC cross section integral by hand. We consider the following simplified scenario: CM frame, spacelike $\theta^{\mu\nu}$, and $s\ll4\rm\Lambda^2_{NC}$. We then obtain the following approximative formula for the cross section
\begin{equation}
\sigma^{\rm C}_{\rm NC}\Big|_{\rm CM}
\simeq\frac{\pi\alpha^2s}{32\Lambda^4_{\rm NC}}\int^1_{-1}dx \;
\bigg[3+x+2\frac{(1+x)^2}{1-x}\bigg]
C^2\big(1+x\big).
\label{NCTCrossSectCnCMSmall}
\end{equation}
We then introduce a small dimensionless shift $\delta$ to the $1-x$ denominator \cite{Peskin:1995ev} in Eq. \eqref{NCTCrossSectCnCMSmall}, making it
\begin{equation}
\sigma^{\rm C}_{\rm NC}\Big|_{\rm CM}
\simeq\frac{\pi\alpha^2s}{32\Lambda^4_{\rm NC}}\int^1_{-1}dx \;
\bigg[3+x+2\frac{(1+x)^2}{1-x+\delta}\bigg]
C^2\big(1+x\big).
\label{NCTCrossSectCnCMSmallmodified}
\end{equation}
Evaluating this integral and keeping only terms till $\mathcal O(\ln\delta)$, we get
\begin{equation}
\sigma^{\rm C}_{\rm NC}\Big|_{\rm CM}\sim\frac{\pi\alpha^2C^2s}{8\Lambda^4_{\rm NC}}\left(4\ln\frac{2}{\delta}-\frac{17}{3}\right).
\end{equation}
Based on the divergence order one may then choose $\delta=s/\Lambda^2_{\rm NC}$ to ensure that
\begin{equation}
\lim_{\Lambda_{\rm NC}\to \infty}\sigma^{\rm C}_{\rm NC}\Big|_{\rm CM}=0.
\label{NCTCrossSectCCMvanish}
\end{equation}
On the other hand this regulator $\delta$, appropriate or not, would limit the validity of $\sigma^{\rm C}_{\rm NC}$ to well below $\Lambda_{\rm NC}$. Thus finding a more natural and/or less limited way to regulate the $t$-channel singularity of NCQED Compton scattering would remain a relevant open problem for future research.

\subsection{Dilepton pair annihilation cross section}

Using the crossing symmetry in accord with the left plot in Fig.\ref{fig:FD2} we replace momenta $(k_1\to-k_4)$, $(k_2\to k_1)$, $(k_3\to-k_2)$, and $(k_4\to k_3)$ in the Compton amplitude square (\ref{NCampsquer}) and obtain an unpolarized 
differential cross section in terms of Mandelstam variables
\begin{equation}
\begin{split}
\frac{d\sigma^{\rm A}}{d\Omega}=\frac{d\sigma_{\rm QED}^{\rm A}}{d\Omega}+\frac{d\sigma_{\rm NC}^{\rm A}}{d\Omega}&=\frac{\alpha^2}{2s}\bigg[\frac{t}{u}+\frac{u}{t}-2\frac{t^2+u^2}{s^2}
\big(1-\cos k_3\theta k_4\big)
\bigg],
\end{split}
\label{DCrossSectAnn}
\end{equation}
with the first two terms out of three representing QED contributions in agreement with Ref. \cite{Peskin:1995ev}, while the third one is coming form the NCQED triple-photon coupling phase factors \cite{Hewett:2000zp}.

Next following the same procedure as in the Compton section by using the NCM frame defined in Eq. (\ref{SphCoordN}), from Eq. (\ref{E4}), we shall find contributions for the general case of the NC space-time. 
 After variable change and integration over $\varphi$, we obtain destructive noncommutative contribution to the unpolarized pair annihilation differential cross section in the NCM frame as a function of incoming energies $\omega_{1,2}$, the NC parameters $c_{03}$ and $A$--given in (\ref{E4})-- and $\Lambda_{\rm NC}$, respectively:
\begin{equation}
\begin{split}
\frac{d\sigma^{\rm A}_{\rm NC}}{dx}&\Bigg|_{\rm NCM}=-2\pi\alpha^2
\frac{\omega_1^2(1-x)^2+\omega_2^2(1+x)^2}
{\big(\omega_1(1-x)+\omega_2(1+x)\big)^4}
\bigg[1-\cos\Big(\frac{\omega_4c_{03}}{\Lambda^2_{\rm NC}}
\big(\omega_1(x-1)+\omega_2(x+1)\big)\Big)
J_0\Big(\frac{\omega_1\omega_4}{\Lambda^2_{\rm NC}}A\sqrt{1-x^2}\Big)\bigg],
\end{split}
\label{NCTCrossSectAnCM}
\end{equation}
describing the pair annihilation cross section showing not only explicit dependence of 
the pair annihilation cross section on the NC scale, 
but also in the head-to-head annihilation geometry revealing all types of space-time noncommutativity, 
just like for Compton scatterings.  

Taking into account two identical particles in the final state we should take 1/2 of the needed integral in forming a total cross section. From Eq. (\ref{NCTCrossSectAnCM}) we next give the pair annihilation cross section in the CM frame of head-to-head collision for the massless case, with a perfectly regular integral which carries pure timelike noncommutativity since $A\to 2\sqrt{c_{01}^2+c_{02}^2}$:
\begin{equation}
\sigma^{\rm A}_{\rm NC}\Big|_{\rm CM}=-\frac{\pi\alpha^2}{s}\int^1_{-1}dx (1+x^2)
\bigg[1-\cos\Big(\frac{sc_{03}}{2\Lambda^2_{\rm NC}}x\Big)J_0\Big(\frac{s A}{4\Lambda^2_{\rm NC}}
\sqrt{1-x^2}\Big)\bigg].
\label{NCTCrossSectACM}
\end{equation}

Interestingly, there is no collinear singularity whatsoever. In the case of pure spacelike noncommutativity, cross section (\ref{NCTCrossSectACM}) vanishes identically, as it should. However the more general conditions (\ref{likelightunitar}) again allow (\ref{NCTCrossSectACM}) to be nonzero. The leading noncommutative contribution at low energies can be expressed as follows:
\begin{equation}
\sigma^{\rm A}_{\rm NC}\Big|_{\rm CM}
\simeq
-\frac{\pi\alpha^2s}{60}\Big(3{\bf E}_\theta^2+{\bf E}_\theta^{\parallel\;2}\Big).
\label{NCTCrossSectACMsmall}
\end{equation}

So for fixed $|{\bf E}_\theta|$, the NC contribution to the cross section at low energies is maximized when ${\bf E}_\theta={\bf E}^\parallel_\theta$, which is opposite to the Bhabha scattering case.

Pair annihilation shows in the NCM frame all types of noncommutativity, while in the CM frame the NC correction becomes pure timelike which could be transferred into the lightlike one, and it is illustrated in the 3D left plot in Fig.\ref{fig:AnnihilProd}. Mild oscillation of the NCQED/QED cross section ratio gets stable by approaching unity for lightlike NC and when 
$\Lambda_{\rm NC}$ and the energy pass $\sim$1 TeV values. 
In the case of CM frame at small energies the NC correction to the total cross section is destructive and finite  (\ref{NCTCrossSectACMsmall}). 

\subsection{Dilepton pair production cross section} 

Again using the crossing symmetry in accord with the right plot of Fig.\ref{fig:FD2}, we can replace momenta $(k_1\to k_1)$, $(k_4\to-k_2)$, $(k_2\to -k_4)$, $(k_3\to-k_3)$ in the amplitude square (\ref{NCampsquer}) and obtain a 
differential cross section for the head-to-head unpolarized massless pair production
\begin{equation}
\begin{split}
\frac{d\sigma^{\rm P}}{d\Omega}=\frac{d\sigma_{\rm QED}^{\rm P}}{d\Omega}+\frac{d\sigma_{\rm NC}^{\rm P}}{d\Omega}
&=\frac{\alpha^2}{2s}\bigg[\frac{t}{u}+\frac{u}{t}-2\frac{t^2+u^2}{s^2}
\big(1-\cos k_1\theta k_2\big)
\bigg],
\end{split}
\label{DCrossSectP}
\end{equation}
where the first two represent QED contributions \cite{Peskin:1995ev}, while the third one is coming form the triple-photon coupling phase factors \cite{Godfrey:2001yy}.

The same procedure as in the Compton section in the NCM frame (\ref{SphCoordN}) after variable change gives the noncommutative part of pair production total cross section, and as a function of only incoming energies $\omega_{1,2}$ , the scale $\Lambda_{\rm NC}$, and only the NC parameter $c_{03}$, respectively. So, first, after using the relevant phase factor for pair production from Eqs. (\ref{SphCoordphaseGener}) and (\ref{B2}), we have found the destructive NC contribution to the differential cross section 
\begin{equation}
\begin{split}
\frac{d\sigma^{\rm P}_{\rm NC}}{dx}\Bigg|_{\rm NCM}&=-2\pi\alpha^2
\bigg[1-\cos\Big(\frac{sc_{03}}{2\Lambda^2_{\rm NC}} \Big)\bigg]
\frac{\omega_1^2(1-x)^2+\omega_2^2(1+x)^2}{\big(\omega_1(1-x)+\omega_2(1+x)\big)^4},
\end{split}
\label{NCTCrossSectPrnCM}
\end{equation}
again with no collinear singularity. Notice that the pair production cross section of head-to-head collision for the massless case is fully timelike and depends on ${\bf E}_\theta^\parallel$ only, which is easy to understand, since this is the only possible nonvanishing term in the NC coupling ($k_1\theta k_2$), the two incoming momenta which are collinear in the head-to-head scattering geometry. This coupling vanishes for a pure spacelike NC parameter yet remains allowed by the general unitary constraint~\eqref{likelightunitar}.

Next, from Eq. (\ref{NCTCrossSectPrnCM}) we compute the pair production total cross section in the CM frame of head-to-head collision for the massless case, $\omega_1=\omega_2=\omega_4$. For energies smaller than scale, $s/2\Lambda^2_{\rm NC}  < 1$, the expansion of cosine function in Eq. (\ref{NCTCrossSectPrnCM}) gives asymptotically  
\begin{equation}
\sigma^{\rm P}_{\rm NC}\Big|_{\rm CM}\simeq
-\frac{\pi}{3}\alpha^2s \;{\bf E}_\theta^{\parallel\;2}
\label{NCTCrossSectPrCMsmall}
\end{equation}

\begin{figure}[t]
\begin{center}
\includegraphics[width=8cm,angle=0]{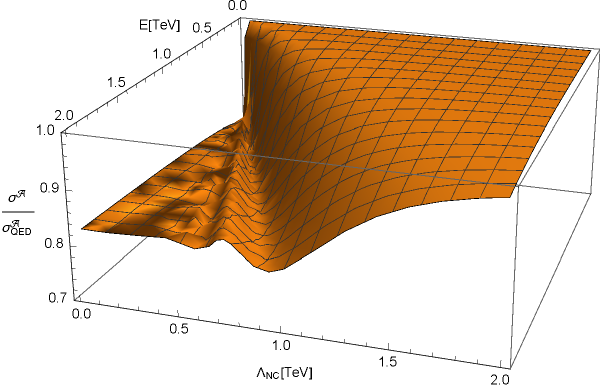}\;\;\;\;
\includegraphics[width=8cm,angle=0]{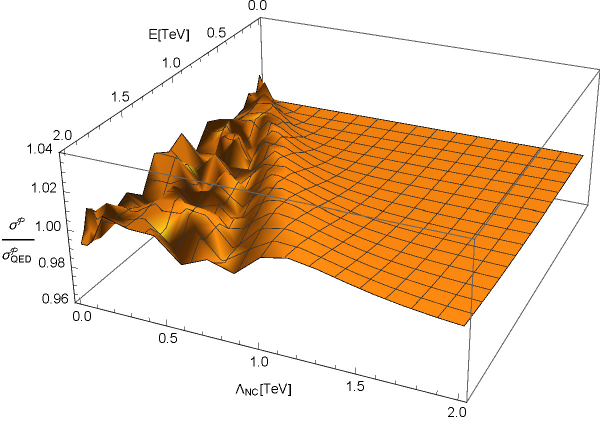}
\end{center}
\caption{Left: pair annihilation, QED+NC cross section (\ref{NCTCrossSectACM}), normalized to the pure QED cross section as a function of incoming energy E and NC scale in the CM frame, for the lightlike case with $|c_{03}|=|-c_{13}|\lesssim1$ and $A=0$. \\
Right: pair production, ratio of QED+NC cross section, with the pure QED cross section as functions of energy E and NC scale, for the lightlike noncommutativity with $|c_{03}|=|-c_{13}|\lesssim1$. }
\label{fig:AnnihilProd}
\end{figure}

Finally we illustrate numerical analyses of both full cross sections in the CM frame, 
i.e., the pair annihilation (\ref{NCTCrossSectACM})  and pair production (\ref{NCTCrossSectPrnCM}) 
as functions of scale $\Lambda_{\rm NC}$, incoming energy and for  
lightlike NC parameters and present them as 3D left and right plots in Fig.\ref{fig:AnnihilProd}, respectively.

Inspecting the left plot in Figs.\ref{fig:AnnihilProd} one realizes that in the CM frame 
the pair annihilation total cross section ratio of the NCQED (\ref{NCTCrossSectACM}) with 
one obtained in QED, as a function of energy, the lightlike noncommutativity, and the NC scale, 
having oscillatory behavior for small E and scale values, than rises and starts to stabilize around unity when both 
$\Lambda_{\rm NC}$ scale and the incoming energy E approach 
$\sim1$ TeV scale. Similar oscillatory behavior one may see in Fig.18 in Ref. \cite{Hewett:2000zp}.

Lepton pair production shows in the NCM frame pure timelike type of noncommutativity, 
which could be transferred into the lightlike one.  
There is no collinear singularity whatsoever, and in the case of CM frame 
and at small incoming energies the NC correction to 
the cross section has a destructive finite value (\ref{NCTCrossSectPrCMsmall}). 
It is illustrated in the right plot in Fig.\ref{fig:AnnihilProd}, 
which due to cosine shows wild oscillatory behavior at small NC scale and large energy values  
and up to 0.5 TeV agrees with Fig. 2 in Ref. \cite{Godfrey:2001yy}. 
However, the right plot in Fig.\ref{fig:AnnihilProd} does show double oscillations, than dropping and stabilization of the NCQED/QED cross section ratio around unity for lightlike NC when $\Lambda_{\rm NC}$ and energy pass 
$\sim$1 TeV values.  

\section{CROSS SECTION OF LIGHT-BY-LIGHT SCATTERINGS}

\subsection{Helicity amplitudes of NCQED photon-photon  scattering}

Since the NCQED $\gamma\gamma\to\gamma\gamma$ is invariant under the SW map, the calculation of the scattering cross sections can follow the general results of tree-level photon scattering amplitudes in NCQED without the SW map. The basic fact is that there exists a one-to-one correspondence between the color ordering in QCD and a $\star$-product ordering in NCQED~\cite{Raju:2009yx,Huang:2010fc}: 
\begin{equation}
\tr \;\prod_{i=1}^n T^{\alpha_i}\; \Longleftrightarrow \;
\exp\Big({-\frac{i}{2} \sum_{i=1}^n p_{2i-1}\theta p_{2i}}\Big).
\label{Helicity1}
\end{equation}
The tree-level $\star$-product ordered NCQED photon scattering amplitudes can then be shown to be identical to their QCD gluon scattering counterpart. The full helicity amplitudes of NCQED $\gamma\gamma\to\gamma\gamma$ scattering can be obtained by summing over all $\star$-product orders. The nonvanishing helicity amplitudes up to a total reflection of all photon helices shall be expressed as follows:
\begin{gather}
\begin{split}
M_{\rm NC}^{++++}(s,t,u)&
=32\pi\alpha
\left(\frac{s}{u}\sin\frac{k_1\theta k_2}{2}\sin\frac{k_3\theta k_4}{2}-\frac{s^2}{tu}\sin\frac{k_1\theta k_4}{2}\sin\frac{k_2\theta k_3}{2}\right),
\\
M_{\rm NC}^{+-+-}(s,t,u)&
=32\pi\alpha
\left(\frac{t^2}{su}\sin\frac{k_1\theta k_2}{2}\sin\frac{k_3\theta k_4}{2}-\frac{t}{u}\sin\frac{k_1\theta k_4}{2}\sin\frac{k_2\theta k_3}{2}\right),
\\
M_{\rm NC}^{++--}(s,t,u)&
=32\pi\alpha
\left(\frac{u}{s}\sin\frac{k_1\theta k_2}{2}\sin\frac{k_3\theta k_4}{2}-\frac{u}{t}\sin\frac{k_1\theta k_4}{2}\sin\frac{k_2\theta k_3}{2}\right).
\label{Helicity24}
\end{split}
\end{gather}

Employing identities $s+t+u=0\,\to\, s^4+t^4+u^4=2(s^2t^2+s^2u^2+t^2u^2)$, and the $\star$-product Jacobi identity (\ref{JIstu}), due to the conjugation relation from (\ref{Helicity24}) we obtain the following sum of amplitude absolute squares
\begin{equation}
\begin{split}
\big|{\cal M}^\gamma_I\big|^2&={\sum\limits_{helicity}}\big|{\cal M}^\gamma_{helicity}\big|^2
=2\Big(\left|M_{\rm NC}^{++++}\right|^2+\left|M_{\rm NC}^{+-+-}\right|^2
+\left|M_{\rm NC}^{++--}\right|^2\Big)
\\
&=2(-32\pi\alpha)^2\cdot(-2)\bigg[\sin^2\frac{k_1\theta k_2}{2}
\sin^2\frac{k_3\theta k_4}{2}\left(\frac{t}{u}+\frac{u}{t}+\frac{tu}{s^2}\right)
+\sin^2\frac{k_1\theta k_3}{2}\sin^2\frac{k_2\theta k_4}{2}\left(\frac{s}{t}+\frac{t}{s}+\frac{st}{u^2}\right)
\\
&\phantom{2(-32\pi\alpha)^2\cdot(-2)X}+\sin^2\frac{k_1\theta k_4}{2}\sin^2\frac{k_2\theta k_3}{2}\left(\frac{s}{u}+\frac{u}{s}+\frac{su}{t^2}\right)\bigg].
\end{split}
\label{Helicity25}
\end{equation}
It is clear that in the above formula each of terms $(k_1\theta k_2,k_3\theta k_4)$, $(k_1\theta k_3,k_2\theta k_4)$ and $(k_1\theta k_4,k_2\theta k_3)$ contains the contributions from all three, the $s$-, $u$-, and $t$-channels, respectively.  Above phase factors are given in Appendix C, where integration over $d\varphi$ has been performed too. This result we shall use next to show the equivalence of the NCQED and the QCD free theories.

Here we notice two remarkable properties: First, the amplitude square~\eqref{Helicity25} is closely connected to its $gg\to gg$ counterpart in QCD, as expected. A conversion to the latter (up to an overall normalization factor) can be achieved by replacing three NC factors with three identical QCD factors in accordance with Eq. \eqref{Helicity1}:
\begin{equation}
\sin^2\frac{k_i\theta k_j}{2}\sin^2\frac{k_k\theta k_l}{2}\;\longrightarrow\;\sum\limits_{\{\alpha_{i},\alpha_{j},\alpha_{k},\alpha_{l}\}}\Big(\sum\limits_{\mu}f^{\alpha_i\alpha_j\mu}f^{\alpha_k\alpha_l\mu}\Big)^2,\;i\not=j\not=k\not=l=1,2,3,4.
\label{Helicity27}
\end{equation}
Second, since in each of the color assignments cases $\{\alpha\}$’s run through all possible combinations, the above three color factors have to be equal to each other. Hence, from (\ref{Helicity25}) we obtain 
\begin{equation}
(-)\left(\frac{t}{u}+\frac{u}{t}+\frac{tu}{s^2}+\frac{s}{t}+\frac{t}{s}+\frac{st}{u^2}+\frac{s}{u}+\frac{u}{s}+\frac{su}{t^2}\right)=\;3-\frac{tu}{s^2}-\frac{su}{t^2}-\frac{st}{u^2},
\label{Helicity28}
\end{equation}
which is exactly the same kinematic factor that occurs in the differential cross section of QCD exclusive free gluon ($gg\to gg$) scattering~\cite{Peskin:1995ev}.

On the other hand, unlike QCD, all the collinear singularities from the factions of Mandelstam variables in 
Eq. \eqref{Helicity25} are canceled by the corresponding collinear zeros in the NC factors. This property can also be shown at helicity amplitude level with some help from the star product Jacobi identity~\eqref{JIstu},
which allows us to transform the amplitudes \eqref{Helicity24} into the following form
\begin{gather}
\begin{split}
M_{\rm NC}^{++++}(s,t,u)&
=32\pi\alpha
\left(\frac{s}{u}\sin\frac{k_1\theta k_3}{2}\sin\frac{k_2\theta k_4}{2}+\frac{s}{t}\sin\frac{k_1\theta k_4}{2}\sin\frac{k_2\theta k_3}{2}\right),
\\
M_{\rm NC}^{++--}(s,t,u)&
=32\pi\alpha
\left(\frac{u}{s}\sin\frac{k_1\theta k_2}{2}\sin\frac{k_3\theta k_4}{2}-\frac{u}{t}\sin\frac{k_1\theta k_4}{2}\sin\frac{k_2\theta k_3}{2}\right)
\\
M_{\rm NC}^{+-+-}(s,t,u)&
=-32\pi\alpha
\left(\frac{t}{s}\sin\frac{k_1\theta k_2}{2}\sin\frac{k_3\theta k_4}{2}+\frac{t}{u}\sin\frac{k_1\theta k_3}{2}\sin\frac{k_2\theta k_4}{2}\right).
\label{Helicity241}
\end{split}
\end{gather}
The cancellation of collinear singularities is then straightforward.

\subsection{Light-by-light cross section in NCQED} 

Using a helicity amplitude decomposition method or signature, we shall determine the NC contributions to the total cross section of the exclusive $\gamma\gamma\to \gamma\gamma$ process starting with the NC amplitudes (\ref{Helicity24}) and than adding on the SM amplitudes \cite{Passarino:1978jh,Dong:1992iz,Jiang:1992sm,Dong:1992fa,Dong:1992hg,Jikia:1993tc,Gounaris:1998qk,Gounaris:1999gh,Davoudiasl:1999di,Bern:2001dg,Baur:2001jj,Hagiwara:1994pw,KlusekGawenda:2010kx,Klusek-Gawenda:2016euz,Harland-Lang:2018iur}. 
Averaging over initial and summing over final helicities in the NCM frame, from Eq. (\ref{TCrossSectC}) we obtain a 
cross section containing three terms: the SM one $\sigma_{\rm SM^2}^\gamma$, the SM interference with NCQED $\sigma_{\rm SM\times NC}^\gamma$, and pure NCQED term $\sigma_{\rm NC^2}^\gamma$, respectively. 
Note that, since the SM contribution dependence on energy is rather tedious \cite{Passarino:1978jh,Dong:1992iz,Jiang:1992sm,Dong:1992fa,Dong:1992hg}, there exist various approximations in the literature \cite{Gounaris:1998qk,Gounaris:1999gh,Davoudiasl:1999di}.  Namely, a very high-energy regime is relevant for further LC$_{\gamma\gamma}$ experiments, while relatively low-energy limit is interesting because the ATLAS Collaboration actually measured a diphoton final state in the central detector with only less than 30 GeV diphoton invariant mass. 

\subsubsection{Pure NCQED contributions to the $\gamma\gamma\to \gamma\gamma$ cross section}

We compute the pure NC contribution to the $\gamma\gamma\to \gamma\gamma$ cross section starting with the helicity amplitude square (\ref{Helicity25}).\footnote{Because of  the similarity or complementarity of NCQED  with QCD,  there exist only three maximal helicity violating (MHV) independent noncommutative  amplitudes $(++++,+-+-,++--)$ given in (\ref{Helicity24}). } Generally, after taking into account two identical particles in the final state, we have obtained for arbitrary energies in the NCM frame the following pure NCQED exclusive cross section $\sigma_{\rm NC^2}^{\gamma}$:
\begin{equation}
\begin{split}
\sigma_{\rm NC^2}^{\gamma}\Big|^{Exclu.}_{\rm NCM}&=
-4\alpha^2\;\int^1_{-1}dx\frac{\omega_4^2}{s^2}\;
\bigg[ \Big(2 \hat I_0^{1234}+ \hat I_+^{1234}+ \hat I_-^{1234}\Big)\Big(\frac{t}{u}+\frac{u}{t}+\frac{tu}{s^2}\Big) 
+\Big(2 \hat I_0^{1324}+ \hat I_+^{1324}+ \hat I_-^{1324}\Big)\Big(\frac{s}{t}+\frac{t}{s}+\frac{st}{u^2}\Big) 
\\&\phantom{XXXXXx}
+\Big(2 \hat I_0^{1423}+ \hat I_+^{1423}+ \hat I_-^{1423}\Big)\Big(\frac{s}{u}+\frac{u}{s}+\frac{su}{t^2}\Big)\bigg],
\end{split}
\label{NCGammaTCrossnCMpure}
\end{equation}
with above integrals over $d\varphi$ being given in Eqs. (\ref{B10}).  
The above cross section (\ref{NCGammaTCrossnCMpure}) is valid for an arbitrary energy regime and for arbitrary scattering angles, which, together with the SM$\times$NC interference term,  
give noncommutatative correction to the total cross section in the laboratory of  NCM frame. The above integrals $\hat I_0^{1324}$, and $\hat I_0^{1423} $ contain new NC constant $G$ given in (\ref{E4}). 

Choosing the CM frame with $\omega_1=\omega_2=\omega_4$, and the pure spacelike noncommutativity where $c_{i0}=0, \forall i=1,2,3$,  from Eq. (\ref{E4}) we have $A\to0$, $G=C=M/2$, and $M\to2\tilde M$. The linear combinations of $\varphi$-integrated phase factors (\ref{C7}) and (\ref{B10}) satisfy the following relations:
\begin{equation}
\begin{split}
2 \hat I_0^{1234}+ \hat I_+^{1234}+ \hat I_-^{1234}&=0,
\\
2 \hat I_0^{1324}+ \hat I_+^{1324}+ \hat I_-^{1324}&=2 \hat I_0^{1423}+
\hat I_+^{1423}+ \hat I_-^{1423}\,
=2\pi\Big[3-4J_0\Big(\frac{s\tilde M}{4\Lambda^2_{\rm NC}}\sqrt{1-x^2}\Big)
+J_0\Big(\frac{s\tilde M}{2\Lambda^2_{\rm NC}}\sqrt{1-x^2}\Big)\Big],
\end{split}
\label{IntegralsNC2CM}
\end{equation}
and we finally have expression for cross section
\begin{equation}
\begin{split}
\sigma_{\rm NC^2}^{\gamma}\Big|^{Exclu.}_{\rm CM}&=
\alpha^2\,s\int_{-1}^1 dx\;
\Big(\frac{1}{s^2}+\frac{1}{t^2}+\frac{1}{u^2}\Big)
\Big(2 \hat I_0^{1423}+ \hat I_+^{1423}+ \hat I_-^{1423}\Big),
\end{split}
\label{NCGammaTCrossCMpure1}
\end{equation}
which vanishes in the absence of spacelike noncommutativity [see (\ref{E4})], as it should. 

Performing expansions in Bessel functions (\ref{IntegralsNC2CM}) we have found that all terms proportional to the $(1/\Lambda_{\rm NC})^4$ cancel out, while the next terms of $(1/\Lambda_{\rm NC})^8$ order of expansion contribute. 
So the first nonvanishing contribution than reads
\begin{equation}
\begin{split}
\sigma_{\rm NC^2}^{\gamma}\Big|^{\rm Expand}_{\rm CM}&
\lesssim \frac{21\pi\alpha^2}{640}\frac{s^3\tilde M^4}{\Lambda^8_{\rm NC}},
\;\tilde M^4=(c_{13}^2+c_{23}^2)^2 \lesssim1,
\end{split}
\label{NCGammaTCrossCMpureExpan}
\end{equation} 
and shows correctness of our computations of Eqs. (\ref{NCGammaTCrossnCMpure}) and  (\ref{IntegralsNC2CM}). 

Note that, after using CM frame and NC phase integrals (\ref{B10}) one can easily see that the above $x$-integrals contain no collinear singularities.

\subsubsection{SM$\times$NCQED interference contributions to the $\gamma\gamma\to \gamma\gamma$ cross section}

Since SM loop diagram contributions do not show the 
 $\varphi$ dependence \cite{Jikia:1993tc,Gounaris:1998qk,Gounaris:1999gh,Davoudiasl:1999di} 
and that ${M^\gamma_{\rm NC}}^\dagger=M^\gamma_{\rm NC}$, 
to determine the SM $\times$ NC interference term  
denoted as $\sigma_{\rm SM\times NC}^\gamma$, 
after using two identical particles in the final state argument, we found   
\begin{equation}
\begin{split}
&\sigma_{\rm SM\times NC}^{\gamma}\Big|^{Exclu.}_{\rm NCM}=
\frac{1}{64\pi^2 }\int^1_{-1}dx\frac{\omega^2_4}{s^2}\int_0^{2\pi} d\varphi\,\bigg[
\Big(M^{++++}_{\rm SM}+{M^{++++}_{\rm SM}}^\dagger\Big) M^{++++}_{\rm NC}
\\&\phantom{XXXXXXXXXXXXXXXXXx}
+\Big(M^{+-+-}_{\rm SM}+{M^{+-+-}_{\rm SM}}^\dagger\Big)  M^{+-+-}_{\rm NC}
+ \Big(M^{++--}_{\rm SM}+{M^{++--}_{\rm SM}}^\dagger\Big)  M^{++--}_{\rm NC}
\bigg],
\end{split}
\label{NCQED-SM}
\end{equation}
where the $\varphi$-integration of the NC helicity amplitudes (\ref{Helicity24}) gives:
\begin{equation}
\begin{split}
\int_0^{2\pi}d\varphi\; M_{\rm NC}^{++++}&=16\pi\alpha
\bigg(\frac{s}{u}\Big(I_- -I_+\Big)^{1234}-\frac{s^2}{tu}\Big(I_- -I_+\Big)^{1423}\bigg)
\\
\int_0^{2\pi}d\varphi \;M_{\rm NC}^{+-+-}&=16\pi\alpha
\bigg(\frac{t^2}{su}\Big(I_- -I_+\Big)^{1234}-\frac{t}{u}\Big(I_- -I_+\Big)^{1423}\bigg),
\\
\int_0^{2\pi}d\varphi \;M_{\rm NC}^{++--}&=16\pi\alpha
\bigg(\frac{u}{s}\Big(I_- -I_+\Big)^{1234}-\frac{u}{t}\Big(I_- -I_+\Big)^{1423}\bigg).
\end{split}
\label{NCGammaInterfIntphinCM1}
\end{equation} 

We define the structures of $I_{\pm}$ integrals in Eq. (\ref{NCGammaInterfIntphinCM1}) in Eq. (\ref{B10}) with respect to the NC type, and with respect to the NCM frame. Taking the above integrals from (\ref{B10}) for the spacelike noncommutativity only, one see that the deference of integrals $(I_--I_+)^{1234}$ vanishes. 

Since the interference contributions for spacelike NC arises from $I_\pm^{1423}$ integrals only, we take the 
$\varphi$-integrated NC amplitudes from Eq. (\ref{NCGammaInterfIntphinCM1}) and the real part of SM amplitudes from 
\cite{Gounaris:1998qk,Gounaris:1999gh,Davoudiasl:1999di,Bern:2001dg} and find the interference terms 
for the unpolarized cross section at high energies:
\begin{equation}
\begin{split}
\sigma_{\rm SM\times NC}^{\gamma}\Big|^{Exclu.}_{\rm NCM}&=
\frac{-\alpha}{2\pi}\int^1_{-1}dx\frac{\omega_4^2}{ s^2 tu}\big(I_- - I_+\big)^{1423}Re\Big[s^2 M^{++++}_{\rm SM}+t^2 M^{+-+-}_{\rm SM}+u^2 M^{++--}_{\rm SM}\Big],
\\
I^{1423}_{\pm}\Big|_{\rm NCM}&=2\pi J_0\bigg[\frac{\omega_1\omega_4}{2\Lambda^2_{\rm NC}}\Big(1\pm\frac{\omega_2}{\omega_1}\Big){\tilde M}\sqrt{1-x^2}\bigg].
\end{split}
\label{NCGammaInterfIntphinCM2}
\end{equation} 
In the NCM frame for spacelike noncommutativity, $\tilde M\not=0$ (\ref{NCTCrossSectMCM}), the high-energy regime expression relevant for the further experiments is good enough for numerical integrations. 
\begin{figure}[t]
\begin{center}
\includegraphics[width=8.5cm,angle=0]{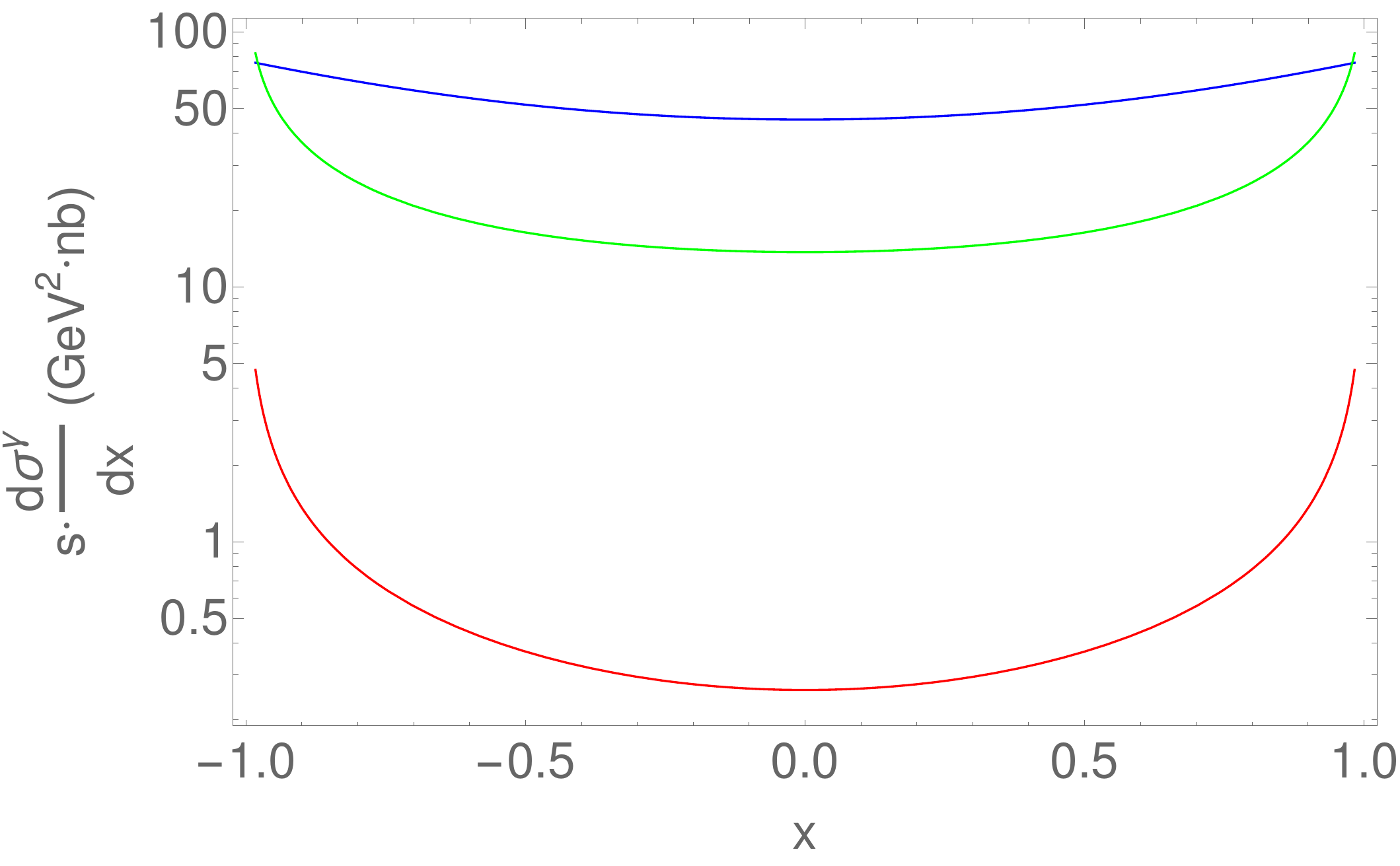}
\includegraphics[width=8.5cm,angle=0]{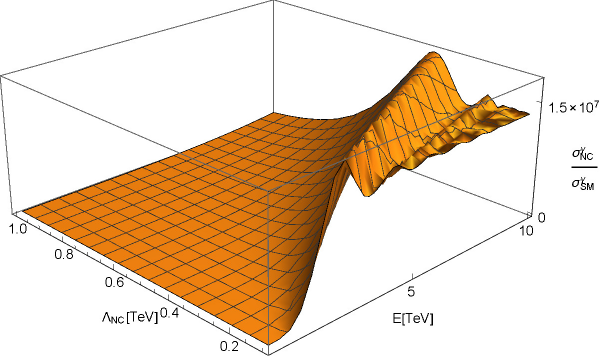}
\end{center}
\caption{Left: the SM and the NCQED incoming energy scaled LbyL angular distributions as functions of $x=\cos\vartheta$ (where $\vartheta$ is the polar angle) in the CM frame. The blue line is pure NC (\ref{NCGammaTCrossCMpure1}), the green line is interference SM$\times$NC (\ref{NCGammaInterfIntphiCM}), and the red line is the pure SM contribution, at $\sqrt{s}=\Lambda_{\rm NC}=100$ GeV. Right: 3D plot in the CM frame of exclusive numerically integrated LbyL ratio of the NCQED contributions with the SM $\rm\sigma_{NC}^\gamma/\sigma^\gamma_{\rm SM}$ using (\ref{NCGammaTCrossCMpure1})+(\ref{NCGammaInterfIntphiCM}), and SM, as a function of incoming energy $\rm E=\sqrt s=\sqrt{4\omega_1\omega_2}$  and scale $\Lambda_{\rm NC}$.}
\label{LbyLNC/SMtotcrossect}
\end{figure}

Like for the pure NC$^2$ case we chose the CM frame and from (\ref{B9}) and (\ref{NCGammaInterfIntphinCM2}) obtain  
\begin{equation}
\sigma_{\rm SM\times NC}^{\gamma}\Big|^{Exclu.}_{\rm CM}=-\frac{\alpha}{4}\int_{-1}^{1}\frac{dx}{stu}
\bigg[1-J_0\Big(\frac{s\tilde M}{4\Lambda^2_{\rm NC}}\sqrt{1-x^2}\Big)\bigg]
Re\Big[s^2 M^{++++}_{\rm SM}+t^2 M^{+-+-}_{\rm SM}+u^2 M^{++--}_{\rm SM}\Big],
\label{NCGammaInterfIntphiCM}
\end{equation} 
which for the small argument ($s\tilde M/4\Lambda^2_{\rm NC} < 1$) expansion of the Bessel function eliminates $(1-x^2)$ in denominator:
\begin{equation}
\begin{split}
\sigma_{\rm SM\times NC}^{\gamma}\Big|^{Exclu.}_{\rm CM}&=\frac{-\alpha\; s \;{\tilde M}^2}{64\;\Lambda^4_{\rm NC}}\int_{-1}^{1}dx\;
Re\Big[M^{++++}_{\rm SM}+\frac{t^2}{s^2} M^{+-+-}_{\rm SM}+\frac{u^2}{s^2}  M^{++--}_{\rm SM}\Big].
\end{split}
\label{NCGammaInterfIntphiCM3Expand}
\end{equation} 
The top left plot in Fig.\ref{LbyLNC/SMtotcrossect}--log scale--shows the angular distributions as functions of $x$ in the CM frame. How ``flat''  the NCQED contribution is depends on the ratio $\sqrt{s}/\Lambda_{\rm NC}$. This ratio varies roughly from one to three in the scenarios considered in the PbPb part. The plots are evaluated at $\sqrt{s}=\Lambda_{\rm NC}=100$ GeV. In this case the NC contributions are already significantly larger than the standard model. Note that the pure NCQED contribution and interference contributions are at similar scale. 

In the SM, LbyL scattering \cite{Gounaris:1999gh,Bern:2001dg} is tree-level forbidden and goes only via quantum loops,\footnote{ The main contributions comes from the first loop  \cite{Passarino:1978jh,Dong:1992iz,Jiang:1992sm,Dong:1992fa,Dong:1992hg,Jikia:1993tc,Gounaris:1998qk,Gounaris:1999gh,Davoudiasl:1999di,Bern:2001dg,Baur:2001jj}, while the second-loop QED and QCD contributions to the amplitudes are small \cite{Bern:2001dg}, contributing to the cross section on the level of a few percent only.} while in the NCQED it exists at tree level.  In the 3D plot of Fig.\ref{LbyLNC/SMtotcrossect}, we  illustrated the exclusive ratio of LbyL cross sections $\rm\sigma^\gamma_{NC}/\sigma^\gamma_{SM}$ of the NC contributions (\ref{NCGammaTCrossCMpure1})+(\ref{NCGammaInterfIntphiCM}) with the SM one in the CM frame, as functions of incoming energy E and the NC scale $\Lambda_{\rm NC}$. Integration over $x=\cos\vartheta$ was performed with cutoff regulator $\epsilon=10^{-7}$. The absence of cosine and presence of Bessel functions only produce mild oscillatory behavior which starts to show up gradually. To see transparently such behavior, we take a wide range of energy and relatively low NC scale. Thus, in the upper 3D plot in Fig.\ref{LbyLNC/SMtotcrossect}, right corner, there are two peaks when energy grows very high and for the small NC scales, where the first corresponds to the pure NC contributions (\ref{NCGammaTCrossCMpure1})--large due to the small value of the NC scale--while the second peak corresponds to the interference (\ref{NCGammaInterfIntphiCM}), respectively.  The 3D plot in Fig.\ref{LbyLNC/SMtotcrossect} shows behavior similar as demonstrated further by the lines of Figs.\ref{figure1} independently whether we are in the fiducial phase space or not. Also the next peaks are actually negligible.  

$\phantom{Thus all XXxxxxxand given as (\ref{A.45-5}), and given as 
(\ref{A.45-5})and given as (\ref{A.45-5}) respectively respectivelyrespectivel}$Q.E.D.  

 
\section{NCQED AND THE ATLAS $^{208}\rm PB-ION$ COLLISION EXPERIMENTS: \protect\\
$\rm PbPb(\gamma\gamma)\to Pb^{\ast}Pb^{\ast}\gamma\gamma$ AND $\rm PbPb(\gamma\gamma)\to Pb^{\ast}Pb^{\ast}\ell^+\ell^-$}

\begin{figure}[t]
\begin{center}
\includegraphics[width=15cm,angle=0]{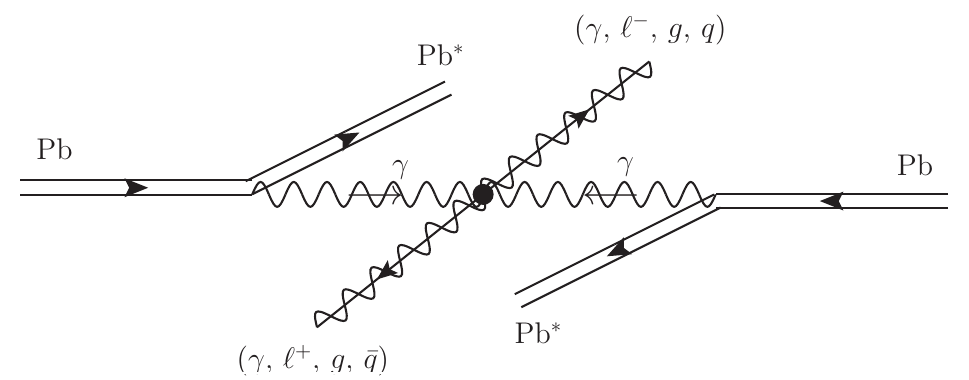} 
\end{center}
\caption{Diagrammatic sketch of the ${\rm PbPb}(\gamma\gamma)\to {\rm Pb}^{\ast}{\rm Pb}^{\ast}(\gamma\gamma,\ell^+\ell^-,gg,\bar qq)$ scatterings in ultraperipheral $^{208}$Pb-ion collisions in the ATLAS experiment at LHC, 
with a black dot representing any point like and/or short-distance (loop) interactions. 
The wavy-solid line corresponds to the outgoing final state pairs; otherwise notations are self-evident.}
\label{fig:Pb}
\end{figure}

In this section we continue searching for the physical phenomena of noncommutativity of spac-etime coordinates via ATLAS $^{208}$Pb-ion experiments producing $\gamma\gamma$ and $\ell^+\ell^-$ final states, sketched diagrammatically in Fig.\ref{fig:Pb}. Particularly, in an attempt to estimate a bound to 
the scale of spacelike NCQED, we take the convoluted exclusive cross sections 
\begin{equation}
\sigma^f_{Theory}=\sigma^f_{\rm SM^2}+
\sigma^f_{\rm SM\times NC}+\sigma^f_{\rm NC^2},\;f=\gamma\gamma,\ell^+\ell^-,gg,\bar q q,
\label{TOTALLbyL}
\end{equation}  
and, under the assumption $\sigma^f_{Theory}\simeq\sigma^f_{Experiment}$ for dominant 
$f=\gamma\gamma,\ell^+\ell^-$ channels, 
compare with convoluted ATLAS measured cross section. Note that produced $gg$ and $\bar q q$ 
exclusive but virtual states in lead experiments get realized as real jets and real meson pair final states captured in detectors, respectively.

In the $\rm PbPb(\gamma\gamma)\to Pb^{\ast}Pb^{\ast}\gamma\gamma$ reaction the incoming $^{208}$Pb ions have survived the electromagnetic interaction, with a possible electromagnetic excitation due to the energy loss via photon emissions, denoted by $(\ast)$. Hence, the final state consists of two low-energy photons and no further activity in the detector--in particular, no reconstructed charged-particle tracks originating from the IP; see \cite{Aad:2019ock}. Although the cross section for exclusive 
$\gamma\gamma\to\gamma\gamma$ process is tiny, various techniques can 
be used to study it indirectly \cite{dEnterria}.

To apply NCQED to the data for LbyL scattering we use the lepton-, quark- and W$^\pm$-boson-loop diagrams 
from Refs. \cite{Gounaris:1998qk,Gounaris:1999gh,Davoudiasl:1999di,Bern:2001dg,Baur:2001jj}, 
combine them as the full SM amplitudes together with NC amplitudes (\ref{Helicity24}), and use Eq. (\ref{TCrossSectC}) to obtain the total exclusive $\gamma\gamma\to \gamma\gamma$  differential cross section:
\begin{equation}
\frac{d\sigma^\gamma}{d\Omega}\Bigg|^{Exclu.}_{Theory}
=\Big(\frac{\omega_4}{4\pi s}\Big)^2\frac{1}{4}\sum\limits^5_{i=1} \Big(\big|M_{\rm SM}^{h_i}\big|^2 + {M_{\rm SM}^{h_i}}^\dagger 
M_{\rm NC}^{h_i} +M_{\rm SM}^{h_i} 
{M^{h_i}_{\rm NC}}^\dagger + \big| {M^{h_i}_{\rm NC}}\big|^2\Big),
\label{SMNCTotCrossSect}
\end{equation}  
where $h_1=++++$, $h_2=++--$, $h_3=+-+-$, $h_4=+--+$, and $h_5=+---$. The pure SM contributions do arise from all five amplitudes $h_i$, however, due to the complementarity of NCQED with QCD, only three MHV NC amplitudes with helicities $h_1,h_2$, and $h_3,$ from Eq. (\ref{Helicity24}) match with the SM in the interference terms (\ref{NCGammaInterfIntphinCM2}) giving rise to the cross section in Eq. (\ref{SMNCTotCrossSect}).

\subsection{Kinematics of the ultraperipheral ion scatterings}

In the ultraperipheral scattering one considers the electromagnetic fields of the incoming 
heavy ions to be a spectrum of real photons moving along the beam axis. This simplifies the incoming photon kinematics in the Laboratory (nCM) frame to be a head-to-head collision. 
Thus, we take the $\rm PbPb(\gamma\gamma)\to Pb^{\ast}Pb^{\ast}\gamma\gamma$ cross section 
as a convolution of exclusive theoretical differential cross section (\ref{SMNCTotCrossSect}) [(\ref{NCGammaTCrossnCMpure}) and (\ref{NCGammaInterfIntphinCM2}) plus the SM one from 
Refs. \cite{Gounaris:1998qk,Gounaris:1999gh,Davoudiasl:1999di,Bern:2001dg,Baur:2001jj}] 
with the incoming photon flux factors, integrated over the range of observed solid angle and measured outgoing photon energies \cite{Aaboud:2017bwk,Aad:2019ock}.

The incoming photon energy spectrum space $(\omega_1,\omega_2)$ is usually reparametrized by the diphoton invariant mass $m_{\gamma\gamma}$, rapidity $Y$, and the volume element transformation:
\begin{equation} 
m_{\gamma\gamma}=\sqrt{s}=\sqrt{4\omega_1\omega_2} , \;\;
Y=\frac{1}{2}\ln\frac{\omega_1}{\omega_2}, \;\;
d\omega_1 d\omega_2=\frac{m_{\gamma\gamma}}{2}dm_{\gamma\gamma} dY.
\label{diphinvmassrapid}
\end{equation}

To determine the rapidity $Y$ of the produced outgoing diphoton mass which, for symmetric systems, is maximal at $Y= 0$ when $\omega_1^{\rm max} = \omega_2^{\rm max} \approx \gamma/b_{\rm min}$ with $b_{\rm min}$ being the minimum separation between the two equal charged nuclei of radius $R_N$, we use for the lead ion spectrum, the impact 
parameter-dependent expression integrated from $b_{\rm min}$ to infinity, with the requirement $b_{\rm min}=R_N$ plus a correction equivalent to the geometrical condition $|\vec b_1-\vec b_2|>R_{\rm min}=2R_N$ to ensure that all collisions occur without hadronic overlap and breakup of the colliding beams. Propagated uncertainties to the final cross sections are of the order of $\pm20\%$ for lead-lead ion collisions, covering different form-factor parametrization and the convolution of the nuclear photon fluxes  \cite{dEnterria:2013zqi}. 

\subsection{Criteria of the ATLAS events selection denoted as the ATLAS cuts}

In the ATLAS  experiment ultraperipheral $\rm PbPb(\gamma\gamma)\to Pb^{\ast}Pb^{\ast}\gamma\gamma$ events \cite{Aaboud:2017bwk,ATLAS-CONF-2019-002,Aad:2019ock} the two incoming photons emitted by the $^{208}$Pb ions are almost collinear, i.e., almost head to head: thus, they are selected by the following rules:

\noindent
-- small diphoton transverse momentum $p_T$  and acoplanarity.\\
-- diphoton invariant mass $m_{\gamma\gamma}$ larger than 6 GeV.\\
-- transverse energy $E_T$ of each photon larger than $E_T\ge E_0=3$ GeV.\\
-- each photon absolute pseudorapidity $|y_i|$ smaller than $y_0= 2.37$ corresponds to constraint on the angular range $10^\circ \lesssim \vartheta\lesssim170^\circ\;(+0.9898 \lesssim \cos\vartheta(=x)\lesssim -0.9898)$
of photons recorded by ATLAS. This ATLAS cutoff   
$-1+\epsilon \lesssim \cos\vartheta \lesssim+1-\epsilon\; {\rm with} \;\epsilon=0.0102$  
is experimentally determined and, thus, different with respect to the numerical dimensionless cutoff $\epsilon=\pm10^{-7}$ used for exclusive $2\to2$ scatterings containing a collinear singularity in $t$ and/or $u$ channels.

The first condition allows us to assume that incoming photons are real and collinear. This assumption simplifies 
the constraint on transverse energy to be only one, since the transverse momentum of outgoing photons must equal each other. The above conditions shall be translated into constraints implied on the theoretical cross section calculation. 

We continue by assuming $k_1$ and $k_2$ as incoming momenta and outgoing momenta being $k_3$ and $k_4=k_1+k_2-k_3$, respectively. Then pseudorapidities $y_{3,4}$ of outgoing depend only on the angle to the z axis:
\begin{equation}
y_3=2Y-y_4,\;
y_4=\frac{1}{2}\ln\frac{1+x}{1-x},
\label{y_3y_4}
\end{equation}
where $Y$ from Eq. (\ref{diphinvmassrapid}) is the rapidity of the total momentum $(k_1+k_2)=(k_3+k_4)$. By this we can express the last constraint
\begin{equation}
|y_3|\le y_0\,\;\&\,\; |y_4|\le y_0 \Longrightarrow \left|\ln\frac{1+x}{1-x}\right|\le 2y_0,
\label{y3y4}
\end{equation}
and we have first simpler constraint on $Y$
\begin{equation}
\frac{1}{2}\left(\frac{1}{2}\ln\frac{1+x}{1-x}-y_0\right)\le Y\le\frac{1}{2}\left(y_0-\frac{1}{2}\ln\frac{1+x}{1-x}\right).
\label{xYconstr}
\end{equation}

To include the transverse energy constraint $E_T\ge E_0$ requires a bit of care. Using diphoton invariant mass 
$m_{\gamma\gamma}$,  
we can express this constraint as
\begin{equation}
E_T=\omega_4\sqrt{1-x^2}=\frac{m_{\gamma\gamma}\sqrt{1-x^2}}{(e^Y+e^{-Y})-(e^Y-e^{-Y})x}\ge E_0.
\label{transvE}
\end{equation}
For a typical cut which requires transverse energy $E_T\ge E_0$ and both outgoing particles to bear pseudorapidity $y_{3,4}\le y_0$, the full sophisticated integral domain is defined as follows:
\begin{equation}
\begin{split}
\varphi&\in [0, 2\pi],\; |x|\le\frac{e^{2y_0}-1}{e^{2y_0}+1},\; m_{\gamma\gamma}\ge 2E_0,
\\
Y&\in \left[\frac{1}{2}\left(\frac{1}{2}\ln\frac{1+x}{1-x}-y_0\right), \frac{1}{2}\left(y_0-\frac{1}{2}\ln\frac{1+x}{1-x}\right)\right]
\\&\cap\left[\frac{1}{2}\ln\frac{1+x}{1-x}+\ln\left(\frac{m_{\gamma\gamma}}{2E_0}-\sqrt{\frac{m_{\gamma\gamma}^2}{4E_0^2}-1}\right), \right.
\left.\frac{1}{2}\ln\frac{1+x}{1-x}+\ln\left(\frac{m_{\gamma\gamma}}{2E_0}+\sqrt{\frac{m_{\gamma\gamma}^2}{4E_0^2}-1}\right)\right],
\end{split}
\label{Constraints}
\end{equation}
and that was used in all of our further numerical integrations in fiducial phase space.

\subsection{Equivalent photon approximation and fiducial cross section}

Photons originating from the high-energy $^{208}$Pb-ion nuclei can be viewed as the photon beam in the equivalent photon approximation (EPA) \cite{Fermi:1925fq,vonWeizsacker:1934nji,Williams:1934ad}. Because of the coherent action of all the protons in the nucleus the electromagnetic field surrounding 
each fast-moving nucleus with the charge $Ze$ is very strong and it is approximated by a distribution of (almost) real photons moving along the beam direction. The simplest version of this approximation is to introduce a photon number function $n(\omega)$, where $\omega$ is the monophoton energy. Under such a convention convolution of exclusive cross section (\ref{SMNCTotCrossSect})  with the photon number function in the fiducial phase space, was expressed in Refs. \cite{Hagiwara:1994pw,KlusekGawenda:2010kx,Harland-Lang:2018iur} as
\begin{equation}
\frac{d\sigma^\gamma}{dm_{\gamma\gamma}}=\int\frac{2}{m_{\gamma\gamma}}  dY d\Omega n(\omega_1)n(\omega_2)\bigg[\frac{d\sigma^\gamma(\omega_1,\omega_2,\vartheta,\varphi)}{d\Omega}\bigg]^{Exclu}_{Theory},
\label{TotFiduc}
\end{equation}
with simplest photon number functions $n(\omega_i)$ from the monopole form factor. 
However this did not quite take into account the ultraperipheral nature of the process:
\begin{equation}
n(\omega)=\frac{2Z^2\alpha}{\pi}\left(\frac{2\omega^2+\gamma^2\xi^2}{2\gamma^2\xi^2}\ln\left(1+\frac{\gamma^2\xi^2}{\omega^2}\right)-1\right),\;
\xi=\sqrt{\frac{6}{\big<r^2\big>}}. 
\label{bnumbfun}
\end{equation}
Here, $\rm\gamma=\sqrt{s_{NN}}/(2m_{N})$ is the Lorentz relativistic factor for the Nth ion nucleus, and $\big<r^2\big>$ is the mean squared radius of that Nth nucleus.

An improved version of the monopole photon number function $N(\omega, |\vec b|)$ is constructed, 
including the ``$\vec b$-impact parameter''  which is a two-dimensional vector that marks 
the position of the nucleus from the position of impact on the plane perpendicular to the beam direction  
\cite{Baur:2001jj}.  Photon number function $N(\omega,b)$, being expressed in terms of the 
second kind modified Bessel function $K_1$, is determined by the nucleus elastic form factor:
\begin{equation}
N(\omega,b)=\frac{Z^2\alpha}{\pi^2}\left(\frac{\omega}{\gamma}K_1\left(b\frac{\omega}{\gamma}\right)-\sqrt{\frac{\omega^2}{\gamma^2}+\xi^2}\cdot K_1\left(b\cdot\sqrt{\frac{\omega^2}{\gamma^2}+\xi^2}\right)\right)^2
, \;\gamma=\frac{\sqrt{s_{\rm NN}}}{2 m_u}.
\label{bnumbfunct}
\end{equation}
In the ATLAS experiment, from $^{208}$Pb-ion nuclear data \cite{DeJager:1987qc} one finds $\xi=0.088$ GeV, 
$\gamma$ is the $^{208}$Pb-ion Lorenz factor, and $m_u=0.931$ GeV is the atomic mass unit. 

The differential cross section in the fiducial phase space is then expressed as the following sevenfold integral:
\begin{equation}
\begin{split}
\frac{d\sigma^\gamma}{dm_{\gamma\gamma}}&=2\int\bigg[\frac{d\sigma(\omega_1,\omega_2,\vartheta,\varphi)}{d\Omega}\bigg]^{Exclu.}_{Theory}d\Omega\frac{dY}{m_{\gamma\gamma}} 
d^2 b_1 d^2 b_2 \Theta(|\vec{b}_1-\vec{b}_2|-R_{\rm min}) N(\omega_1, |\vec b_1|)N(\omega_2, |\vec b_2|),
\end{split}
\label{TotFidu}
\end{equation}
with $R_{\rm min}=2R_{\rm (^{208}Pb)}=14$ $ \rm fm\simeq 71 \; GeV^{-1}$, where in practice 
the fourfold integration over impact parameters $\vec b_1$ and $\vec b_2$ can be performed by 
defining the improved luminosity function
$\frac{d^2L_{\gamma\gamma}}{dm_{\gamma\gamma}dY}$:
\begin{equation}
\begin{split}
&\frac{d^2 L_{\gamma\gamma}}{dm_{\gamma\gamma} dY}=\frac{2}{m_{\gamma\gamma}}\int d^2 b_1 d^2 b_2 \Theta(|\vec{b}_1-\vec{b}_2|-R_{\rm min}) N(\omega_1, |\vec b_1|)N(\omega_2, |\vec b_2|)
\\&=\frac{4\pi}{m_{\gamma\gamma}}\int d |\vec b_1| d |\vec b_2| \int\limits_0^{2\pi}d\phi N(\omega_1, |\vec b_1|)N(\omega_2, |\vec b_2|)
\Theta\left({\sqrt{|\vec b_1|^2+|\vec b_2|^2-2|\vec b_1||\vec b_2|\cos\phi}-R_{\rm min}}\right).
\end{split}
\label{intb1b2}
\end{equation}
Taking this into account the cross section of $\rm PbPb(\gamma\gamma)\to Pb^{\ast}Pb^{\ast}\gamma\gamma$ is expressed in the fiducial phase space by convoluting (\ref{SMNCTotCrossSect}) 
with luminosity function (\ref{intb1b2}) in the following form:
\begin{equation}
\frac{d\sigma^\gamma}{dm_{\gamma\gamma}}=\frac{1}{2}\int dY  \frac{d^2 L_{\gamma\gamma}}
{d m_{\gamma\gamma}\,dY}d\Omega
\bigg[\frac{d\sigma^\gamma(\omega_1,\omega_2,\vartheta,\varphi)}{d\Omega}\bigg]^{Exclu}_{Theory}.  
\label{TotFidu1}
\end{equation}

Using this photon number function from one-loop and monopole approximations we first compute numerically 
the fiducial total cross section of the $\rm PbPb(\gamma\gamma)\to Pb^{\ast}Pb^{\ast}\gamma\gamma$ collision in the SM for the ATLAS cuts and found it to be 57 nb, 
which is comparable to the previously reported  values~\cite{Aaboud:2017bwk, Aad:2019ock}. 
Two other ATLAS kinematic cuts discussed in the literature were also listed in Table I. 
In the same Table I we have also presented additional two cross sections, 
that is the 120 and 12 nb coming from the ALICE and CMS experiments, respectively.
\begin{table}
\begin{center}
\begin{tabular}{|c|c|c|}
\hline
Kinematic cuts & Collaboration & $\sigma^\gamma_{\rm SM}(1-{\rm loop})_{monopole}$ (nb)\\
\hline
$m_{\gamma\gamma}>5$ GeV, $|y_i|<$7 &ATLAS & 382 \\
$m_{\gamma\gamma}>5$ GeV, $p_t>2$ GeV,  $|y_i|<$7 &ATLAS& 190 \\
$p_t>$0.9 GeV, $|y_i|<$0.7 &ALICE & 120\\
$p_t>$5.5 GeV, $|y_i|<$2.5 &CMS & 12\\
\hline
\end{tabular}
\caption{A few total SM fiducial $\rm PbPb(\gamma\gamma)\to Pb^{\ast}Pb^{\ast}\gamma\gamma$ 
cross sections calculated at $\sqrt{\rm s_{ NN}}=$5.5 TeV (Lorentz factor $\gamma=2930$) 
\cite{Klusek-Gawenda:2016euz}, and for various collaboration experimental cuts.}
\end{center}
\label{table1}
\end{table}
In the case of monopole approximation our integrated cross section at the SM one loop given in Table I relatively well agree with numbers published in Table I in Ref. \cite{Klusek-Gawenda:2016euz}. 

\subsection{NCQED contribution to $\rm PbPb(\gamma\gamma)\to Pb^{\ast}Pb^{\ast}\gamma\gamma$ reaction for ATLAS cuts}

\begin{figure}[t]
\begin{center} 
\includegraphics[width=8.5cm,angle=0]{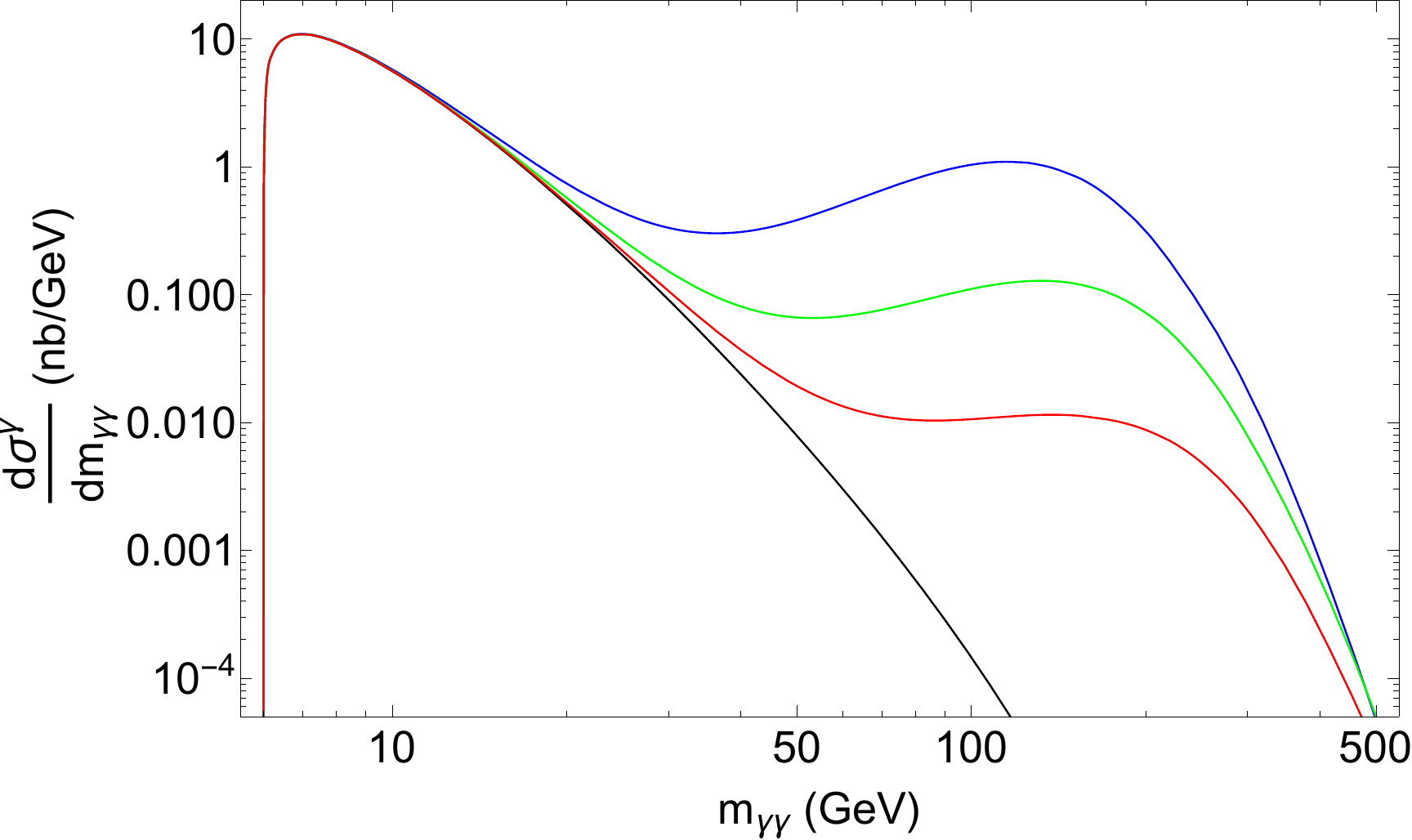}
\includegraphics[width=8.5cm,angle=0]{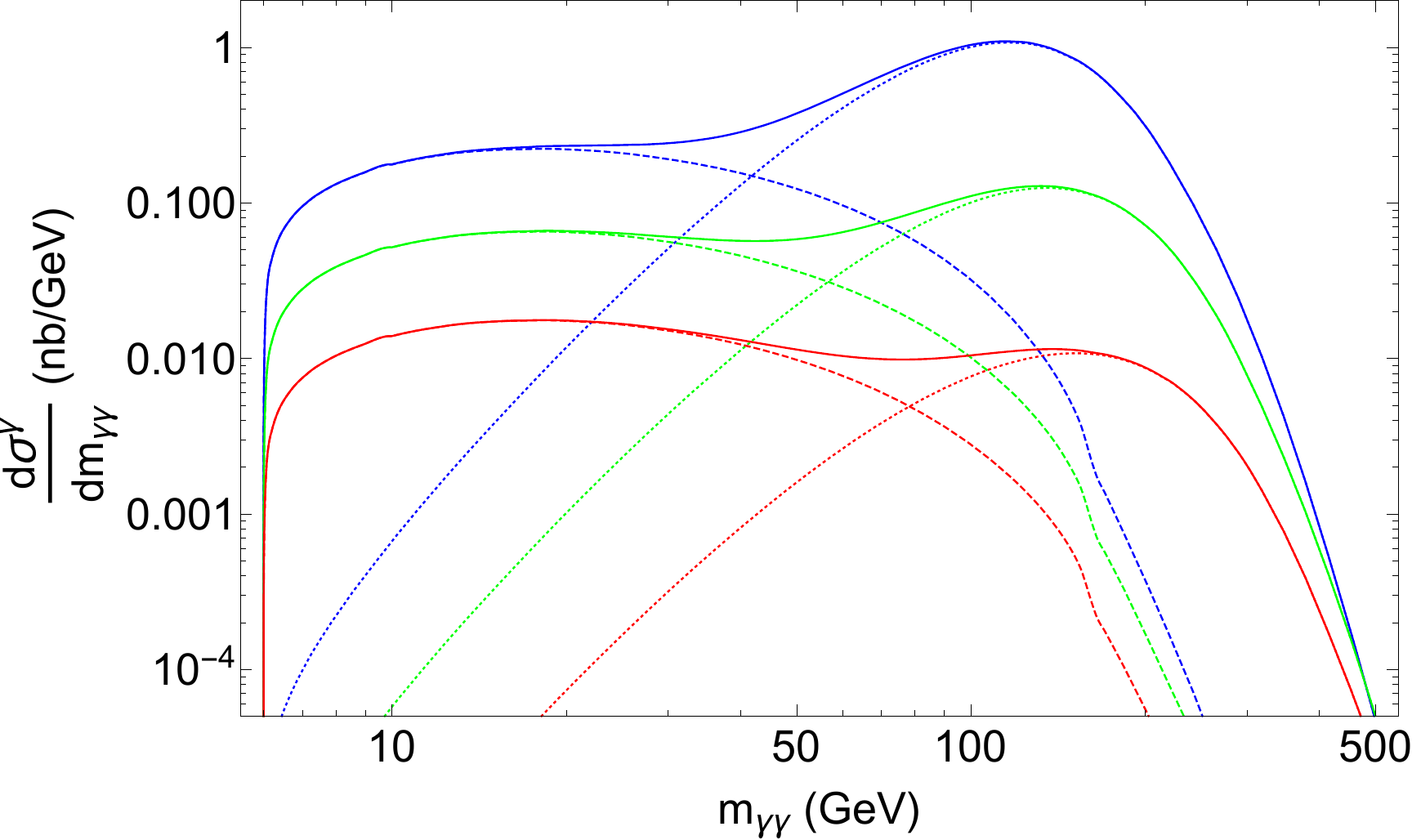}
\end{center}
\caption{Left panel: convoluted differential cross section versus diphoton invariant mass distribution of the 
$\rm PbPb(\gamma\gamma)\to Pb^{\ast}Pb^{\ast}\gamma\gamma$ collision in fiducial phase space of the PbPb system in the ATLAS $\rm PbPb(\gamma\gamma)\to Pb^{\ast}Pb^{\ast}\gamma\gamma$ experiment with the ATLAS cuts and $\sqrt{\rm s_{NN}}=5.02$ TeV, for the SM (black) as well as the SM+($\theta$-exact NCQED) with $\rm\Lambda_{NC}$ values 53 (blue line), 72 (green line) and 100 GeV (red line). Right panel: convoluted  NCQED related contributions to the fiducial cross section versus diphoton invariant mass distribution under the ATLAS cuts conditions for $\Lambda_{\rm NC}$ values 53 (blue line), 72 (green line) and 100 GeV (red line). Dotted lines are pure NC contribution $\sigma^{\gamma}_{\rm NC^2}$, dashed are interference terms $\sigma^{\gamma}_{\rm SM\times NC}$, and solid lines are $\sigma^{\gamma}_{\rm NC}=\sigma^{\gamma}_{\rm NC^2}+\sigma^{\gamma}_{\rm SM\times NC}$.}
\label{figure1}
\end{figure}

Applying the protocol tested in the former subsection, we calculate the combined contribution from both the SM one-loop and NCQED tree-level $\gamma\gamma\to\gamma\gamma$ amplitudes in the ATLAS $\rm PbPb(\gamma\gamma)\to Pb^{\ast}Pb^{\ast}\gamma\gamma$ experiment, with $^{208}$Pb-ion center of mass energy $\sqrt{\rm s_{NN}}=5.02$ TeV \cite{Aad:2019ock}, which is sensitive to most of the $4\pi$ solid angle, implying that the integral (\ref{TotFidu},\ref{TotFidu1}) ranges over the ATLAS cuts: $\vartheta$ cutoff with $\epsilon=0.0102$, with $0 \le\varphi\le 2\pi$,  $m_{\rm Pb}$=0.9315 GeV, and $\gamma=2693$, giving the data span as a function of diphoton invariant mass $m_{\gamma\gamma}$, presented in Fig.\ref{figure1} and Table~II.

It seems that in the LbyL scattering measurements ATLAS has recorded only events with less than 30 GeV diphoton invariant mass [Fig.3(a) in Ref. \cite{ATLAS-CONF-2019-002}, or the same figure as Fig.2(b) in Ref. \cite{Aad:2019ock}]. Inspecting Fig. 4 in the Appendix in Ref. \cite{ATLAS-CONF-2019-002}, which displays event 453765663 from run 366994 with diphoton invariant mass of $m_{\gamma\gamma}=\sqrt{s}=29$ GeV, we conclude that this LbyL  scattering event and other recorded events certainly belong to the class of exclusive relatively low-energy (with respect to $m_t$) processes,  which is in accord with the conclusion in Ref. \cite{Klusek-Gawenda:2016euz} that the cross section for elastic $\gamma\gamma\to\gamma\gamma$ scattering could be measured in the present $^{208}$Pb-ion collisions only for subprocess energies smaller than 30 GeV. 
\begin{table}
\begin{center}
\begin{tabular}{|c|c|c|c|c|c|}
\hline
$\rm\Lambda_{NC}$ (GeV) &  $\sigma^{\gamma}_{\rm SM}$ (nb) & $\sigma^{\gamma}_{\rm SM\times NC}$ (nb) & $\sigma^{\gamma}_{\rm NC^2}$ (nb)  &  $\rm\frac{d\sigma^{\gamma}_{\rm NCQED}}{dm_{\gamma\gamma}}\big|_{max} \big( \frac{nb}{GeV}\big)$  & $\rm m_{\gamma\gamma}|_{max}(GeV)$\\
\hline
53 & 57&12.1 & 125.5 & 1.09 & 115 \\
72 & 57&3.6 & 17.6 & 0.13 & 132 \\
100 & 57&1.0 & 1.8 & 0.011 & 138 \\
\hline
\end{tabular}
\caption{Summary of the predicted NCQED related contribution to the $\rm PbPb(\gamma\gamma)\to Pb^{\ast}Pb^{\ast}\gamma\gamma$ fiducial cross sections $\sigma^{\gamma}_{\rm SM+NC}=
\sigma^{\gamma}_{\rm SM}+\sigma^{\gamma}_{\rm SM\times NC}+\sigma^{\gamma}_{\rm NC^2}$ for ATLAS cuts at $\sqrt{s_{\rm NN}}=5.02$ TeV and various $\Lambda_{\rm NC}$ values.}
\end{center}
\label{table2}
\end{table}

We first consider matching the reported experimental fiducial cross section of 78 nb \cite{Aad:2019ock} 
by combination (\ref{SMNCTotCrossSect})-(\ref{TotFidu1}), which yields an unimpressive scale $\rm\Lambda_{NC}$ of about 72 GeV, a value which had been ruled out some time ago. 
Since the relevant NC scale is small in the ATLAS $\rm PbPb(\gamma\gamma)\to Pb^{\ast}Pb^{\ast}\gamma\gamma$ scenario, we notice that pure NC  amplitude $\big| {M^\gamma_{\rm NC}}\big|^2$ gives the main contribution to the fiducial cross section diphoton invariant mass distribution peak value of $\frac{d\sigma^{\gamma}}{dm_{\gamma\gamma}}$ (right  panel in Fig.\ref{figure1}), which occurs moderately above $\rm\Lambda_{NC}$ as expected from the Bessel function $J_0$ dependence [see Eqs. (\ref{NCGammaTCrossnCMpure}) and (\ref{NCGammaInterfIntphinCM2})] on the scale 
$\rm\Lambda_{NC}$. So the high-energy peak value of $\frac{d\sigma^{\gamma}}{dm_{\gamma\gamma}}$ in the left panel in Fig.~\ref{figure1} is a direct consequence of the pure NC amplitudes and determines the possibility of observing excessive events due to the NCQED. 

\subsection{Recent ATLAS lead experiments}

Proposing that the difference between LHC-ATLAS measured $\rm PbPb(\gamma\gamma)\to Pb^{\ast}Pb^{\ast}\gamma\gamma$ scattering cross section in ultraperipheral lead-lead collisions and the SM prediction is mainly due to the noncomutativity of space-time (\ref{TOTALLbyL}),  by using Eqs. (\ref{Helicity24}), (\ref{TOTALLbyL}) and (\ref{Constraints})--(\ref{bnumbfunct}) we obtain the total cross section in the fiducial phase space to be $59.8$ nb, relatively close to the experimental mean value, however only at  $\rm\Lambda_{NC}=100$ GeV. Inspecting Table II one may clearly see that present ATLAS experiment basically gives only the NC scale smaller than 100 GeV. Namely, from the left panel in Fig. \ref{figure1} and Table II we see  that second peak values of $\rm\sim (1,0.1,0.01)$ nb/GeV exist for $\Lambda_{\rm NC}\gtrsim (53,72,100)$ GeV, respectively. In addition, after checking the distribution of cross section with respect to diphoton invariant mass $m_{\gamma\gamma}$, we conclude that such a scenario does not explain the observation of the ATLAS experiment, since the cross section induced by NCQED amplitude comes from a wide band at $\rm m_{\gamma\gamma}\gtrsim 100$ GeV, as nicely presented by the ATLAS Collaboration in Fig.2(b) in Ref. \cite{Aad:2019ock}. Given the current integrated luminosity accumulation speed ($\rm\sim 1$ $\rm nb^{-1}$/yr) for current LHC, we doubt that any stronger bound of $\rm\Lambda_{NC}\simeq 100$ GeV can be achieved from the present-day ATLAS $\rm PbPb(\gamma\gamma)\to Pb^{\ast}Pb^{\ast}\gamma\gamma$ experiment.

The lepton pair production in the framework of ATLAS lead experiments shall be presented and discussed further. 

\subsection{LbyL scatterings in the next-generation collider experiments}

\subsubsection{Convoluted cross section versus diphoton invariant mass distribution}

\begin{figure}[t]
\begin{center}
\includegraphics[width=8.5cm,angle=0]{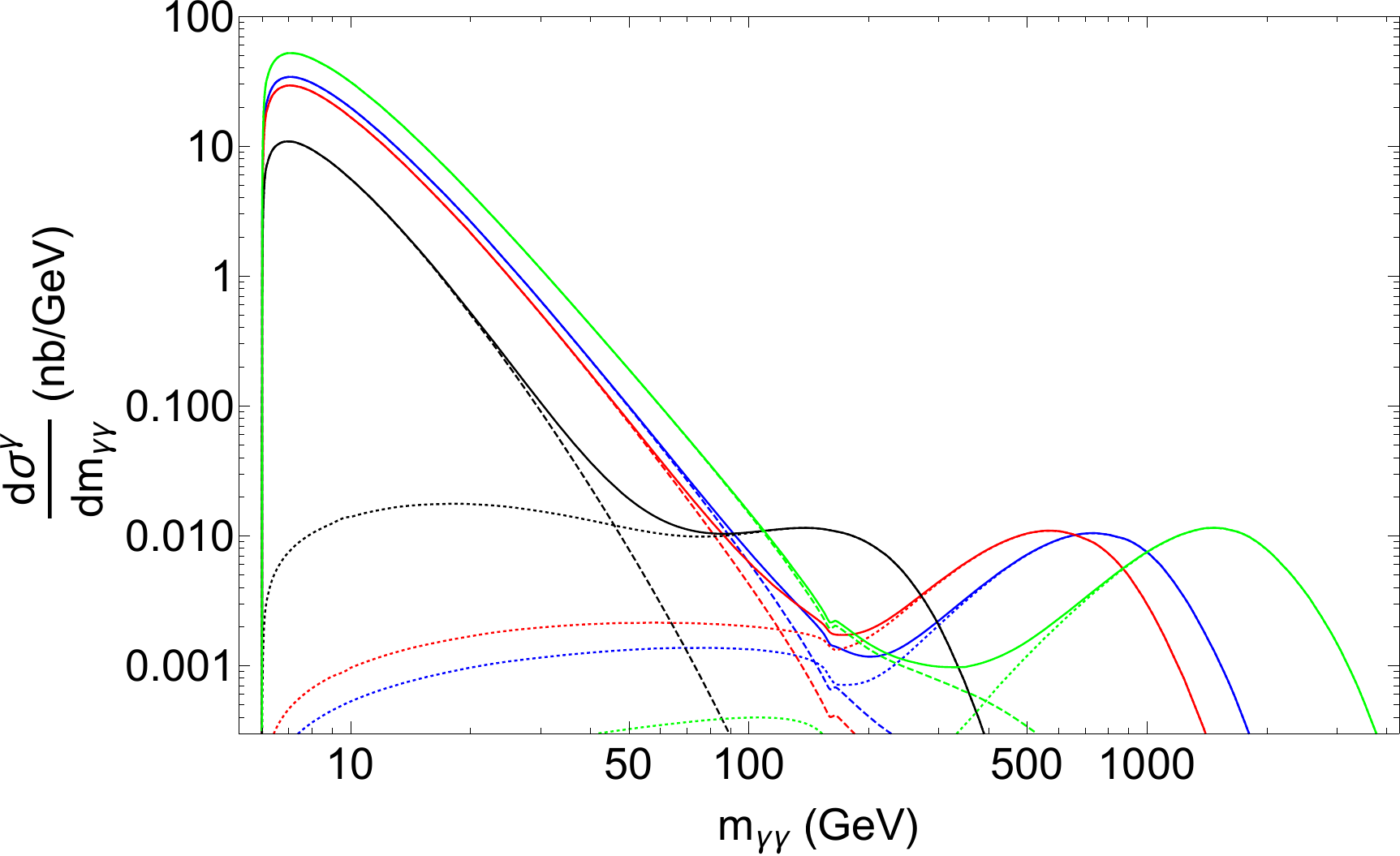}
\includegraphics[width=8.5cm,angle=0]{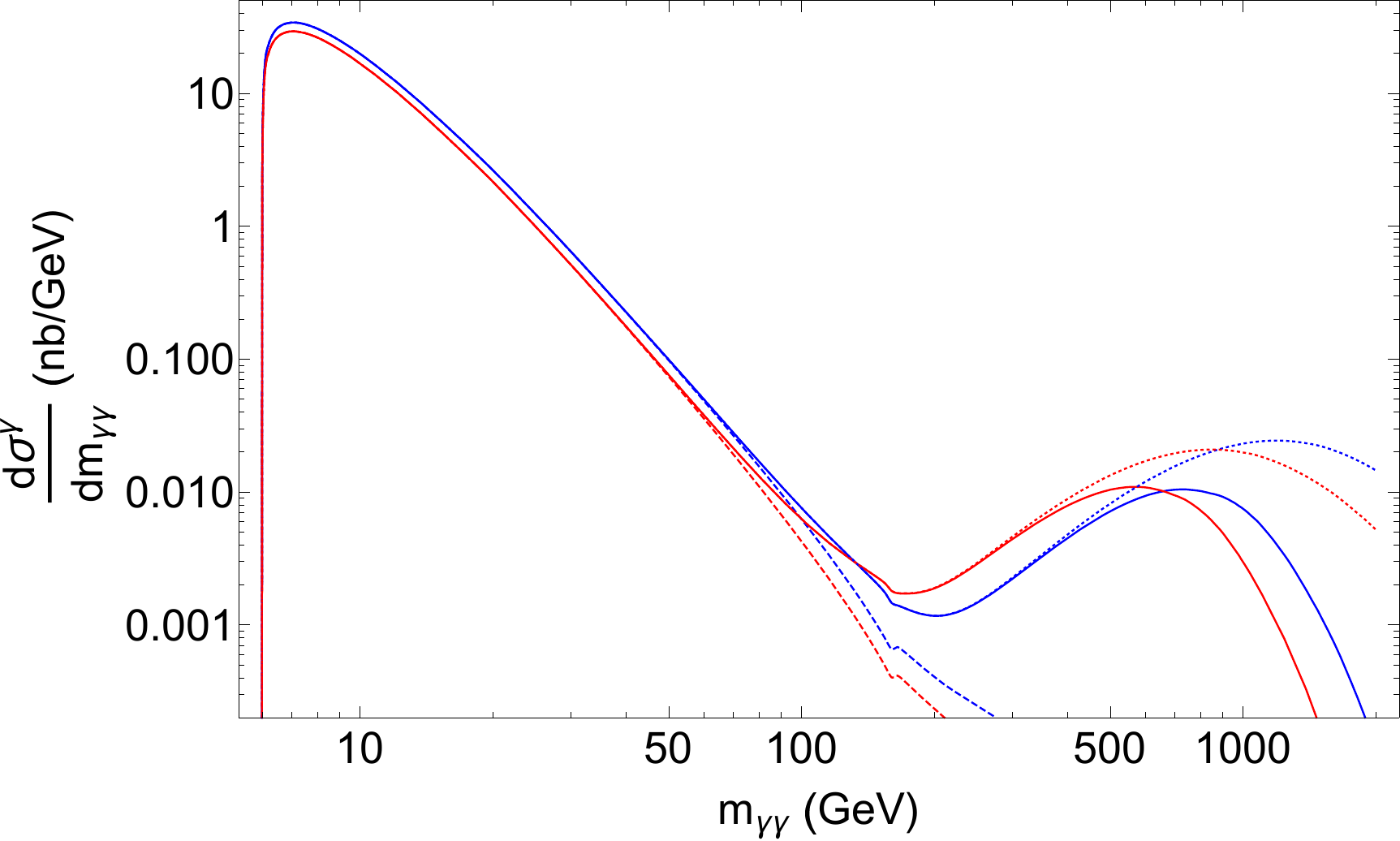}
\end{center}
\caption{Left panel: convoluted differential cross section versus diphoton invariant mass distribution of the $\rm PbPb(\gamma\gamma)\to Pb^{\ast}Pb^{\ast}\gamma\gamma$ collision in fiducial phase space of the PbPb system for the future case of higher-energy ATLAS-like experiments. Dashed lines show the SM contributions, dotted are for $\theta$-exact NCQED, and solid for the SM+($\theta$-exact)NCQED, respectively. Black lines are for $\rm\sqrt{s_{NN}}=5.02$ TeV and 
$\rm\Lambda_{NC}=100$ GeV; red for $\rm\sqrt{s_{NN}}=25.10$ TeV and $\rm\Lambda_{NC}=257$ GeV; 
blue are for $\rm\sqrt{s_{NN}}=35.14$ TeV and $\rm\Lambda_{NC}=311$ GeV, 
and green curves correspond to $\rm\sqrt{s_{NN}}=100.40$ TeV, $\rm\Lambda_{NC}=523$ GeV. 
Right panel: convoluted differential cross section versus diphoton invariant mass distribution of the $\rm PbPb(\gamma\gamma)\to Pb^{\ast}Pb^{\ast}\gamma\gamma$ collision in fiducial phase space of the PbPb system for the future case of higher-energy ATLAS-like experiments. Dashed lines are for SM and solid are for the SM+($\theta$-exact)NCQED  contributions. Dotted lines are SM plus leading order in $\theta$-expanded NCQED results. Red and blue lines are for the same energies as in the left plot.
}
\label{figure5}
\end{figure}

Because of the importance of the possibility to discover noncommutativity of space-time, we shall now discuss Fig.\ref{figure5} in some more detail. The first left dashed and solid line large peaks, up to $m_{\gamma\gamma}\sim100$ GeV, correspond to the $\frac{d\sigma_{\rm SM}^{\gamma}}{dm_{\gamma\gamma}}$ and the $\frac{d\sigma_{\rm SM+NC}^{\gamma}}{dm_{\gamma\gamma}}$ diphoton fiducial cross section distributions, while the second solid lines peaks correspond to the sum of interference and the pure NCQED terms, respectively. This outstanding second peak reflects the evolution of NC factors with respect to energy scales: When energy scales become larger than $\Lambda_{\rm NC}$, the NC factors become oscillatory and bounded. Consequently, the NC amplitudes are very small at very low energies where the SM contribution dominates and then increase quickly and compete with the exponentially decreasing luminosity factor~\cite{Baur:2001jj} to give the rising side of the NC second peak. Such new feature is a genuine peculiarity of our $\theta$-exact NCQED.  
\begin{table}
\begin{center}
\begin{tabular}{|c|c|c|c|c|c|c|}
\hline
$\rm\sqrt{s_{NN}}$ (TeV) & $\gamma$ & $\Lambda_{\rm NC}$ (GeV) & $\sigma^{\gamma}_{\rm SM}$ (nb) & $\sigma^{\gamma}_{\rm NC}$ (nb) & $\rm m_{\gamma\gamma}|_{max}$ (GeV)
\\
\hline
5.02 & 2693 & 100 & 57 & 2.8 &  138
\\
25.10 & 13465 & 257 & 178 & 6.6 &  567
\\
35.14 & 18851 & 311 & 211 & 7.9 &   737
\\
100.40 & 53860 & 523 & 336 & 16.9 &  1480
\\
\hline
\end{tabular}
\caption{Estimations for ATLAS $\rm PbPb(\gamma\gamma)\to Pb^{\ast}Pb^{\ast}\gamma\gamma$ like experiments at higher energies. Here we made adjustments of the NC scale in a way to get $\rm\frac{d\sigma^{\gamma}_{\rm NCQED}}{dm_{\gamma\gamma}}\big|_{max} $ $\rm\simeq 0.01$ (nb/GeV). }
\end{center}
\label{table3}
\end{table}

Since the future higher-energy experiments could increase the experimental sensitivity to the noncommutative effects drastically, we also estimate potential improvements 
one would expect from the next generation of colliders, in particular, upgraded CERN HL-LHC 
\footnote{The High-Luminosity Large Hadron Collider (HL-LHC) project aims to crank up the performance of the LHC in order to increase the potential for discoveries after 2027. The objective is to increase luminosity by a factor of 10 beyond the LHC's design value.}  
up to $\rm\sqrt{s_{NN}}=$14 TeV, the proposal for next-generation Chinese hadron collider SppC with energy $\rm\sqrt{s_{NN}}=$70 TeV \cite{SppCPbPb,Canbay:2017rbg}, and from the Future Circular Collider (FCC) proposal up to $\rm\sqrt{s_{NN}}=$100 TeV \cite{Armesto:2014iaa,Dainese:2016gch,Acar:2016rde,Abada:2019zxq,Abada:2019lih,Benedikt:2018csr,Abada:2019ono}. Assuming that all the kinematic ATLAS cuts remains the same, while only the energy scale or Lorentz factor scales up 5, 7, or 20 times with respect to the current ATLAS value $\rm\sqrt{s_{NN}}=$5.02 TeV, we estimate the noncommutative scale $\rm\Lambda_{NC}$ corresponding to a high-energy $\frac{d\sigma^{\gamma}}{dm_{\gamma\gamma}}$ maximum with $\rm\sim 0.01$ nb/GeV magnitude and present that in Table III and Fig.\ref{figure5}. We find that $\rm\Lambda_{NC}$ are about 2.5, 3.1, or 5.2 times $\rm\sim 100$ GeV, for the nowadays ATLAS energy scale 5.02 TeV, respectively. Also the left plot in Fig.\ref{figure5} and Table III ($\rm\sqrt{s_{\rm NN}}\sim$ 5, 25, 35, 100 TeV), both show a new feature in the form of second peaks at diphoton invariant mass $m_{\gamma\gamma}\sim$ 140, 570, 740, 1500 GeV  induced by the huge relativistic effect at higher energies (see the much larger $\gamma$ in Table III), shifted with respect to that in the left plot in Fig.\ref{figure1}.  

Such improvements are considerable, yet still insufficient to make this kind of experiment(s) convenient for bounding $\rm\Lambda_{NC}$ when comparing to other known bounds, unless the integrated luminosity can be further improved by multiple scales of the next-generation hadron colliders. Because of the SM W loops the left and right panels of Fig.\ref{figure5} at $\rm m_{\gamma\gamma}$ between 150 and 200 GeV show a small dent arising from the W-loop contributions to the pure SM and the SM$\times$NC interference terms, respectively. Namely, W contribution starts to show up by inducing a sharp drop or turn between 100 and 200 GeV presented in Fig. 1 in \cite{Bern:2001dg}. That dent is unfortunately experimentally invisible.

The existence of a second peak, with maximum  $\rm m_{\gamma\gamma}|_{max}$ given in Table III, reflect the evolution of NC factors with respect to energy scales:  When energy scales are much smaller than $\rm\Lambda_{\rm NC}$, the NC factors are increasing as monomials with high power. Once the energy scales become larger than $\rm\Lambda_{\rm NC}$ the NC factors become oscillatory and bounded. Consequently, the NC amplitudes are very small at very low energies where the SM contribution dominates and then increase fast and compete with the exponentially decreasing luminosity factor~\cite{Baur:2001jj} to give the rising side of the NC second peak. So when the energy scale goes beyond  $\rm\Lambda_{\rm NC}$ the NC amplitudes start to deviate from monomial increase, so that $\rm\frac{d\sigma^{\gamma}}{dm_{\gamma\gamma}}$ falls down quickly. The subsequent oscillatory behaviors of the NC factors are fully suppressed by the luminosity function and cannot be seen in this process. 

In the case of the $\theta$-expanded NCQED model cross sections presented in the right panel in Fig.\ref{figure5} (dotted lines for $\theta$-expanded model) show the behavior of $\frac{d\sigma^{\gamma}}{dm_{\gamma\gamma}}$  with the same two peaks. Here we give example for two higher energies, red lines correspond to $\rm\sqrt{s_{NN}}=25.10$ TeV ($\rm\Lambda_{NC}=257$ GeV), and the blue lines are for $\rm\sqrt{s_{NN}}=35.14$ TeV ($\rm\Lambda_{NC}=311$ GeV), respectively. So the same NC peak shows up but a bit higher and shifted from about 700 to about 1050 GeV, that is shifted up for about 30\% in diphoton energy. The same shall work for other energy cases, like in Fig.\ref{figure5}, showing this way explicitly that the important NC peak shows up in both the expanded and unexpanded theory, showing that it is an genuine feature arising from the Moyal-Weyl manifold of the NCQED. 

Our estimate show that like in the original ATLAS case in Fig. \ref{figure1}, the NCQED contribution in the next-generation hadron collider ultraperipheral heavy ion scattering scenarios also manifests as a second peak of 
$\rm\frac{d\sigma^{\gamma}}{dm_{\gamma\gamma}}$ at a diphoton mass range moderately higher than $\rm\Lambda_{NC}$ (see Fig. \ref{figure5}).  Also, both panels in Fig.\ref{figure5} transparently show that the point at which NCQED starts decoupling \footnote{Interestingly, the NCQED decoupling from SM is at about the same energy as the maximal exceeding energy of the recent ATLAS experiment at $\rm\sqrt{s_{NN}}=5.02$ TeV, of $\lesssim30$ GeV \cite{Aad:2019ock}.}
 from the SM, and the NCQED contribution second peaks, as functions of ($\rm\sqrt{s_{NN}},\Lambda_{NC}$) follows more or less a similar sort of pattern.

It is also important to note that the SM $\rm PbPb(\gamma\gamma)\to Pb^{\ast}Pb^{\ast}\gamma\gamma$ background in the NC second peak region is a few orders of magnitude lower than the NC second peak. This simply states that at each of the NC second peaks, i.e. at each $\rm m_{\gamma\gamma}|_{max}$ from Tables II and III, in the left plot in Fig.\ref{figure5}, the corresponding $\rm \frac{d\sigma^{\gamma}_{SM}}{dm_{\gamma\gamma}}$ values are a number of orders of magnitude below the $\rm \frac{d\sigma^{\gamma}_{SM+NCQED}}{dm_{\gamma\gamma}} \simeq \frac{d\sigma^{\gamma}_{NCQED}}{dm_{\gamma\gamma}}$ second peaks.
 
Inspection of our results presented in Figs. \ref{figure1} and Fig.\ref{figure5}, right panel, represent a new result with respect to our recent paper \cite{Horvat:2020ycy}-- uncover that novel, beyond the SM, behavior starting to show up gradually and become visible when approaching $m_{\gamma\gamma}\sim$ 30 GeV, a maximal value of the diphoton invariant mass reached in the ATLAS experiment at $\rm\sqrt{s_{NN}}=5.02$ TeV, see Fig.2(b) in Ref. \cite{Aad:2019ock}.  Note that NC peaks at $m_{\gamma\gamma}\sim$ 150-2000 GeV in both panels in Figs. \ref{figure5} could be a bit misleading because of the log-log scales. It is a quite wide band if the diphoton invariant mass scale is linear. The hardest problem is still the absolute value of the peak which will require having totally 1000 $\rm nb^{-1}$ integrated luminosity. To be absolutely clear to have a signal would require about ten events at the peak. According to the CERN ATLAS and CMS Collaborations, further planes of upgrading the LHC to HL-LHC, and the Chinese proposal for SppC \cite{Canbay:2017rbg}, as well as within the FCC \cite{Acar:2016rde,Abada:2019zxq,Abada:2019lih,Benedikt:2018csr,Abada:2019ono,Jowett:2015dmf,Jowett:2019jni} proposal, to increase luminosity by at least a factor of 10 beyond today’s LHC is quite possible, and ten events at the peak are, in fact, accessible.

\subsubsection{The $\vartheta$ and $\varphi$ angular distributions of convoluted LbyL scattering}

\begin{figure}[t]
\begin{center}
\includegraphics[width=17cm,angle=0]{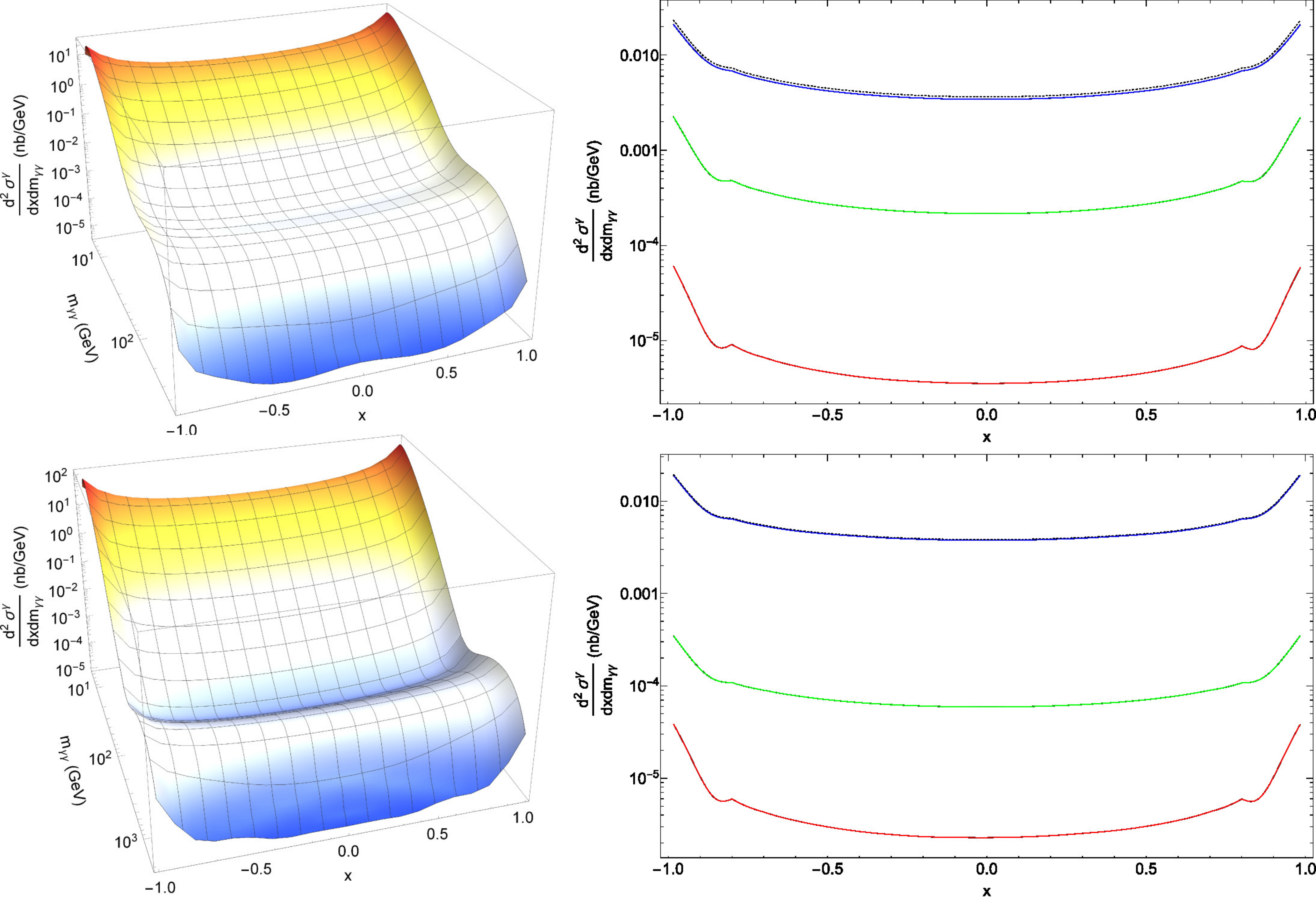}
\end{center}
\caption{The $\rm PbPb(\gamma\gamma)\to Pb^{\ast}Pb^{\ast}\gamma\gamma$ differential cross section 
$\frac{d^2\sigma^\gamma}{dxdm_{\gamma\gamma}}$ in the ATLAS ($\rm\sqrt{s_{NN}}$=5.02 TeV, $\rm\Lambda_{NC}=100$ GeV, upper plots) and FCC ($\rm\sqrt{s_{NN}}$=35.14 TeV, $\rm\Lambda_{NC}=311$ GeV, lower plots). In each scenario, the left 3D plot shows a total differential cross section with respect to $x$ and $m_{\gamma\gamma}$. The right plot compares the behaviors of the SM (red line), interference (green line), and NCQED (blue line) contributions as well as the total differential cross section (black dashed line) with respect to $x$ when $m_{\gamma\gamma}$ is at the NC peak position listed in Table III.}
\label{xdifferential}
\end{figure}

\begin{figure}[t]
\begin{center}
\includegraphics[width=17cm,angle=0]{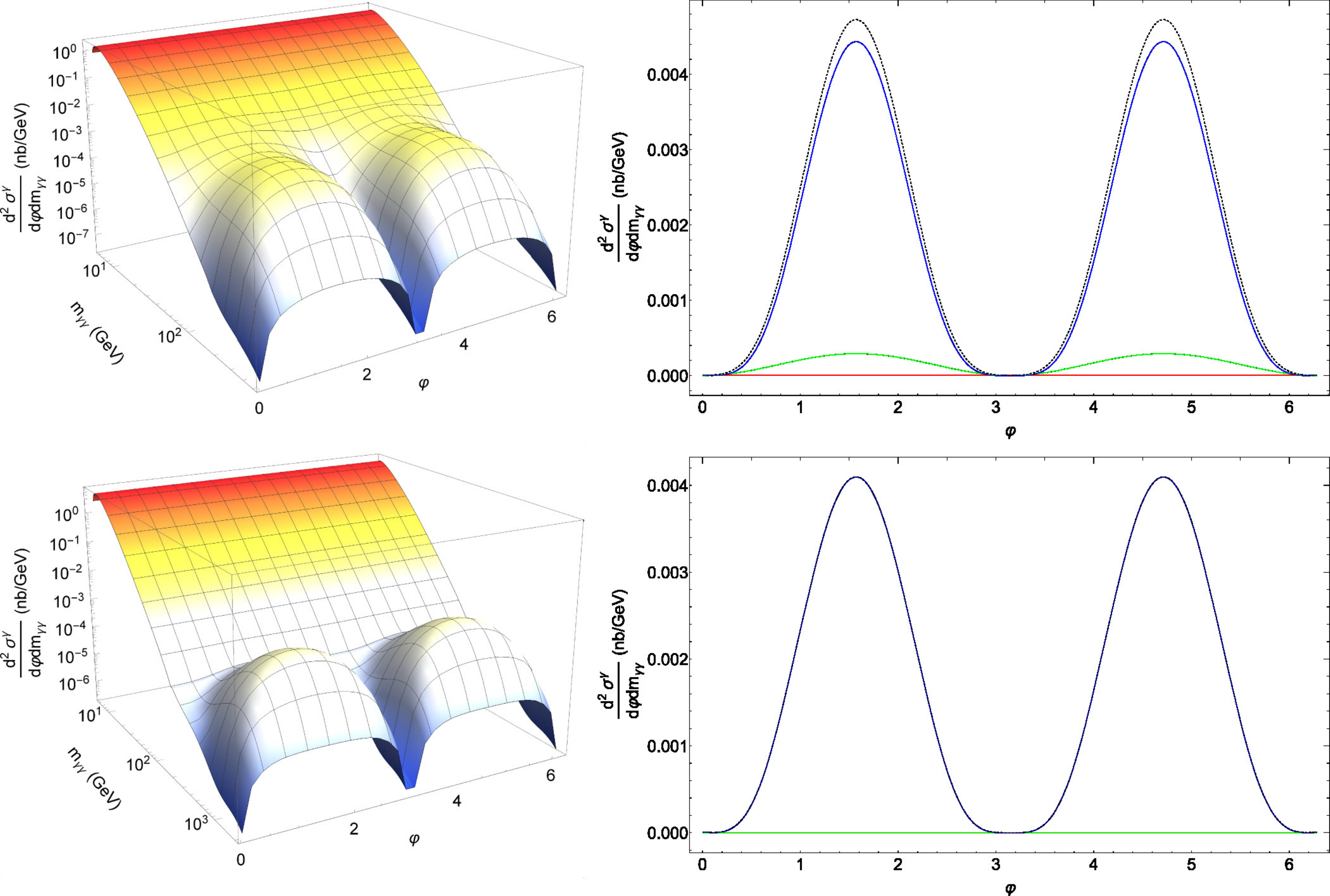}
\end{center}
\caption{The $\rm PbPb(\gamma\gamma)\to Pb^{\ast}Pb^{\ast}\gamma\gamma$ differential cross section $\frac{d^2\sigma^\gamma}{d\varphi dm_{\gamma\gamma}}$ in the ATLAS ($\rm\sqrt{s_{NN}}$=5.02 TeV, $\rm\Lambda_{NC}=100$ GeV, upper plots) and FCC ($\rm\sqrt{s_{NN}}$=35.14 TeV, $\rm\Lambda_{NC}=311$ GeV, lower plots). In each scenario, the left 3D plot shows the total differential cross section with respect to $\varphi$ and $m_{\gamma\gamma}$. The right plot compares the behaviors of the SM (red line), interference (green line) and NCQED (blue line) contributions as well as the total differential cross section (black dashed line) with respect to $\varphi$ when $m_{\gamma\gamma}$ is at the NC peak position listed in Table III. The transverse angle $\varphi$ is chosen to be zero at the direction of the projection of ${\bf B}_\theta$ onto the transverse plane.}
\label{phidifferential}
\end{figure}

The broad NC peak discussed in the last subsection raises one additional question on how to distinguish it from other (currently unknown) physical processes within the same regime. It is long known that NCQED breaks Lorentzian symmetry and induces nontrivial dependence of the differential cross section to the transverse angle $\varphi$. We have also learned from the left plot in Fig.\ref{LbyLNC/SMtotcrossect} that the $\gamma\gamma\to\gamma\gamma$ differential cross section also depends on the longitudinal angle $\vartheta$ in a unique way. So we investigated the $\rm PbPb(\gamma\gamma)\to Pb^{\ast}Pb^{\ast}\gamma\gamma$ differential cross section with respect to $x=\cos\vartheta$ and $\varphi$ for ATLAS and its FCC analogy scenarios. The results are summarised in Figs. \ref{xdifferential} and \ref{phidifferential}. As we can see from the plots, the $x$ dependence of the $\rm PbPb(\gamma\gamma)\to Pb^{\ast}Pb^{\ast}\gamma\gamma$ differential cross section follows the same trend as the $\gamma\gamma\to\gamma\gamma$ process in the left plot in Fig.\ref{LbyLNC/SMtotcrossect}. The NC contributions are flatter than the SM contribution yet the difference is not significant within the angular range considered.


The $\varphi$ dependence of the different cross section is, on the other hand, considerably more significant. The NC contributions vanish at the orientations where the transverse momentum is parallel to the transverse projection of ${\bf B}_{\theta}$ while reaching the maximum at the orientations perpendicular to the transverse projection of ${\bf B}_{\theta}$. This can be easily explained by observing the fact that in a head-to-head collision geometry, the NC factor $k_{1|2}\theta k_{3|4}$ is proportional to $|\vec{k}_{1,2}\cdot({\bf B}^\bot_{\theta}\times \vec{k}^\bot)|$. We therefore conclude that the $\varphi$ variation of the differential cross section is likely an appropriate signature for the NCQED contributions, given that enough events could be available for determining the anisotropy.

\subsection{Noncommutative background processes in the LbyL scatterings}

In the ATLAS Letter \cite{Aad:2019ock}, a large portion of material is, as it should be, devoted to the very important analysis of background processes. Namely the dominant background processes, i.e., the central exclusive productions (CEPs) $gg\to\gamma\gamma$ and $\gamma\gamma\to e^+e^-$ as well as other fake-photon background contributions,\footnote{Fake diphoton events such as cosmic-ray muons, the $\gamma\gamma\to \ell^+\ell^-,\bar q q$, as well as vector meson  dominance mechanisms driven meson pair productions and cascade annihilations, etc., are all analyzed in Ref. \cite{Aad:2019ock} showing negligible impact on the overall LbyL results. The same was found for a  single-bremstralung photon production contributions within the $^{208}$Pb-ion collision cross section measurements  \cite{Aad:2019ock}.} are in the ATLAS experiment estimated from data, and the statistical significance against the background-only hypothesis is found to be 8.2 standard deviations \cite{Aad:2019ock}. From the perspective of our NCQED contributions to the exclusive $\gamma\gamma\to\gamma\gamma$ process, 
we discuss sources to the noncommutative background contributions in the LbyL collisions which are represented by the following exclusive three-level  processes.

i) The first two are dilepton pair annihilation and productions in the CM frame. Our Fig. \ref{fig:AnnihilProd} shows oscillatory behavior of both processes, while from Eqs. (\ref{NCTCrossSectAnCM}) and  (\ref{NCTCrossSectPrnCM}) we have  
$\sigma^{\rm A;P}_{\rm QED}>0$ (see \cite{Peskin:1995ev}), and $\sigma^{\rm A;P}_{\rm NC}<0$, i.e. both NCQED annihilation and production, are destructive with respect to QED contributions. We display them as a 2-cascade  of production-annihilation process $\gamma\gamma\longrightarrow \ell^+\ell^-\longrightarrow\gamma\gamma$, contributing to the $\gamma\gamma$ final state.

ii) Third, in Bhabha cross sections, QED and NCQED contributions are both constructive: 
$\sigma^{\rm B}_{\rm QED}>0$ \cite{Peskin:1995ev}, and $\sigma^{\rm B}_{\rm NC}>0$ (\ref{NCTCrossSectBnCM}), and together with the above two, through 3-cascades of  production-scattering-annihilation 
$\gamma\gamma\longrightarrow \ell^+\ell^-\longrightarrow \ell^+\ell^-\longrightarrow\gamma\gamma$, 
also contribute to the $\gamma\gamma$ final state.

iii) Both, the 2-cascade and the 3-cascade processes shall be understood as a kind of long-distance effect, 
contributing to the $\gamma\gamma$ final state.  However, due to the sign switch in the NCQED contributions they should partially cancel general CEPs as dominant background contributions to the fake $\gamma\gamma$ final state, see the discussion in Ref. \cite{Aad:2019ock}.

iv) Exclusive Compton and M\o ller scatterings cannot produce a fake $\gamma\gamma$ final state at all. 

v) All the above is much welcome since that way the noncommutative backgrounds become even more suppressed. Therefore, we conclude that NCQED background contributions to the LbyL scatterings should be in the heavy ion collision scenario(s) experimentally invisible, i.e., the NCQED framework contribution to the $\gamma\gamma$ final state, as a true single final state,  becomes more exposed and dominated; see Figs. \ref{figure1} and \ref{figure5}. 

\subsection{Lepton pair production}
\begin{figure}
\begin{center}
\includegraphics[width=17cm,angle=0]{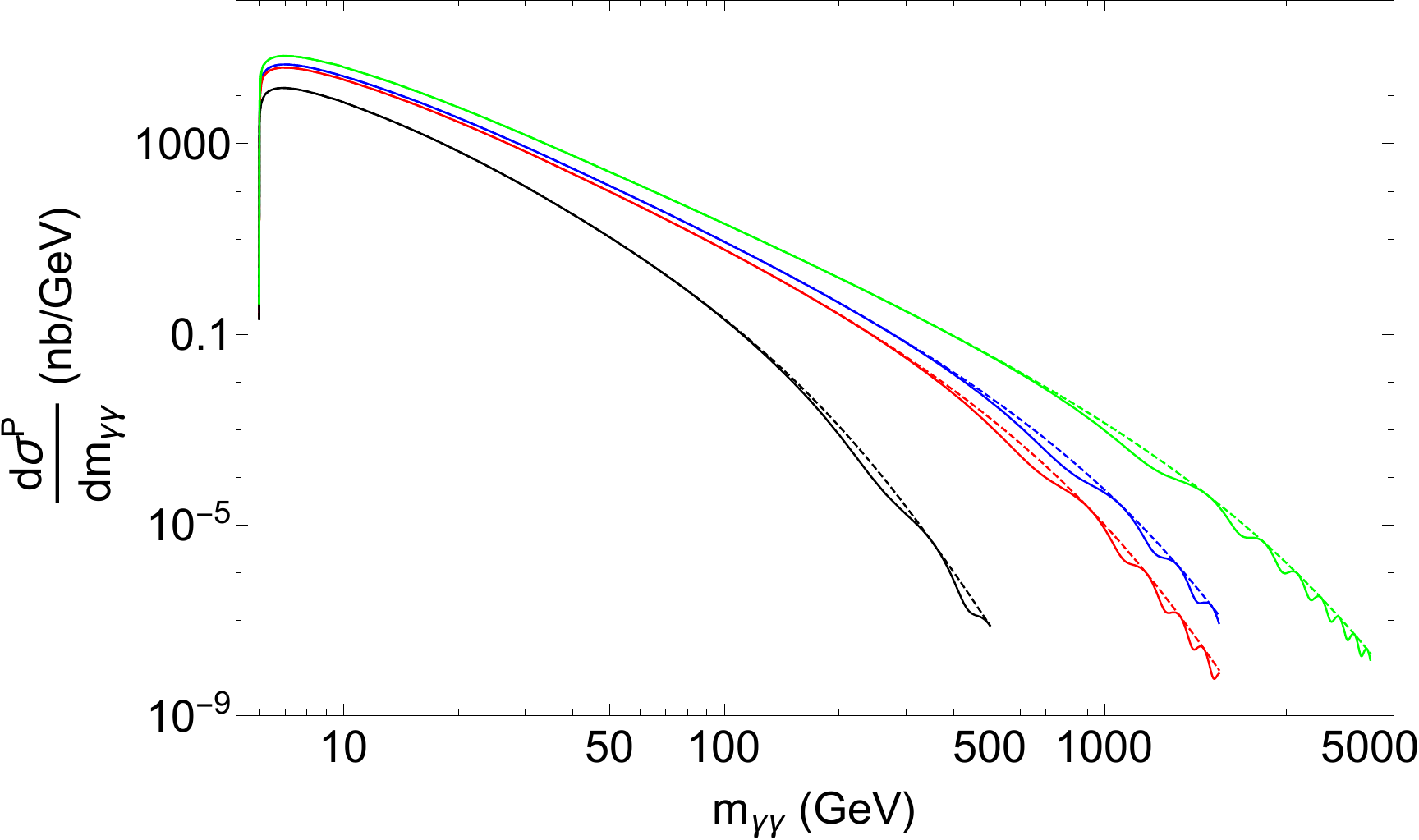}
\end{center}
\caption{Convoluted differential cross section versus diphoton invariant mass distribution of the $\rm PbPb \to Pb^{\ast}Pb^{\ast}\ell^+\ell^-$ collision in fiducial phase space of PbPb system for current and future higher-energy ATLAS-like experiments. Dashed lines are for QED contributions, and solid lines correspond to the QED+($\theta$-exact)NCQED, respectively. Black lines are for $\rm\sqrt{s_{NN}}=5.02$ TeV and $\rm\Lambda_{NC}=100$ GeV; red for $\rm\sqrt{s_{NN}}=25.10$ TeV and $\rm\Lambda_{NC}=257$ GeV; blue for $\rm\sqrt{s_{NN}}=35.14$ TeV and $\rm\Lambda_{NC}=311$ GeV, and green are for $\rm\sqrt{s_{NN}}=100.40$ TeV and $\rm\Lambda_{NC}=523$ GeV values from Table III. }
\label{figure6}
\end{figure}
One may also consider NCQED contributions to the $\rm PbPb(\gamma\gamma) \to Pb^{\ast}Pb^{\ast}\ell^+\ell^-$ reaction via tree-level exclusive pair production processes $\gamma\gamma\to\ell^+\ell^-$. 
The full spin averaged exclusive differential cross section in NCQED is given in Eq. (\ref{NCTCrossSectPrnCM}).
We notice that the NC factor is sensitive to $\theta^{03}$ component only in the geometry of the 
exclusive $\gamma\gamma\to\ell^+\ell^-$ processes on an ATLAS $\rm PbPb(\gamma\gamma) \to Pb^{\ast}Pb^{\ast}\ell^+\ell^-$ setting. Here we use lightlike noncommutativity and set $|\theta^{03}|=|-\theta^{13}|=\Lambda_{\rm NC}^{-2}$ and by convolution estimate the NCQED correction to the $\rm PbPb(\gamma\gamma) \to Pb^{\ast}Pb^{\ast}\ell^+\ell^-$ process. We assume exactly the same ATLAS kinematic cuts and energy scales to calculate convoluted lepton pair production from (\ref{NCTCrossSectPrnCM}) in analogy to the left panel in Fig. \ref{figure5}. Resulting diphoton fiducial cross section distributions $\frac{d\sigma_{\rm QED}^{\rm P}}{dm_{\gamma\gamma}}$ and the $\frac{d\sigma_{\rm QED+NC}^{\rm P}}{dm_{\gamma\gamma}}$ are displayed in Fig.\ref{figure6}, which shows that NCQED plots give a negative correction to the $\rm PbPb(\gamma\gamma) \to Pb^{\ast}Pb^{\ast}\ell^+\ell^-$ process. 
The relative magnitudes tend to be large at high energies and not much different from the QED background. 
At the same time, the absolute magnitudes in the same range are inherently very small.  Also, it is interesting that theoretical decouplings of NCQED from ordinary QED at the ATLAS experiment energy of 5.02 TeV (black lines in Fig.\ref{figure6}) start to show up gradually at maximal excess of diphoton invariant mass $\sim 100$ GeV, 3 times higher than in the case of LbyL scattering. Because of the cosine dependence of energy in dilepton cross section (\ref{NCTCrossSectPrnCM}) oscillatory behavior with respect to the LbyL scattering starts to show up at a bit higher energy, i.e., when approaching  a diphoton invariant mass of $\sim$200 GeV. We therefore conclude that $\rm PbPb(\gamma\gamma) \to Pb^{\ast}Pb^{\ast}\gamma\gamma$ collision producing exclusively two monophotons is a much better probe to the NC scale than  $\rm PbPb(\gamma\gamma) \to Pb^{\ast}Pb^{\ast}\ell^+\ell^-$ process.

\section{CONCLUSIONS}

In this work we present an explicit proof that all tree-level two-by-two scattering amplitudes in NCQED are invariant with respect to an invertible SW map. This surprisingly simple result is in accordance with our prior formal analysis that U(1) NC(S)YM with and without the SW map are equivalent to each other on shell~ \cite{Martin:2016hji,Martin:2016saw}. Our result, like its more formal precedent, is nontrivial because the reversible $\theta$-exact SW maps from noncommutative to commutative fields are highly nonlocal \footnote{We are grateful to C.P. Martin for providing to us important early references \cite{Weinberg:1968de,Kallosh1970,Salam:1971sp,Bergere:1975tr} which, while clarifying the issue of equivalence within the field redefinition of gauge transformations, also discuss locality versus nonlocality issues in QFTs, generally. From that perspective, note that there are also discussions of the standard nonlocal NC SW mapped theory versus the global NC one  \cite{Aschieri:2018vgu}, indicating that there might be a need for a different interpretation of the original Seiberg and Witten paper results.}. For this reason the results on the redefinition of local fields do not automatically apply to NCQED with the SW map. Motivated by this newly found explicit invariance, we revisit the differential and total cross sections of NCQED 
two-by-two processes. 

First, we observe several similarities to the analogous processes in QCD. 
Kinematic structures--fractions of Mandelstam variables--of NCQED contributions to cross sections of two-by-two processes (Bhabha, M\o ller, annihilation, production, Compton and LbyL) given in Refs. \cite{Hewett:2000zp,Mathews:2000we,Godfrey:2001yy} are the same as corresponding fractions of Mandelstam variables  of free quark-gluon processes in QCD, shown in Eqs. (17.70), (17.71), (17.75), (17.76), (17.77), and (17.78) of Peskin and Schroeder \cite{Peskin:1995ev}. 

It is, however,  second to note that the collinear singularities of QCD processes do not match their NCQED counterparts. The NCQED Compton scattering exhibits the same $t$-channel collinear singularity as QCD quark-gluon scattering but at a lower power. And, while NCQED LbyL scattering shares the same color- or star-product-ordered amplitudes with QCD free gluon-gluon scattering amplitude, the extra momentum-dependent noncommutative factors completely cancel the collinear divergences from the fractions of Mandelstam variables and make the NCQED $\gamma\gamma\to\gamma\gamma$ process collinear divergence-free.

Comparing with the upper bounds derived from tree-level NC processes, we consider the collinear singularity in the NCQED Compton scattering as more worthy for further investigations in the near future. If we adopt an analogy to QCD here then this singularity appears to suggest an unknown and nontrivial soft physics of NCQED, which would be instrumental in canceling the IR singularities in the hard process(es). The difference between collinear singularities in NCQED and QCD, in particular the lack of it in the NCQED $\gamma\gamma\to\gamma\gamma$ process, seems to indicate that the conjectured soft physics may be very nontrivial by itself. One would further wish that such completion could result in a theory compatible with the long-known negative one-loop $\beta$ function in U(1) NCYM~\cite{Martin:1999aq} and/or the (in)famous UV-IR mixing~\cite{Minwalla:1999px,Hayakawa:1999yt,Matusis:2000jf,Martin:2020ddo}. On the other hand, the scaling is QCD.  A full investigation along this direction lies beyond the scope of this work. The authors would, nevertheless, be truly grateful if some progress can be achieved soon.

As an application of the theoretical progress made in this work, we investigate the possibility of detecting NCQED signals in the ultraperipheral $\rm PbPb(\gamma\gamma) \to Pb^{\ast}Pb^{\ast}\gamma\gamma$ and $\rm PbPb(\gamma\gamma) \to Pb^*Pb^*\ell^+\ell^-$ future scattering experiments at HL-LHC and FCC, presented in Figs. \ref{figure5} and \ref{figure6}, respectively. We follow the equivalence photon approximation and employ a monopole form factor to calculate the total cross section of both processes. Our results indicate that the $\rm PbPb(\gamma\gamma) \to Pb^{\ast}Pb^{\ast}\gamma\gamma$ channel probes NC scales better than $\rm PbPb(\gamma\gamma) \to Pb^{\ast}Pb^{\ast}\ell^+\ell^-$. Yet neither channel could probe really large scales ($\rm\Lambda_{NC}\gtrsim 0.5$~TeV) even after we extrapolate the ion energy to beyond the next-generation hadron collider proposal(s). 

\acknowledgments{
We are grateful to C.P. Martin for many discussions regarding SW map versus field redefinition issue. J.T. would like to thank ATLAS  collaboration colleagues, Mateusz Dyndal and Matthias Schott, for valuable discussions regarding LbyL scattering in PbPb collisions experiments performed by ATLAS at the LHC.  J.T. would also like to thank Wolfgang Hollik, Dieter L\"ust and Peter Minkowski for many discussions and to acknowledge support of Dieter L\"ust and Max-Planck-Institute for Physics,  M\"unchen,  for hospitality. The work of J.Y. has been supported by Croatian Science Foundation. The work of D.L. is supported under the Serbian Ministry of Education, Science and Technological Development project No. 451-03-9/2021-14/200162. A great deal of computation was done by using {\it Mathematica} \cite{mathematica}.}

\appendix

\section{FEYNMAN RULES}   
\subsection{Matter sector: Non-SW($I$)$|$SW($II$) map-induced terms}
\begin{figure}
\begin{center}
\includegraphics[width=14cm,angle=0]{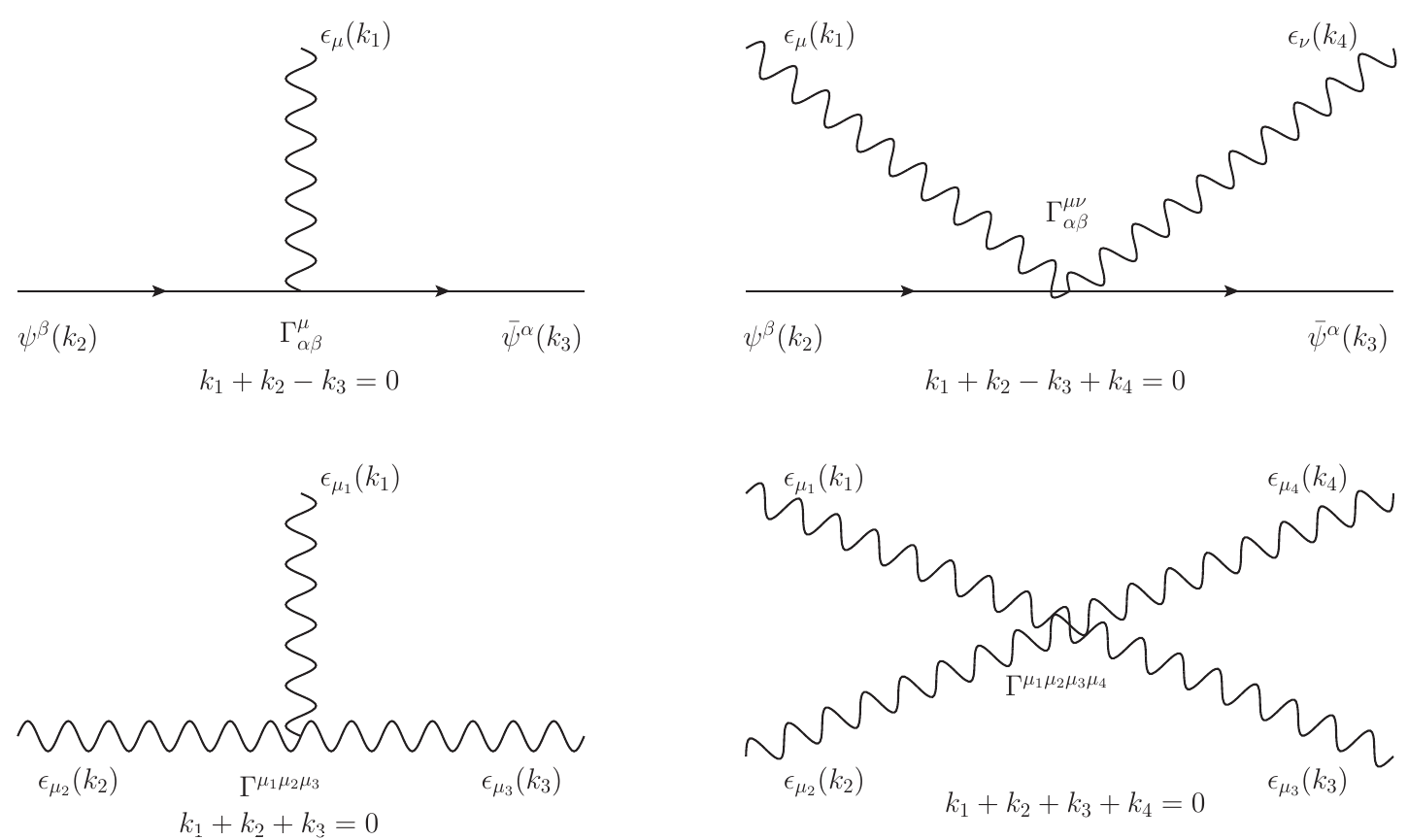}
\end{center}
\caption{NCQED Feynman rules with all gauge fields incoming.}
\label{fig:FR}
\end{figure}

To obtain vertices in momentum space we next follow the regular procedure, with momenta assignment given in Fig.\ref{fig:FR},
which from the first diagram for electron yields the following FRs:
\begin{eqnarray}
\Gamma^\mu_{\alpha\beta}&\equiv&
\Gamma^\mu_{\alpha\beta}\big(\bar u_e(k_3) u_e(k_2)\gamma(k_1)\big)
\equiv\Big[\Gamma^\mu_I+\Gamma^\mu_{II}\Big]_{\alpha\beta}\equiv\Big[\Gamma^\mu_{I+II}\Big]_{\alpha\beta},\;k_1+k_2=k_3,
\nonumber\\
\Big[\Gamma^\mu_I+\Gamma^\mu_{II}\Big]_{\alpha\beta}
&=&ie\Big[\gamma^\mu-\frac{i}{2}F_\bullet(k_1,k_2)\Big((k_1\theta k_2)\gamma^\mu- \slashed{k_1}(\theta k_2)^\mu +(\slashed{k_2}-m_e)(\theta k_1)^\mu\Big)\Big]_{\alpha\beta},
\label{FRpee1}\\
\big[\Gamma^\mu_I\big]_{\alpha\beta}&=&ie\Big[ e^{-i\frac{k_1\theta k_2}{2}}\gamma^\mu\Big]_{\alpha\beta},\;
\big[\Gamma^\mu_{II}\big]_{\alpha\beta}=-\frac{e}{2}F_\bullet(k_1,k_2)\Big[\Big(\slashed{k_1}(\theta k_2)^\mu-(\slashed{k_2}-m_e)(\theta k_1)^\mu\Big)\Big]_{\alpha\beta},
\label{FRpee2}\\
F_\bullet(k_1,k_2)&=&\frac{e^{-i\frac{k_1\theta k_2}{2}}-1}{-i\frac{k_1\theta k_2}{2}}=F_\bullet(k_2,k_1)^\ast
\equiv e^{-i\frac{k_1\theta k_2}{4}}\frac{\sin\frac{k_1\theta k_2}{4}}{\frac{k_1\theta k_2}{4}}.
\label{bulletFR}
\end{eqnarray}
The first term $\Gamma^\mu_{I}$ in Eq. (\ref{FRpee2}) represents FRs for the NCQED without the SW map and is in agreement with Refs. \cite{Mathews:2000we,Hewett:2000zp,Godfrey:2001yy}, while additional terms denoted as 
$\Gamma^\mu_{II}$, are arising due to the $\theta$-exact SW map.  Note that the Lorentz structure of the second term in Eq. (\ref{FRpee2}) is the same as the NC neutrino-photon coupling, in the case of massive neutrinos \cite{Schupp:2008fs,Horvat:2011iv}. 

From the two-photon--two-electron diagram in Fig.\ref{fig:FR} the regular recipe yields:
\begin{equation}
\begin{split}
\Gamma^{\mu\nu}_{\alpha\beta}&=
\Gamma^{\mu\nu}_{\alpha\beta}\big(\bar u_e(k_3) u_e(k_2)\gamma(k_4)\gamma(k_1)\big)
\equiv(\Gamma_{I}^{\mu\nu}+\Gamma_{II}^{\mu\nu})_{\alpha\beta}\equiv\Big[\Gamma^{\mu\nu}_{I+II}\Big]_{\alpha\beta},
\\
(\Gamma_{I}^{\mu\nu})_{\alpha\beta}&=0,
\\
(\Gamma_{II}^{\mu\nu})_{\alpha\beta}&=
\frac{e^2}{2}\Big[\frac{i}{2}\Big(F_\bullet(k_1,k_2)F_\bullet(k_4,k_3)(\theta k_2)^{\mu}(\theta k_3)^{\nu} \slashed{k_1}+F_\bullet(k_1,k_3)F_\bullet(k_4,k_2)(\theta k_2)^{\nu}(\theta k_3)^{\mu} \slashed{k_4}\Big)
\\&
+\Big(\gamma^{\mu}(\theta k_3)^{\nu}e^{-i\frac{k_1\theta k_2}{2}}F_\bullet(k_4,k_3)+\gamma^{\nu}(\theta k_3)^{\mu}e^{-i\frac{k_4\theta k_2}{2}}F_\bullet(k_1,k_3)\Big)
\\&
-\Big((\theta k_2)^{\mu}\gamma^{\nu}e^{-i\frac{k_4\theta k_3}{2}}F_\bullet(k_1,k_2)+(\theta k_2)^{\nu}\gamma^{\mu}e^{-i\frac{k_1\theta k_3}{2}}F_\bullet(k_4,k_2)\Big)
\\&
-\Big(2\big(\gamma^{\mu}(\theta k_1)^{\nu}+\gamma^{\nu}
(\theta k_4)^{\mu}\big)+\theta^{\mu\nu}\big(\slashed{k_1}-\slashed{k_4}\big)\Big)e^{-i\frac{k_3\theta k_2}{2}}F_{\star_2}(k_4,k_1)\Big]_{\alpha\beta},
\end{split}
\label{FReepp}
\end{equation}
which is symmetric under simultaneous pair exchanges $(\nu, k_4)\leftrightarrow(\mu,k_1)$. 
Here, $\Gamma_{I}^{\mu\nu}$ represents a term irrelevant for the on-shell tree-level computations; thus, we indicate that fact by equating it with zero. This FR is coming from the action (\ref{Sppee}) and/or (\ref{Sppeer}) induced by the $\theta$-exact SW maps for gauge field strength (\ref{2.2}) and fermions (\ref{Psi}), respectively.

\subsection{Gauge sector: Non-SW($I$)$|$SW($II$) map-induced terms}

\subsubsection{Triple-photon coupling: General FRs}

Considering the gauge sector by employing a straightforward reading-out procedure from $S_{a^3}$ (\ref{f3}) for the third diagram in Fig. \ref{fig:FR} we have the Feynman rule $\Gamma^{\mu_1\mu_2\mu_3}$ for the triple-photon vertex in momentum space:
\begin{eqnarray}
\Gamma^{\mu_1\mu_2\mu_3}_{I+II}&\equiv&\Gamma^{\mu_1\mu_2\mu_3}(\gamma(k_1),
\gamma(k_2),\gamma(k_3))=eV^{\mu_1\mu_2\mu_3}_{I+II}(k_1,k_2,k_3)F_{\star_2}(k_2,k_3),\;
F_{\star_2}(p,q)=\frac{\sin\frac{p\theta q}{2}}{\frac{p\theta q}{2}},
\label{Fg}
\end{eqnarray}
with vertex function $V^{\mu_1\mu_2\mu_3}_{I+II}(k_1,k_2,k_3)$ and $F_{\star_2}(k_2,k_3)$ being taken from Refs. \cite{Horvat:2013rga,Horvat:2015aca}:
\begin{eqnarray}
V^{\mu_1\mu_2\mu_3}_I(k_1,k_2,k_3)=&-&(k_1\theta k_2)\big[(k_1-k_2)^{\mu_3}g^{\mu_1\mu_2}+(k_2-k_3)^{\mu_1} g^{\mu_2\mu_3}+(k_3-k_1)^{\mu_2}g^{\mu_1\mu_3}\big],
\nonumber\\
V^{\mu_1\mu_2\mu_3}_{II}(k_1,k_2,k_3)=&-&\theta^{\mu_1\mu_2}\big[k_1^{\mu_3}(k_2 k_3)-k_2^{\mu_3}(k_1 k_3)\big]-\theta^{\mu_2\mu_3}\big[k_2^{\mu_1}(k_1 k_3)-k_3^{\mu_1}(k_1 k_2)\big]
-\theta^{\mu_3\mu_1}\big[k_3^{\mu_2}(k_1 k_2)-k_1^{\mu_2}(k_2 k_3)\big]
\nonumber\\
&+&(\theta k_1)^{\mu_2}\big[g^{\mu_1\mu_3}k_3^2-k_3^{\mu_1} k_3^{\mu_3}\big]
+(\theta k_1)^{\mu_3}\big[g^{\mu_1\mu_2}k_2^2-k_2^{\mu_1} k_2^{\mu_2}\big]+(\theta k_2)^{\mu_1}\big[g^{\mu_2\mu_3}k_3^2-k_3^{\mu_2} k_3^{\mu_3}\big]
\nonumber\\
&+&(\theta k_2)^{\mu_3}\big[g^{\mu_1\mu_2}k_1^2-k_1^{\mu_1} k_1^{\mu_2}\big]+(\theta k_3)^{\mu_2}\big[g^{\mu_1\mu_3}k_1^2-k_1^{\mu_1} k_1^{\mu_3}\big]+(\theta k_3)^{\mu_1}\big[g^{\mu_2\mu_3}k_2^2-k_2^{\mu_2} k_2^{\mu_3}\big].
\label{FgA}
\end{eqnarray}
Here again, only the first line of (\ref{FgA}), with function $2\sin\frac{k_2\theta k_3}{2}$ instead of $F_{\star_2}(k_2,k_3)$, from SW NCQED triple-photon FR (\ref{Fg}) and (\ref{FgA}), represents the FRs for the NCQED theory without the SW map, in agreement with Refs. \cite{Hewett:2000zp,Buric:2007qx}, respectively. 

\subsubsection{Four-photon coupling: General considerations and FRs}

Finally, the $e^3$ order SW map for the gauge field strength produces the four-photon interaction diagram in Fig. \ref{fig:FR} arising from the action $S_{f^4}$ (\ref{f4}). The Feynman rule $\Gamma^{\mu\nu\rho\tau}$ for fourth diagram in Fig.\ref{fig:FR} is much more complicated, even with additional gauge freedom $\kappa$'s fixed to unity. 
Those FRs are generally given in detail in Eqs. (16), (B.1), (B.2), (B.4)--(B.6) in Ref. \cite{Horvat:2015aca}. After the inspection of tadpole contributions to the photon polarization tensor for two different SW maps (I) and (II), represented with gauge freedom parameters $(\kappa,\kappa_i)$'s and $(\kappa,\kappa_i^\prime)$’s, respectively, in Ref. \cite{Horvat:2015aca} we have found that for 
$\kappa=\kappa_1=\kappa_2=\kappa_3=\kappa_4=\kappa'_1=\kappa'_2=\kappa'_3=\kappa'_4=1$ the sum of bauble and tadpole contributions to polarization tensor (I) given by Eq. (37) and contributions to polarization tensor (II) (38) are equal. Thus, we shall use the FR from Ref. \cite{Horvat:2015aca} for the SW maps (I) and (II) with all $\kappa$'s=1 and with momenta $k_i$ in Fig.\ref{fig:FR} being the incoming ones. Because of number of typos in FRs from Appendix B in Ref.  \cite{Horvat:2015aca}, next we repeat all, but corrected, FR equations: 
\begin{equation}
\begin{split}
&\Gamma_{I+II}^{\mu_1\mu_2\mu_3\mu_4}\left(k_1,k_2,k_3,k_4\right)=\frac{ie^2}{4}S_4\Big[\Gamma_{\rm A}^{\mu_1\mu_2\mu_3\mu_4}\left(k_1,k_2,k_3,k_4\right)
+\Gamma_{\rm B}^{\mu_1\mu_2\mu_3\mu_4}\left(k_1,k_2,k_3,k_4\right)
\\&+\Gamma_1^{\mu_1\mu_2\mu_3\mu_4}\left(k_1,k_2,k_3,k_4\right)+\Gamma_2^{\mu_1\mu_2\mu_3\mu_4}\left(k_1,k_2,k_3,k_4\right)
+\Gamma_3^{\mu_1\mu_2\mu_3\mu_4}\left(k_1,k_2,k_3,k_4\right)
\\&+\Gamma_4^{\mu_1\mu_2\mu_3\mu_4}\left(k_1,k_2,k_3,k_4\right)+\Gamma_5^{\mu_1\mu_2\mu_3\mu_4}\left(k_1,k_2,k_3,k_4\right)
\Big]_{I+II}\delta\left(k_1+k_2+k_3+k_4\right),
\end{split}
\label{A.1}
\end{equation}
where $S_4$ denotes permutations over all $(\{k_i, \mu_i\}, \forall i=1,2,3,4),$ pairs simultaneously. Capital indices $I+II$, as before from Eqs. (\ref{FRpee1})--(\ref{FgA}), indicate a splitting of total contribution into the one $I$ arising from the action without the SW map and the second $II$, as a contribution induced by the SW mapping. 

First we split $\Gamma_A$ into five and $\Gamma_B$ into three pieces, respectively:
\begin{gather}
\begin{split}
S_4\Gamma_A^{\mu_1\mu_2\mu_3\mu_4}&(k_1,k_2,k_3,k_4)\big|_{I+II}=
\sum_{i=1}^{5}S_4\Gamma_{A_i}^{\mu_1\mu_2\mu_3\mu_4}(k_1,k_2,k_3,k_4)\big|_{I+II},
\\
S_4\Gamma_{A_i}^{\mu_1\mu_2\mu_3\mu_4}&(k_1,k_2,k_3,k_4)\big|_{I+II}=
S_4\Big(V_{A_i}^{\mu_1\mu_2\mu_3\mu_4}(k_1,k_2,k_3,k_4)\big|_{I+II}\,
F_{\star_2}(k_1,k_2)F_{\star_2}(k_3,k_4)\Big),
\\
S_4\Gamma_B^{\mu_1\mu_2\mu_3\mu_4}&(k_1,k_2,k_3,k_4)
\big|_{I+II}=
\sum_{i=1}^{3}S_4\Gamma_{B_i}^{\mu_1\mu_2\mu_3\mu_4}(k_1,k_2,k_3,k_4)\big|_{I+II},
\\
S_4\Gamma_{B_i}^{\mu_1\mu_2\mu_3\mu_4}&(k_1,k_2,k_3,k_4)\big|_{I+II}=
S_4\Big(V_{B_i}^{\mu_1\mu_2\mu_3\mu_4}
(k_1,k_2,k_3,k_4)\big|_{I+II}\;
F_{\star_2}(k_1,k_2)F_{\star_2}(k_3,k_4)\Big),
\end{split}
\label{A.2}
\end{gather}
\begin{gather}
\begin{split}
V_{A_1}^{\mu_1\mu_2\mu_3\mu_4}\big|_{I+II}&\equiv
V_{A_1}^{\mu_1\mu_2\mu_3\mu_4}\big|_{I}=
(k_1\theta k_2)(k_3\theta k_4)g^{\mu_1\mu_3}g^{\mu_2\mu_4},
\\
V_{A_2}^{\mu_1\mu_2\mu_3\mu_4}\big|_{I+II}&\equiv
V_{A_2}^{\mu_1\mu_2\mu_3\mu_4}\big|_{II}=
\theta^{\mu_1\mu_2}\theta^{\mu_3\mu_4}(k_1 k_3)(k_2k_4),
\\
V_{A_3}^{\mu_1\mu_2\mu_3\mu_4}\big|_{I+II}&=
(k_1 k_3)\Big[
(\theta k_2)^{\mu_1}\Big((\theta k_4)^{\mu_3}g^{\mu_2\mu_4}-\theta^{\mu_3\mu_4}k_4^{\mu_2}\Big)
-\theta^{\mu_1\mu_2}(\theta k_4)^{\mu_3}k_2^{\mu_4}
\Big]
\\&
\phantom{.}+(k_2 k_4)\Big[(\theta k_1)^{\mu_2}\Big((\theta k_3)^{\mu_4}g^{\mu_1\mu_3}+\theta^{\mu_3\mu_4}k_3^{\mu_1}\Big)+\theta^{\mu_1\mu_2}(\theta k_3)^{\mu_4}k_1^{\mu_3}\Big],
\\
V_{A_4}^{\mu_1\mu_2\mu_3\mu_4}\big|_{I+II}&=
(k_1\theta k_2)\Big[\theta^{\mu_3\mu_4}k_3^{\mu_1}k_4^{\mu_2}-(\theta k_4)^{\mu_3}k_3^{\mu_1}g^{\mu_2\mu_4}+k_4^{\mu_2}(\theta k_3)^{\mu_4}g^{\mu_1\mu_3}\Big]
\\&\phantom{.}+(k_3\theta k_4)\Big[
\theta^{\mu_1\mu_2}k_1^{\mu_3}k_2^{\mu_4}
-(\theta k_2)^{\mu_1}k_1^{\mu_3}g^{\mu_2\mu_4}
+(\theta k_1)^{\mu_2}k_2^{\mu_4}g^{\mu_1\mu_3}\Big],
\\
V_{A_5}^{\mu_1\mu_2\mu_3\mu_4}\big|_{I+II}&=
-\Big[(\theta k_1)^{\mu_2}(\theta k_4)^{\mu_3}k_2^{\mu_4}k_3^{\mu_1}
+(\theta k_2)^{\mu_1}(\theta k_3)^{\mu_4}k_1^{\mu_3}k_4^{\mu_2}\Big],
\end{split}
\label{A.3}
\end{gather}
where clearly only the first term in Eq. (\ref{A.3}), the $V_{A_1}$, belongs to the class of non-SW map-induced terms, since that term was induced due to 
the $\star$-commutator term of the NCYM gauge Lagrangian 
$F^{\mu\nu}\star F_{\mu\nu}\sim[a^\mu\stackrel{\star}{,}a^\nu][a_\mu\stackrel{\star}{,}a_\nu]$, with $A_\mu(x)\equiv a_\mu(x)$; see details in \cite{Hewett:2000zp}. What follows are additional, but irrelevant terms:  
\begin{gather}
\begin{split}
V_{B_1}^{\mu_1\mu_2\mu_3\mu_4}\big|_{I+II}&=
2(\theta k_1)^{\mu_1}\Big[(k_2 k_3)\big(k_4^{\mu_2}\theta^{\mu_3\mu_4}
-(\theta k_4)^{\mu_3}g^{\mu_2\mu_4}\big)
-(k_2 k_4)\big(k_3^{\mu_2}\theta^{\mu_3\mu_4}+(\theta k_3)^{\mu_4}g^{\mu_2\mu_3}\big)\Big]
\\
V_{B_2}^{\mu_1\mu_2\mu_3\mu_4}\big|_{I+II}&=
2(\theta k_1)^{\mu_1}(k_3\theta k_4)\Big(k_2^{\mu_3}g^{\mu_2\mu_4}-g^{\mu_2\mu_3}k_2^{\mu_4}
\Big),
\\
V_{B_3}^{\mu_1\mu_2\mu_3\mu_4}\big|_{I+II}&=
2(\theta k_1)^{\mu_1}\Big(k_4^{\mu_2}k_2^{\mu_3}(\theta k_3)^{\mu_4}+k_3^{\mu_2}(\theta k_4)^{\mu_3}k_2^{\mu_4}
\Big).
\end{split}
\label{A.4}
\end{gather}
Tensor structures of remaining $\Gamma_i, i=1,2,3,4,5$, terms are listed next
\begin{gather}
\begin{split}
\Gamma_1^{\mu_1\mu_2\mu_3\mu_4}\left(k_1,k_2,k_3,k_4\right)\big|_{II}=&2F_{\star_{3'}}\left(k_2,k_3,k_4\right)V_1^{\mu_1\mu_2\mu_3\mu_4}\left(k_1,k_2,k_3,k_4\right)\big|_{II},
\\
\Gamma_2^{\mu_1\mu_2\mu_3\mu_4}\left(p_1,p_2,p_3,p_4\right)\big|_{II}=&\Big(2F_{\star_2}\left(k_1,k_2\right)F_{\star_2}\left(k_3,k_4\right)-F_{\star_{3'}}\left(k_2,k_3,k_4\right)
-F_{\star_{3'}}\left(k_4,k_3,k_2\right)\Big)
\\&\cdot 
V_2^{\mu_1\mu_2\mu_3\mu_4}\left(k_1,k_2,k_3,k_4\right)\big|_{II},
\\
\Gamma_3^{\mu_1\mu_2\mu_3\mu_4}\left(k_1,k_2,k_3,k_4\right)\big|_{II}=&F_{\star_{3'}}(k_4,k_2,k_3)V_3^{\mu_1\mu_2\mu_3\mu_4}\left(k_1,k_2,k_3,k_4\right)\big|_{II},
\\
\Gamma_4^{\mu_1\mu_2\mu_3\mu_4}\left(k_1,k_2,k_3,k_4\right)\big|_{II}=&F_{\star_2}(k_1,k_2)F_{\star_2}(k_3,k_4)V_4^{\mu_1\mu_2\mu_3\mu_4}\left(k_1,k_2,k_3,k_4\right)\big|_{II},
\\
\Gamma_5^{\mu_1\mu_2\mu_3\mu_4}\left(k_1,k_2,k_3,k_4\right)\big|_{II}=&F_{\rm (I)}(k_2,k_3,k_4)V_5^{\mu_1\mu_2\mu_3\mu_4}\left(k_1,k_2,k_3,k_4\right)\big|_{II},
\end{split}
\label{A.5}
\end{gather}
where the above $\star$ products induced functions in momentum space: $F_{\star_2}$, $F_{\star_{3'}}$ and $F_{(\rm I)}$ are  defined in Refs. \cite{Trampetic:2015zma,Horvat:2015aca}, respectively. Tensor structures of $V_i$'s are given next:
\begin{gather}
\begin{split}
V_1^{\mu_1\mu_2\mu_3\mu_4}&\left(k_1,k_2,k_3,k_4\right)\big|_{II}
\\&\phantom{XX}
=2(k_1 k_2)\Big(k_3^{\mu_1}(\theta k_4)^{\mu_2}\theta^{\mu_3\mu_4}-k_3^{\mu_1}(\theta k_4)^{\mu_3}\theta^{\mu_2\mu_4}
+g^{\mu_1\mu_3}(\theta k_4)^{\mu_2}(\theta k_3)^{\mu_4}+(k_3\theta k_4)g^{\mu_1\mu_3}\theta^{\mu_2\mu_4}\Big)
\\&
\phantom{XXXx}-2k_1^{\mu_2}\Big((k_2\theta k_4)k_3^{\mu_1}\theta^{\mu_3\mu_4}+k_3^{\mu_1}(\theta k_4)^{\mu_3}(\theta k_2)^{\mu_4}
+(k_2\theta k_4)g^{\mu_1\mu_3}(\theta k_3)^{\mu_4}-(k_3\theta k_4)g^{\mu_1\mu_3}(\theta k_2)^{\mu_4}\Big)
\\&\phantom{XX.}
+2(k_1 k_3)\Big(k_2^{\mu_1}(\theta k_4)^{\mu_3}\theta^{\mu_2\mu_4}-k_2^{\mu_1}(\theta k_4)^{\mu_2}\theta^{\mu_3\mu_4}
+g^{\mu_1\mu_2}(\theta k_4)^{\mu_3}(\theta k_2)^{\mu_4}+(k_2\theta k_4)g^{\mu_1\mu_2}\theta^{\mu_3\mu_4}\Big)
\\&\phantom{XXXx}
-2k_1^{\mu_3}\Big((k_3\theta k_4)k_2^{\mu_1}\theta^{\mu_2\mu_4}+k_2^{\mu_1}(\theta k_4)^{\mu_2}(\theta k_3)^{\mu_4}
+(k_3\theta k_4)g^{\mu_1\mu_2}(\theta k_2)^{\mu_4}-(k_2\theta k_4)g^{\mu_1\mu_2}(\theta k_3)^{\mu_4}\Big),
\\
V_2^{\mu_1\mu_2\mu_3\mu_4}&\left(k_1,k_2,k_3,k_4\right)\big|_{II}=(\theta k_4)^{\mu_3}\Big[(k_1 k_2)\Big(k_4^{\mu_1}\theta^{\mu_2\mu_4}-(\theta k_4)^{\mu_2}g^{\mu_1\mu_4}\Big)
+k_1^{\mu_2}\Big((\theta k_2)^{\mu_4}k_4^{\mu_1}+(k_2\theta k_4)g^{\mu_1\mu_4}\Big)
\\&\phantom{XXXXXXXXXXXXx}
-(k_1k_4)\Big(k_2^{\mu_1}\theta^{\mu_1\mu_4}+(\theta k_2)^{\mu_4}g^{\mu_1\mu_2}\Big)
+k_1^{\mu_4}\Big(k_2^{\mu_1}(\theta k_4)^{\mu_2}
-(k_2\theta k_4)g^{\mu_1\mu_2}\Big)\Big],
\\
V_3^{\mu_1\mu_2\mu_3\mu_4}&\left(k_1,k_2,k_3,k_4\right)\big|_{II}=\Big(k_1^{\mu_4}k_4^{\mu_1}-(k_1 k_4)g^{\mu_1\mu_4}\Big)\Big((\theta k_2)^{\mu_3}(\theta k_3)^{\mu_2}+\theta^{\mu_2\mu_3}(k_2\theta k_3)\Big),
\\
V_4^{\mu_1\mu_2\mu_3\mu_4}&\left(k_1,k_2,k_3,k_4\right)\big|_{II}=\Big((k_1k_4)g^{\mu_1\mu_4}-k_1^{\mu_4}k_4^{\mu_1}\Big)(\theta k_2)^{\mu_2}(\theta k_3)^{\mu_3},
\\
V_5^{\mu_1\mu_2\mu_3\mu_4}&\left(k_1,k_2,k_3,k_4\right)\big|_{II}=\Big(k_1^{\mu_4}k_4^{\mu_1}-(k_1 k_4)g^{\mu_1\mu_4}\Big)
(\theta k_2)^{\mu_2}
\Big((\theta k_2)^{\mu_3}(k_3\theta k_4)+(\theta k_4)^{\mu_3}(k_2\theta k_3)\Big).
\end{split}
\label{A.6}
\end{gather}
For any additional details, we refer to Ref. \cite{Horvat:2015aca}.
However, the irrelevant terms are proportional to the free field equation for {$a_\mu(x)$ }, 
therefore, since they vanish due to the EOM and gauge condition, they are irrelevant to the tree-level scattering processes we are heading for. Thus, we collect now the relevant terms at the $e^2$ order which does not vanish on shell. As we mentioned at the beginning of the paper in Sec. II, we shall work out only the FRs from relevant gauge action given in Sec. II.B. Thus, the irrelevant FRs (\ref{A.4})--(\ref{A.6}) we are giving just for the sake of completeness, but we do not need them further on. However, one should not forget that those irrelevant terms are indeed very much relevant for arbitrary loop and/or non-Abelian computations.

\subsubsection{Four-photon coupling: Relevant terms FRs}

Starting with action (\ref{S}) and obtaining relevant part (\ref{R4p-4}), from where we already have FRs arising from the very first term in Eq. (\ref{R4p-4}) presented in Ref. \cite{Horvat:2015aca} and denoted as $\Gamma_A$ and/or $V_A$ in (\ref{A.1}) and (\ref{A.3}), respectively. Thus FRs from second, third, and fourth terms in Eq. (\ref{R4p-4}) we determine next. Note that only $\star_2$ is present twice,  thus producing at the very end terms proportional only to $F_{\star_2}(k_1,k_2)F_{\star_2}(k_3,k_4)$, etc. The relevant four-photon vertex in accord with fourth diagram in Fig.\ref{fig:FR} is
\begin{equation}
\begin{split}
&\Gamma_{\rm relevant}^{\mu_1\mu_2\mu_3\mu_4}(k_1,k_2,k_3,k_4)
=\frac{ie^2}{4}\Big[W^{\mu_1\mu_2\mu_3\mu_4}(k_1,k_2,k_3,k_4)
F_{\star_2}(k_1,k_2)F_{\star_2}(k_3,k_4)
\\&
+{\rm all\; S_4\; permutations\; over}\; \{k_i, \mu_i\}{\rm \;pairs\; simutaneously}\Big]_{I+II}\delta\left(k_1+k_2+k_3+k_4\right),
\end{split}
\label{F.9}
\end{equation}
\begin{equation}
W^{\mu_1\mu_2\mu_3\mu_4}(k_1,k_2,k_3,k_4)
=V_A^{\mu_1\mu_2\mu_3\mu_4}+W_1^{\mu_1\mu_2\mu_3\mu_4}+W_2^{\mu_1\mu_2\mu_3\mu_4}+W_3^{\mu_1\mu_2\mu_3\mu_4},
\label{F.10}
\end{equation}
where $V_A$ is exactly the same as terms given in Eqs. (\ref{A.1})--(\ref{A.3}). The $W_1$ is coming from the second term in the first line, while $W_2$, and $W_3$ are coming from first and second terms in second line of Eq. (\ref{R4p-4}) respectively.  After we split the above $W_1$ into three terms, we have:
\begin{equation}
\begin{split}
W_1^{\mu_1\mu_2\mu_3\mu_4}&(k_1,k_2,k_3,k_4)=
\sum_{i=1}^3W^{\mu_1\mu_2\mu_3\mu_4}_{1_i}(k_1,k_2,k_3,k_4)
\\
W^{\mu_1\mu_2\mu_3\mu_4}_{1_1}&(k_1,k_2,k_3,k_4)
=2(\theta k_2)^{\mu_1}\theta^{\mu_3\mu_4}
\Big[k_3^{\mu_2}(k_2 k_4)-k_4^{\mu_2}(k_2 k_3)\Big],
\\
W^{\mu_1\mu_2\mu_3\mu_4}_{1_2}&(k_1,k_2,k_3,k_4)
=2(\theta k_2)^{\mu_1}
\Big[(\theta k_3)^{\mu_4}
\big(g^{\mu_2\mu_3}(k_2 k_4)-k_2^{\mu_3}k_4^{\mu_2}\big)
+(\theta k_4)^{\mu_3}
\big(g^{\mu_2\mu_4}(k_2 k_3)-k_2^{\mu_4}k_3^{\mu_2}\big)\Big],
\\
W^{\mu_1\mu_2\mu_3\mu_4}_{1_3}&(k_1,k_2,k_3,k_4)=
2(\theta k_2)^{\mu_1}(k_3\theta k_4)\Big[k_2^{\mu_4}g^{\mu_2\mu_3}-k_2^{\mu_3}g^{\mu_2\mu_4}\Big],
\label{F.11}
\end{split}
\end{equation} 
\begin{equation}
\begin{split}
&W_2^{\mu_1\mu_2\mu_3\mu_4}(k_1,k_2,k_3,k_4)=2(\theta k_2)^{\mu_1}(\theta k_4)^{\mu_3}
\Big[g^{\mu_2\mu_4}(k_2 k_4)-k_2^{\mu_4}k_4^{\mu_2}\Big],
\label{F.12}
\end{split}
\end{equation}
\begin{equation}
W_3^{\mu_1\mu_2\mu_3\mu_4}(k_1,k_2,k_3,k_4)
=2\big(k_1\theta k_2\big)\Big[2(\theta k_4)^{\mu_3}
\big(k_1^{\mu_4}g^{\mu_1\mu_2}-k_1^{\mu_2}g^{\mu_1\mu_4}\big)
+\theta^{\mu_3\mu_4}\big(k_1^{\mu_2}k_4^{\mu_1}-g^{\mu_1\mu_2}(k_1 k_4)
\big)\Big].
\label{F.W3}
\end{equation}

\section{THE $\varphi$-INTEGRATIONS NEEDED IN ALL NCQED AMPLITUDES}

\subsection{Noncommutative phase factors}

Choosing the general frame for ${\bf1}(k_1)+{\bf2}(k_2)\to {\bf3}(k_3)+{\bf4}(k_4)$ collision, which we called the non-center-of-mass--NCM--frame, we start with 4-momenta=(energy, 3-momenta), for the incoming first particle $k_1=(\omega_1,\vec k_1)$, and for the incoming second one as $k_2=(\omega_2,\vec k_2)$, where both momenta to lie on the $z$ axis. The scattered outgoing particle 4-momenta are $k_3=(\omega_3,\vec k_3)$ and $k_4=(\omega_4,\vec k_4)$, respectively. In the spherical coordinate system for the massless case with 3-momentum $\vec k_1+\vec k_2=\vec k_3+\vec k_4$ and energy $\omega_1+\omega_2=\omega_3+\omega_4$ conservations, respectively, we have
\begin{equation}
\begin{split}
k_1&=(\omega_1,0,0,\omega_1), \;k_2=(\omega_2,0,0,-\omega_2), 
\;k_4=(\omega_4,\omega_4\sin\vartheta\cos\varphi,\omega_4\sin\vartheta\sin\varphi,\omega_4\cos\vartheta),
\\
k_3&
=(\omega_1+\omega_2-\omega_4,-\omega_4\sin\vartheta\cos\varphi,-\omega_4\sin\vartheta\sin\varphi, \omega_1-\omega_2-\omega_4\cos\vartheta),
\end{split}
\label{SphCoordN}
\end{equation}
while using respecting Mandelstam variable $s,t,u$ definitions (\ref{Mandelstam}) we have  
\begin{equation}
s=4\omega_1\omega_2,\;t=-2\omega_1\omega_4(1-\cos\vartheta),\;u=-2\omega_2\omega_4(1+\cos\vartheta).
\label{Mandelstam1}
\end{equation}
Taking the case of photons there is $s+t+u=0$ from where, together with energy conservation, we express a pair of incoming energies $\omega_{1,2}$ in terms of pair of outgoing energies  $\omega_{3,4}$, and vice versa: 
\begin{equation}
\begin{split} 
\omega_1&=
\frac{1}{2}\bigg({\omega_3+\omega_4\big(1+x\big)
\pm\sqrt{\omega^2_3-\omega^2_4(1-x^2)}}\bigg),\;\;
\omega_2=
\frac{1}{2}\bigg({\omega_3+\omega_4\big(1-x\big)
\mp\sqrt{\omega^2_3-\omega^2_4(1-x^2)}}\bigg),
\\
\omega_3&=\frac{(\omega^2_1+\omega^2_2)-
(\omega^2_1-\omega^2_2)x}{(\omega_1+\omega_2)-
(\omega_1-\omega_2)x},\;\;\;\;
\omega_4=\frac{2\omega_1\omega_2}{\omega_1(1-x)+\omega_2(1+x)}\;,\;x=\cos\vartheta.
\end{split}
\label{Mandelstam2}
\end{equation}

Now, by using noncommutative matrix $\theta$ and kinematics, 
from Eqs. (\ref{Cmatrix}) and (\ref{SphCoordN}), respectively, 
we write down analytically all phase factors generated by the NCQED 
introduced in the previous section, and in the NCM frame of 
head-to-head $2\to2$ collision necessary further to compute and 
analyze NC contributions to all electron and/or photon scattering 
and/or annihilation processes:
\begin{equation}
\begin{split}
k_1\theta k_2&=-\frac{2\omega_1\omega_2}{\Lambda^2_{\rm NC}} c_{03},
\\
k_1\theta k_3&=\frac{-\omega_1}{\Lambda^2_{\rm NC}}
\Big[c_{03}\Big(2\omega_2+\omega_4(\cos\vartheta-1)\Big)+\omega_4\sin\vartheta
\Big((c_{01}-c_{13})\cos\varphi+(c_{02}-c_{23})\sin\varphi\Big)\Big],
\\
k_1\theta k_4&=\frac{\omega_1\omega_4}{\Lambda^2_{\rm NC}}
\Big[c_{03}(\cos\vartheta-1)+\sin\vartheta\Big((c_{01}-c_{13})\cos\varphi
+(c_{02}-c_{23})\sin\varphi\Big)\Big],
\\
k_2\theta k_3&
=\frac{\omega_2}{\Lambda^2_{\rm NC}}
\Big[c_{03}\Big(2\omega_1-\omega_4(\cos\vartheta+1)\Big)-\omega_4\sin\vartheta\Big((c_{01}+c_{13})\cos\varphi+(c_{02}+c_{23})\sin\varphi\Big)\Big],
\\
k_2\theta k_4&=\frac{\omega_2\omega_4}{\Lambda^2_{\rm NC}}
\Big[c_{03}(\cos\vartheta+1)+\sin\vartheta\Big((c_{01}+c_{13})\cos\varphi
+(c_{02}+c_{23})\sin\varphi\Big)\Big],
\\
k_3\theta k_4&
=\frac{\omega_4}{\Lambda^2_{\rm NC}}
\Big[c_{03}\Big(\omega_1(\cos\vartheta-1)+\omega_2(\cos\vartheta+1)\Big)
\\&
\phantom{XXXX}
+\sin\vartheta\Big(\big(\omega_1+\omega_2)(c_{01}\cos\varphi+c_{02}\sin\varphi)
-(\omega_1-\omega_2)(c_{13}\cos\varphi+c_{23}\sin\varphi)\Big)\Big].
\end{split}
\label{SphCoordphaseGener}
\end{equation}

\subsection{The $\varphi$ integrations over the noncommutative phase factors}

Using Eq. (\ref{SphCoordphaseGener}), after decomposing sinus functions in Eqs. (\ref{Helicity24}) and (\ref{Helicity25}), and for other processes,  with the help of \cite{Gradshteyn} we may neglect terms which in the integration over $d\varphi$ give zeros\footnote{Note that while computing the $\varphi$ distribution of certain processes we have used non-$\varphi$ integrated, i.e. full NC phase factors given in Eq. (\ref{SphCoordphaseGener}).} and obtain all contributing phase factors needed:
\begin{equation}
\begin{split} 
\cos k_1\theta k_2&=\cos\Big(-\frac{2\omega_1\omega_2}{\Lambda_{\rm NC}^2}c_{03}\Big),
\\
\cos k_1\theta k_3&=\cos\Big(-\frac{\omega_1}{\Lambda_{\rm NC}^2}c_{03}\big(2\omega_2+\omega_4(x-1)\big)\Big)\cdot\cos \Big(-\frac{\omega_1\omega_4}{\Lambda_{\rm NC}^2}C\sqrt{1-x^2}\sin(\varphi+\gamma_C)\Big),
\\
\cos k_1\theta k_4&=\cos\Big(\frac{\omega_1\omega_4}{\Lambda_{\rm NC}^2}c_{03}(x-1)\Big)\cdot\cos \Big(\frac{\omega_1\omega_4}{\Lambda_{\rm NC}^2}C\sqrt{1-x^2}\sin(\varphi+\gamma_C)\Big),
\\
\cos k_2\theta k_3&=
\cos\Big(\frac{\omega_2}{\Lambda_{\rm NC}^2}c_{03}\big(2\omega_1-\omega_4(x+1)\big)\Big)
\cdot\cos\Big(
-\frac{\omega_2\omega_4}{\Lambda_{\rm NC}^2}G\sqrt{1-x^2}\sin(\varphi+\gamma_G)
\Big)
\\
\cos k_2\theta k_4&=\cos\Big(\frac{\omega_2\omega_4}{\Lambda_{\rm NC}^2}c_{03}(x+1)\Big)\cdot\cos \Big(\frac{\omega_2\omega_4}{\Lambda_{\rm NC}^2}G\sqrt{1-x^2}\sin(\varphi+\gamma_G)\Big),
\\
\cos k_3\theta k_4&=\cos\Big(\frac{\omega_4}{\Lambda_{\rm NC}^2}c_{03}\big(\omega_1(x-1)+\omega_2(x+1)\big)\Big)
\cdot\cos \Big(\frac{\omega_1\omega_4}{\Lambda_{\rm NC}^2}A\sqrt{1-x^2}\sin(\varphi+\gamma_A)\Big),
\end{split}
\label{B2}
\end{equation}
\begin{equation}
\begin{split}
& 
\cos (k_1\theta k_2\pm k_3\theta k_4)=\cos\Big(\frac{c_{03}}{\Lambda_{\rm NC}^2}\big[-2\omega_1\omega_2\pm\omega_4\big(\omega_1(x-1)+\omega_2(x+1)\big)\big]\Big)
\cdot\cos\Big(\frac{\omega_1\omega_4}{\Lambda_{\rm NC}^2}A\sqrt{1-x^2}\sin(\varphi+\gamma_A)\Big),
\\& 
\cos (k_1\theta k_3+ k_2\theta k_4)=\cos \Big(\frac{\omega_1\omega_4}{\Lambda_{\rm NC}^2}M\sqrt{1-x^2}\sin(\varphi+\gamma_M)\Big),
\\& 
\cos (k_1\theta k_3 -k_2\theta k_4)=\cos\Big(\frac{2\omega_2\omega_4}{\Lambda_{\rm NC}^2}c_{03}(x+1)\Big)\cdot\cos \Big(\frac{\omega_1\omega_4}{\Lambda_{\rm NC}^2}A\sqrt{1-x^2}\sin(\varphi+\gamma_A)\Big),
\\& 
\cos (k_1\theta k_4+ k_2\theta k_3)=\cos \Big(\frac{\omega_1\omega_4}{\Lambda_{\rm NC}^2}M\sqrt{1-x^2}\sin(\varphi+\gamma_M)\Big),
\\& 
\cos (k_1\theta k_4-k_2\theta k_3)=\cos\Big(\frac{2\omega_1\omega_4}{\Lambda_{\rm NC}^2}c_{03}(x-1)\Big)\cdot\cos \Big(\frac{\omega_1\omega_4}{\Lambda_{\rm NC}^2}A\sqrt{1-x^2}\sin(\varphi+\gamma_A)\Big),
\end{split}
\label{B3}
\end{equation}
with the following definitions of the noncommutative coefficients
\begin{equation}
\begin{split}
C=&\sqrt{(c_{01}-c_{13})^2+(c_{02}-c_{23})^2}, \;\sin\gamma_C=\frac{1}{C}(c_{01}-c_{13}), \;\cos\gamma_C=\frac{1}{C}(c_{02}-c_{23}),
\\
G=&\sqrt{(c_{01}+c_{13})^2+(c_{02}+c_{23})^2}, \;\sin\gamma_G=\frac{1}{G}(c_{01}+c_{13}), \;\cos\gamma_G=\frac{1}{G}(c_{02}+c_{23}),
\\
A=&\sqrt{\Big(c_{01}\big(1+\frac{\omega_2}{\omega_1}\big)-c_{13}\big(1-\frac{\omega_2}{\omega_1}\big)\Big)^2
+\Big(c_{02}\big(1+\frac{\omega_2}{\omega_1}\big)-c_{23}\big(1-\frac{\omega_2}{\omega_1}\big)\Big)^2}, 
\\
\sin\gamma_A=&\frac{1}{A}\Big(c_{01}\big(1+\frac{\omega_2}{\omega_1}\big)-c_{13}\big(1-\frac{\omega_2}{\omega_1}\big)\Big), \,
\cos\gamma_A=\frac{1}{A}\Big(c_{02}\big(1+\frac{\omega_2}{\omega_1}\big)-c_{23}\big(1-\frac{\omega_2}{\omega_1}\big)\Big),
\\
M=&\sqrt{\Big(c_{01}\big(1-\frac{\omega_2}{\omega_1}\big)-c_{13}\big(1+\frac{\omega_2}{\omega_1}\big)\Big)^2+\Big(c_{02}\big(1-\frac{\omega_2}{\omega_1}\big)-c_{23}\big(1+\frac{\omega_2}{\omega_1}\big)\Big)^2}, 
\\
\sin\gamma_M=&\frac{1}{M}\Big(c_{01}\big(1-\frac{\omega_2}{\omega_1})-c_{13}\big(1+\frac{\omega_2}{\omega_1}\big)\Big), \,
\cos\gamma_M=\frac{1}{M}\Big(c_{02}\big(1-\frac{\omega_2}{\omega_1}\big)-c_{23}\big(1+\frac{\omega_2}{\omega_1}\big)\Big).
\end{split}
\label{E4}
\end{equation}

Now, since the SM amplitudes $M_{\rm SM}^{h_i}, \forall i=1,...,5$---see \cite{Gounaris:1998qk,Gounaris:1999gh,Davoudiasl:1999di,Bern:2001dg,Baur:2001jj}---do not depend on $\varphi$, the form which contains only the linear type of phase factor products $2\sin\alpha\sin\beta=\cos(\alpha-\beta)-\cos(\alpha+\beta)$ shall be applied for computations of interference terms in Eq. (\ref{SMNCTotCrossSect}). Second, from above general form we have to integrate over $d\varphi$ 
a square phase factor coming from Eq. (\ref{Helicity25}) as $4\sin^2{\frac{\alpha}{2}}\sin^2{\frac{\beta}{2}}
=1-\cos\alpha-\cos\beta+\frac{1}{2}(\cos(\alpha+\beta)+\cos(\alpha-\beta))$. 
Combining both types of phase, the $2\sin\alpha\sin\beta$ and the $4\sin^2{\frac{\alpha}{2}}\sin^2{\frac{\beta}{2}}$, 
we found two types of integrals $I$, and  $\hat I$, i.e. $I_{\pm}$, and $\hat I_{0,\pm}$, respectively:
\begin{gather}
\int_0^{2\pi}d\varphi\;2\sin\alpha\sin\beta\;\;\Longrightarrow\;\;{I_-}(x)-{I_+}(x),\;x=\cos\vartheta, 
\label{C6}\\
\phantom{XXxx..}
\int_0^{2\pi}d\varphi\;4\sin^2{\frac{\alpha}{2}}\sin^2{\frac{\beta}{2}}\;\;\Longrightarrow\;\; \hat I_0(x)+\frac{1}{2}\big(\hat I_+(x)+ \hat I_-(x)\big).
\label{C7}
\end{gather}
Next by performing $\varphi$ integrations of each of three phase terms in $M_{\rm NC}^{++++}$, $M_{\rm NC}^{+-+-}$, and $M_{\rm NC}^{++--}$ amplitudes,  after making simple replacement $\Lambda^2_{\rm NC}\to2\Lambda^2_{\rm NC}$ in (\ref{B3}) phases, and using (\ref{C6}) structure, we obtain the integrals we need:
\begin{equation}
\begin{split}
I^{1234}_\pm(x)&=2\pi\cos\bigg[\frac{c_{03}}{2\Lambda^2_{\rm NC}}\Big(-2\omega_1\omega_2\pm\omega_4\big(\omega_1(x-1)+\omega_2(x+1)\big)\Big)\bigg]J_0\bigg[\frac{\omega_1\omega_4}{2\Lambda^2_{\rm NC}}
A\sqrt{1-x^2}\bigg],
\\
I^{1324}_\pm(x)&=2\pi\cos\bigg[\frac{c_{03}}{2\Lambda^2_{\rm NC}}\Big(-\omega_1\big(2\omega_2+\omega_4(x-1)\big)\pm \omega_2\omega_4(x+1)\Big)\bigg]J_0\bigg[\frac{\omega_1\omega_4}{2\Lambda^2_{\rm NC}}{\tiny{M\choose A}}\sqrt{1-x^2}\bigg],
\\
I^{1423}_\pm(x)&=2\pi\cos\bigg[\frac{c_{03}}{2\Lambda^2_{\rm NC}}\Big(\pm \omega_2\big(2\omega_1-\omega_4(x+1)\big)+\omega_1\omega_4(x-1)\Big)\bigg]J_0\bigg[\frac{\omega_1\omega_4}{2\Lambda^2_{\rm NC}}{M\choose A}\sqrt{1-x^2}\bigg],
\end{split}
\label{B9}
\end{equation} 
and we note that for the spacelike case $c_{03}=0$ the difference $(I^{1234}_- -I^{1234}_+)$ vanishes.

Now we find $d\varphi$ integrals over the (\ref{C7}) structure of phases as:
\begin{equation}
\begin{split}
\hat I^{1234}_0(x)
&=2\pi\bigg(1-\cos\Big[\frac{2c_{03}}{\Lambda^2_{\rm NC}}\omega_1\omega_2\Big]
-\cos\Big[\frac{c_{03}}{\Lambda^2_{\rm NC}}\omega_4
\big(\omega_1(x-1)+\omega_2(x+1)\big)\Big]J_0\Big[\frac{\omega_1\omega_4}{\Lambda^2_{\rm NC}}A\sqrt{1-x^2}\Big]\bigg),
\\
\hat I^{1324}_0(x)
&=2\pi\bigg(1-\cos\Big[\frac{{c_{03}}}{\Lambda^2_{\rm NC}}\omega_1
\big(2\omega_2+\omega_4(x-1)\big)\Big]J_0\Big[\frac{\omega_1\omega_4}{\Lambda^2_{\rm NC}}C\sqrt{1-x^2}\Big]
-\cos\Big[\frac{c_{03}}{\Lambda^2_{\rm NC}}
\omega_2\omega_4(x+1)\Big]J_0\Big[\frac{\omega_2\omega_4}{\Lambda^2_{\rm NC}}G\sqrt{1-x^2}\Big]\bigg),
\\
\hat I^{1423}_0(x)
&=2\pi\bigg(1-\cos\Big[\frac{c_{03}}{\Lambda^2_{\rm NC}}
\omega_1\omega_4(x-1)\Big]J_0\Big[\frac{\omega_1\omega_4}{\Lambda^2_{\rm NC}}C\sqrt{1-x^2}\Big]
-\cos\Big[\frac{ c_{03}}{\Lambda^2_{\rm NC}}\omega_2
\big(2\omega_1-\omega_4(x+1)\big)\Big]J_0\Big[\frac{\omega_2\omega_4}{\Lambda^2_{\rm NC}}G\sqrt{1-x^2}\Big]\bigg),
\\
\hat I^{1234}_\pm(x)&=2\pi\cos\bigg[\frac{c_{03}}{\Lambda^2_{\rm NC}}
\Big(-2\omega_1\omega_2\pm\omega_4
\big(\omega_1(x-1)+\omega_2(x+1)\big)\Big)\bigg]J_0\bigg[\frac{\omega_1\omega_4}{\Lambda^2_{\rm NC}}
A\sqrt{1-x^2}\bigg],
\\
\hat I^{1324}_\pm(x)&=2\pi\cos\bigg[\frac{c_{03}}{\Lambda^2_{\rm NC}}
\Big(-\omega_1\big(2\omega_2+\omega_4(x-1)\big)\pm \omega_2\omega_4(x+1)\Big)\bigg]J_0\bigg[\frac{\omega_1\omega_4}{\Lambda^2_{\rm NC}}{\tiny{M\choose A}}\sqrt{1-x^2}\bigg],
\\
\hat I^{1423}_\pm(x)&=2\pi\cos\bigg[\frac{c_{03}}{\Lambda^2_{\rm NC}}
\Big(\pm \omega_2\big(2\omega_1-\omega_4(x+1)\big)+\omega_1\omega_4(x-1)\Big)
\bigg]J_0\bigg[\frac{\omega_1\omega_4}{\Lambda^2_{\rm NC}}{M\choose A}\sqrt{1-x^2}\bigg].
\end{split}
\label{B10}
\end{equation}

\section{VANISHING OF THE SW$(II)$ MAP-INDUCED AMPLITUDE ${\cal M}^\gamma_{II}$: PROOF OF EQ. (\ref{4gIIstu})}

Explicit cancellation of the SW$(II)$ map-induced contributions to $\gamma\gamma\to\gamma\gamma$ process from Fig.\ref{fig:FD6} in the particular case of Moyal U(1) NCQED is very demanding, because some quite nonstandard tensorial analysis within a sum of relevant Feynman diagrams is needed. That is the reason why lengthy tensorial contractions 
in a sum of $s-, t-$ and $u$-channel 3$\gamma$-vertex diagrams plus 4$\gamma$-vertex diagram from Fig.\ref{fig:FD6} 
executed by hand and computer is given in detail next, showing explicit vanishing of ${\cal M}^\gamma_{II}$, i.e. Eq. (\ref{4gIIstu}). We start with details of $3\gamma$- and $4\gamma$-vertices and relevant amplitudes, in the $(s,t,u)$ channels, generated by both the non-SW($I$)$|$SW($II$) map-induced terms, respectively.

\subsubsection{$3\gamma$-vertices, $(s,t,u)$ channels: Non-SW($I$)$|$SW($II$) map-induced terms}

Using FRs (\ref{Fg}) and (\ref{FgA}), and from Fig.\ref{fig:FD6}, and after splitting the non-SW parts from the SW parts, 
applying momentum conservations and free field equations, 
we obtain the following  twice six non-SW($I$)$|$SW($II$) map-induced $3\gamma$-vertex contributions : 
\begin{equation}
\begin{split}
\Gamma_I^{\mu_1\mu_2\alpha}(k_1,k_2,q_s)&=
e(k_2\theta k_1)\big[2k_2^{\mu_1} g^{\mu_2\alpha}-2{k_1}^{\mu_2} g^{\mu_1\alpha}
+(k_1-k_2)^\alpha g^{\mu_1\mu_2}\big]{F_{\star_2}(k_2,k_1)},
\\
\Gamma_I^{\mu_3\mu_4\beta}(-k_3,-k_4,-q_s)&=
e(k_4\theta k_3)\big[2{k_3}^{\mu_4} g^{\beta\mu_3} - 2{k_4}^{\mu_3} g^{\beta\mu_4}
+ (k_4-k_3)^\beta g^{\mu_3\mu_4}\big]{F_{\star_2}(k_4,k_3)},
\\
\Gamma_I^{\mu_1\alpha\mu_4}(k_1,-q_t,-k_4)&=
e(k_4\theta k_1)\big[2k_1^{\mu_4} g^{\alpha\mu_1}+2{k_4}^{\mu_1} g^{\alpha\mu_4}
-(k_4+k_1)^\alpha g^{\mu_1\mu_4}\big]{F_{\star_2}(k_4,k_1)},
\\
\Gamma_I^{\mu_2\mu_3\beta}(k_2,-k_3,q_t)&=
e(k_3\theta k_2)\big[2k_2^{\mu_3} g^{\beta\mu_2}+2{k_3}^{\mu_2} g^{\beta\mu_3}
-(k_3+k_2)^\beta g^{\mu_2\mu_3}\big]{F_{\star_2}(k_3,k_2)},
\\
\Gamma_I^{\mu_1\alpha\mu_3}(k_1,-q_u,-k_3)&=
e(k_3\theta k_1)\big[2{k_3}^{\mu_1} g^{\mu_3\alpha}+2{k_1}^{\mu_3} g^{\mu_1\alpha}
-(k_3+k_1)^\alpha g^{\mu_1\mu_3}\big]{F_{\star_2}(k_3,k_1)},
\\
\Gamma_I^{\mu_2\mu_4\beta}(k_2,-k_4,q_u)&=
e(k_4\theta k_2)\big[2{k_2}^{\mu_4} g^{\beta\mu_2}+2{k_4}^{\mu_2} g^{\beta\mu_4}
-(k_4+k_2)^\beta g^{\mu_2\mu_4}\big]{F_{\star_2}(k_4,k_2)}.
\end{split}
\label{Vertices3GIFR}
\end{equation}
\begin{eqnarray}
\Gamma_{II}^{\mu_1\mu_2\alpha}(k_1,k_2,q_s)&=&
e\Big\{(k_1k_2)\big[2(\theta k_2)^{\mu_1} g^{\alpha\mu_2}+
2(\theta k_1)^{\mu_2} g^{\alpha\mu_1}
+{\theta}^{\mu_1\mu_2}(k_1-k_2)^\alpha \big]
\nonumber\\
&&\phantom{Xx.}-\big[(\theta k_2)^{\mu_1} {k_1}^{\mu_2}
+(\theta k_1)^{\mu_2} k_2^{\mu_1} \big](k_1+k_2)^\alpha
\Big\}\;{F_{\star_2}(k_1,k_2)},
\nonumber\\
\Gamma_{II}^{\mu_3\mu_4\beta}(-k_3,-k_4,-q_s)&=&
-e\Big\{(k_3k_4)\big[2(\theta k_3)^{\mu_4} g^{\beta\mu_3}+2(\theta k_4)^{\mu_3} g^{\beta\mu_4}+{\theta}^{\mu_3\mu_4}(k_3-k_4)^\beta \big]
\nonumber\\
&&\phantom{Xxx}-\big[(\theta k_3)^{\mu_4} {k_4}^{\mu_3}+(\theta k_4)^{\mu_3} {k_3}^{\mu_4} \big](k_3+k_4)^\beta\Big\}\;{F_{\star_2}(k_3,k_4)},
\nonumber\\
\Gamma_{II}^{\mu_1\alpha\mu_4}(k_1,-q_t,-k_4)&=&
e\Big\{(k_1k_4)\big[2(\theta k_4)^{\mu_1} g^{\alpha\mu_4}-2(\theta k_1)^{\mu_4} g^{\alpha\mu_1}-{\theta}^{\mu_1\mu_4}(k_4+k_1)^\alpha \big]
\nonumber\\
&&\phantom{XX.}-\big[(\theta k_1)^{\mu_4} {k_4}^{\mu_1}
+(\theta k_4)^{\mu_1} k_1^{\mu_4} \big](k_4-k_1)^\alpha
\Big\}\;{F_{\star_2}(k_4,k_1)},
\nonumber\\
\Gamma_{II}^{\mu_2\mu_3\beta}(k_2,-k_3,q_t,)&=&
e\Big\{(k_2k_3)\big[2(\theta k_3)^{\mu_2} g^{\beta\mu_3}-2(\theta k_2)^{\mu_3} g^{\beta\mu_2}-{\theta}^{\mu_2\mu_3}(k_3+k_2)^\beta \big]
\nonumber\\
&&\phantom{XX.}-\big[(\theta k_2)^{\mu_3} {k_3}^{\mu_2}+(\theta k_3)^{\mu_2} k_2^{\mu_3} \big](k_3-k_2)^\beta\Big\}\;{F_{\star_2}(k_3,k_2)},
\nonumber\\
\Gamma_{II}^{\mu_1\alpha\mu_3}(k_1,-q_u,-k_3)&=&
e\Big\{(k_1k_3)\big[2(\theta k_3)^{\mu_1} g^{\alpha\mu_3}-2(\theta k_1)^{\mu_3} g^{\alpha\mu_1}-{\theta}^{\mu_1\mu_3}(k_3+k_1)^\alpha \big]
\nonumber\\
&&\phantom{Xxx}-\big[(\theta k_3)^{\mu_1} k_1^{\mu_3}+(\theta k_1)^{\mu_3} {k_3}^{\mu_1} \big](k_3-k_1)^\alpha\Big\}\;{F_{\star_2}(k_3,k_1)},
\nonumber\\
\Gamma_{II}^{\mu_2\mu_4\beta}(k_2,-k_4,q_u)&=&
e\Big\{(k_2k_4)\big[2(\theta k_4)^{\mu_2} g^{\beta\mu_4}-2(\theta k_2)^{\mu_4} g^{\beta\mu_2}-{\theta}^{\mu_2\mu_4}(k_4+k_2)^\beta \big]
\nonumber\\
&&\phantom{Xxx}-\big[(\theta k_2)^{\mu_4} {k_4}^{\mu_2}
+(\theta k_4)^{\mu_2} {k_2}^{\mu_4} \big](k_4-k_2)^\beta
\Big\}\;{F_{\star_2}(k_4,k_2)}.
\nonumber\\
\label{Vertices3GIIFR}
\end{eqnarray}
needed to show the Ward identity for amplitude (\ref{ForGammaAmplitude2}) and vanishing of the SW map-induced contributions to scattering amplitude (\ref{4gIIstu}).

\subsubsection{$3\gamma$-diagrams, $(s,t,u)$ channels: SW($II$) map-induced amplitudes}

The sum of $3\gamma$-vertex diagrams defined in Fig.\ref{fig:FD6} gives the following amplitude:
\begin{equation}
\begin{split}
&{\cal M}^{3\gamma}_{II}=-i 
\epsilon_{\mu_1}(k_1)\epsilon_{\mu_2}(k_2){\epsilon}^{\ast}_{\mu_3}(k_3)\epsilon^{\ast}_{\mu_4}(k_4) 
\\&\cdot\bigg\{\frac{1}{s}\Big[\Gamma_{I}^{\mu_1\mu_2\alpha}(k_1,k_2,q_s)\Gamma_{II}^{\mu_3\mu_4\alpha}(-k_3,-k_4,-q_s)
+\Gamma_{II}^{\mu_1\mu_2\alpha}(k_1,k_2,q_s)\Gamma_{I}^{\mu_3\mu_4\alpha}(-k_3,-k_4,-q_s)
\\&\phantom{Xx}
+\Gamma_{II}^{\mu_1\mu_2\alpha}(k_1,k_2,q_s)\Gamma_{II}^{\mu_3\mu_4\alpha}(-k_3,-k_4,-q_s)\Big]
\\&\phantom{.}
+\frac{1}{t}\Big[\Gamma_{I}^{\mu_1\alpha\mu_4}(k_1,-q_t,-k_4)\Gamma_{II}^{\mu_2\mu_3\alpha}(k_2,-k_3,q_t)
+\Gamma_{II}^{\mu_1\alpha\mu_4}(k_1,-q_t,-k_4)\Gamma_{I}^{\mu_2\mu_3\alpha}(k_2,-k_3,q_t)
\\&\phantom{Xx}
+\Gamma_{II}^{\mu_1\alpha\mu_4}(k_1,-q_t,-k_4)\Gamma_{II}^{\mu_2\mu_3\alpha}(k_2,-k_3,q_t)\Big]
\\&\phantom{.}
+\frac{1}{u}\Big[\Gamma_{I}^{\mu_1\alpha\mu_3}(k_1,-q_u,-k_3)\Gamma_{II}^{\mu_2\mu_4\alpha}(k_2,-k_4,q_u)
+\Gamma_{II}^{\mu_1\alpha\mu_3}(k_1,-q_u,-k_3)\Gamma_{I}^{\mu_2\mu_4\alpha}(k_2,-k_4,q_u)
\\&\phantom{Xx}
+\Gamma_{II}^{\mu_1\alpha\mu_3}(k_1,-q_u,-k_3)\Gamma_{II}^{\mu_2\mu_4\alpha}(k_2,-k_4,q_u)\Big]\bigg\}, 
\end{split}
\label{3photonsum}
\end{equation} 
which produces all nine terms, from $s, t$ and $u$ channels in Eq. (\ref{3photonsum}) separately, as:
\begin{equation}
\begin{split}
&\frac{-i}{(k_1+k_2)^2}\Big[\Gamma_{I}^{\mu_1\mu_2\alpha}(k_1,k_2,q_s)
\Gamma_{II}^{\mu_3\mu_4\alpha}(-k_3,-k_4,-q_s)
+\Gamma_{II}^{\mu_1\mu_2\alpha}(k_1,k_2,q_s)
\Gamma_{I}^{\mu_3\mu_4\alpha}(-k_3,-k_4,-q_s)
\\&\phantom{XXXX.}
+\Gamma_{II}^{\mu_1\mu_2\alpha}(k_1,k_2,q_s)
\Gamma_{II}^{\mu_3\mu_4\alpha}(-k_3,-k_4,-q_s)\Big]
\\&\phantom{.}
=(-ie^2)\bigg\{(k_1\theta k_2)\bigg[k_1^{\mu_2}\Big( 
2(\theta k_3)^{\mu_4}g^{\mu_1\mu_3}
+2(\theta k_4)^{\mu_3}g^{\mu_1\mu_4}
+\theta^{\mu_3\mu_4}(k_3 - k_4)^{\mu_1}\Big)
\\&\phantom{XXXXXXXXx}
-k_2^{\mu_1}\Big( 
2(\theta k_3)^{\mu_4}g^{\mu_2\mu_3}
+2(\theta k_4)^{\mu_3}g^{\mu_2\mu_4}
+\theta^{\mu_3\mu_4}(k_3 - k_4)^{\mu_2} \Big)
\\&\phantom{XXXXX.}+
g^{\mu_1\mu_2}\Big( 
(\theta k_3)^{\mu_4}(k_2-k_1)^{\mu_3}
+(\theta k_4)^{\mu_3}(k_2-k_1)^{\mu_4}
+\frac{1}{2}\theta^{\mu_3\mu_4}(k_2 - k_1)(k_3 - k_4)\Big)\bigg]
\\&\phantom{XXXX.}
+(k_3\theta k_4)\bigg[k_3^{\mu_4}\Big( 
2(\theta k_1)^{\mu_2}g^{\mu_1\mu_3}
+2(\theta k_2)^{\mu_1}g^{\mu_2\mu_3}
+\theta^{\mu_1\mu_2}(k_1 - k_2)^{\mu_3}\Big)
\\&\phantom{XXXXXXXX..}
-k_4^{\mu_3}\Big( 
2(\theta k_1)^{\mu_2}g^{\mu_1\mu_4}
+2(\theta k_2)^{\mu_1}g^{\mu_2\mu_4}
+\theta^{\mu_1\mu_2}(k_1 - k_2)^{\mu_4} \Big)
\\&\phantom{XXXXX.}+
g^{\mu_3\mu_4}\Big( 
(\theta k_2)^{\mu_1}(k_4-k_3)^{\mu_2}
+(\theta k_1)^{\mu_2}(k_4-k_3)^{\mu_1}
+\frac{1}{2}\theta^{\mu_1\mu_2}(k_1 - k_2)(k_4 - k_3)\Big)\bigg]
\\&\phantom{XXXXx}
+ (k_3k_4)\bigg[(\theta k_1)^{\mu_2}\Big(2(\theta k_3)^{\mu_4} g^{\mu_1\mu_3}+2(\theta k_4)^{\mu_3} g^{\mu_1\mu_4}+ \theta^{\mu_3\mu_4}(k_3-k_4)^{\mu_1}\Big)
\\&\phantom{XXXXXXXX.}
+(\theta k_2)^{\mu_1}\Big(2(\theta k_3)^{\mu_4} g^{\mu_2\mu_3}+2(\theta k_4)^{\mu_2} g^{\mu_2\mu_4}+
\theta^{\mu_3\mu_4}(k_3-k_4)^{\mu_2} \Big)
\\&\phantom{XXXXX}
+\theta^{\mu_1\mu_2}\Big((\theta k_3)^{\mu_4} (k_1-k_2)^{\mu_3}+(\theta k_4)^{\mu_3} (k_1-k_2)^{\mu_4}
+\frac{1}{2}\theta^{\mu_3\mu_4}(k_1-k_2)(k_3-k_4)\Big)\bigg]
\\&\phantom{XXXXXXXXx}
-\Big((\theta k_1)^{\mu_2} k_2^{\mu_1}+(\theta k_2)^{\mu_1} k_1^{\mu_2} \Big)
\Big((\theta k_3)^{\mu_4} k_4^{\mu_3}+(\theta k_4)^{\mu_3} k_3^{\mu_4} \Big)
\bigg\}\stackrel{\star}{S}, 
\end{split}
\label{Sstar}
\end{equation}
\begin{equation}
\begin{split}
&\frac{-i}{(k_1-k_4)^2}\Big[\Gamma_{I}^{\mu_1\alpha\mu_2}(k_1,-q_t,-k_4)\Gamma_{II}^{\mu_3\mu_4\alpha}(k_2,-k_3,q_t)
+\Gamma_{II}^{\mu_1\alpha\mu_2}(k_1,-q_t,-k_4)\Gamma_{I}^{\mu_3\mu_4\alpha}(k_2,-k_3,q_t)
\\&\phantom{XXXX.}
+\Gamma_{II}^{\mu_1\alpha\mu_2}(k_1,-q_t,-k_4)\Gamma_{II}^{\mu_3\mu_4\alpha}(k_2,-k_3,q_t)\Big]
\\&\phantom{x}
=(-ie^2)
\bigg\{(k_1\theta k_4)
\bigg[k_1^{\mu_4}\Big( 
2(\theta k_3)^{\mu_2}g^{\mu_1\mu_3}
-2(\theta k_2)^{\mu_3}g^{\mu_1\mu_2}
-\theta^{\mu_2\mu_3}(k_2 + k_3)^{\mu_1}\Big)
\\&\phantom{XXXXXXXXx.}
+k_4^{\mu_1}\Big( 
2(\theta k_3)^{\mu_2}g^{\mu_3\mu_4}
-2(\theta k_2)^{\mu_3}g^{\mu_2\mu_4}
-\theta^{\mu_2\mu_3}(k_2 + k_3)^{\mu_4} \Big)
\\&\phantom{XXXXXx}
-g^{\mu_1\mu_4}\Big( 
(\theta k_3)^{\mu_2}(k_1+k_4)^{\mu_3}
-(\theta k_2)^{\mu_3}(k_1+k_4)^{\mu_2}
-\frac{1}{2}\theta^{\mu_2\mu_3}(k_2 + k_3)(k_1 + k_4)\Big)\bigg]
\\&\phantom{XXXXx}
+(k_2\theta k_3)
\bigg[k_2^{\mu_3}\Big( 
2(\theta k_4)^{\mu_1}g^{\mu_2\mu_4}
-2(\theta k_1)^{\mu_4}g^{\mu_1\mu_2}
-\theta^{\mu_1\mu_4}(k_1 + k_4)^{\mu_2}\Big)
\\&\phantom{XXXXXXXXx}
+k_3^{\mu_2}\Big( 
2(\theta k_4)^{\mu_1}g^{\mu_3\mu_4}
-2(\theta k_1)^{\mu_4}g^{\mu_1\mu_3}
-\theta^{\mu_1\mu_4}(k_1 + k_4)^{\mu_3} \Big)
\\&\phantom{XXXXXx}
-g^{\mu_2\mu_3}\Big( 
(\theta k_4)^{\mu_1}(k_2+k_3)^{\mu_4}
-(\theta k_1)^{\mu_4}(k_2+k_3)^{\mu_1}
-\frac{1}{2}\theta^{\mu_1\mu_4}(k_1 + k_4)(k_2+ k_3)\Big)\bigg]
\\&\phantom{XXXXxx}
+ (k_2k_3)\bigg[(\theta k_4)^{\mu_1}\Big(2(\theta k_3)^{\mu_2} g^{\mu_3\mu_4}
-2(\theta k_2)^{\mu_3} g^{\mu_2\mu_4}
-\theta^{\mu_2\mu_3}(k_2+k_3)^{\mu_4} \Big)
\\&\phantom{XXXXXXXXx.}
-(\theta k_1)^{\mu_4}\Big(2(\theta k_3)^{\mu_2} g^{\mu_1\mu_3}
-2(\theta k_2)^{\mu_3} g^{\mu_1\mu_2}- \theta^{\mu_2\mu_3}(k_2+k_3)^{\mu_1}\Big)
\\&\phantom{XXXXXx} 
-\theta^{\mu_1\mu_4}\Big((\theta k_3)^{\mu_2} 
(k_1+k_4)^{\mu_3}-(\theta k_2)^{\mu_3} (k_1+k_4)^{\mu_2}
-\frac{1}{2}\theta^{\mu_2\mu_3}(k_1+k_4)(k_3+k_2)\Big)\bigg]
\\&\phantom{XXXXXXXX.}
-\Big((\theta k_1)^{\mu_4} k_4^{\mu_1}
+(\theta k_4)^{\mu_1} k_1^{\mu_4} \Big)
\Big((\theta k_2)^{\mu_3} k_3^{\mu_2}+(\theta k_3)^{\mu_2} k_2^{\mu_3} \Big)
\bigg\}\stackrel{\star}{T}, 
\end{split}
\label{Tstar}
\end{equation}
and
\begin{equation}
\begin{split}
&\frac{-i}{(k_1-k_3)^2}\Big[
\Gamma_{I}^{\mu_1\alpha\mu_2}(k_1,-q_u,-k_3)\Gamma_{II}^{\mu_3\mu_4\alpha}(k_2,-k_4,q_u)
+\Gamma_{II}^{\mu_1\alpha\mu_2}(k_1,-q_u,-k_3)\Gamma_{I}^{\mu_3\mu_4\alpha}(k_2,-k_4,q_u)
\\&\phantom{XXXX}
+\Gamma_{II}^{\mu_1\alpha\mu_2}(k_1,-q_u,-k_3)\Gamma_{II}^{\mu_3\mu_4\alpha}(k_2,-k_4,q_u)
\Big]
\\&\phantom{x}
=(-ie^2)\bigg\{(k_1\theta k_3)
\bigg[k_1^{\mu_3}\Big( 
2(\theta k_4)^{\mu_2}g^{\mu_1\mu_4}
-2(\theta k_2)^{\mu_4}g^{\mu_1\mu_2}
-\theta^{\mu_2\mu_4}(k_2 + k_4)^{\mu_1} \Big)
\\&\phantom{XXXXXXXXX}
+k_3^{\mu_1}\Big( 
2(\theta k_4)^{\mu_2}g^{\mu_3\mu_4}
-2(\theta k_2)^{\mu_4}g^{\mu_2\mu_3}
-\theta^{\mu_2\mu_4}(k_2 + k_4)^{\mu_3}\Big)
\\&\phantom{XXXXXx}
-g^{\mu_1\mu_3}\Big( 
(\theta k_4)^{\mu_2}(k_1+ k_3)^{\mu_4}
-(\theta k_2)^{\mu_4}(k_1+ k_3)^{\mu_2}
-\frac{1}{2}\theta^{\mu_2\mu_4}(k_1 + k_3)(k_2 + k_4)\Big)\bigg]
\\&\phantom{XXXXx}
+(k_2\theta k_4)
\bigg[k_2^{\mu_4}\Big( 
2(\theta k_3)^{\mu_1}g^{\mu_2\mu_3}
-2(\theta k_1)^{\mu_3}g^{\mu_1\mu_2}
-\theta^{\mu_1\mu_3}(k_1 + k_3)^{\mu_2}\Big)
\\&\phantom{XXXXXXXXx.}
+k_4^{\mu_2}\Big( 
2(\theta k_3)^{\mu_1}g^{\mu_2\mu_3}
-2(\theta k_1)^{\mu_3}g^{\mu_1\mu_4}
-\theta^{\mu_1\mu_3}(k_1 + k_3)^{\mu_4} \Big)
\\&\phantom{XXXXXx}
-g^{\mu_2\mu_4}\Big( 
(\theta k_3)^{\mu_1}(k_2+k_4)^{\mu_3}
-(\theta k_1)^{\mu_3}(k_2+k_4)^{\mu_1}
-\frac{1}{2}\theta^{\mu_1\mu_3}(k_1 + k_3)(k_2 + k_4)\Big)\bigg]
\\&\phantom{XXXXxx}+ (k_2k_4)
\bigg[ (\theta k_3)^{\mu_1}\Big(2(\theta k_4)^{\mu_2} g^{\mu_3\mu_4}
-2(\theta k_2)^{\mu_4} g^{\mu_2\mu_3}
-\theta^{\mu_2\mu_4}(k_2+k_4)^{\mu_3} \Big)
\\&\phantom{XXXXXXXXx.}
-(\theta k_1)^{\mu_3}
\Big(2(\theta k_4)^{\mu_2} g^{\mu_1\mu_4}
-2(\theta k_2)^{\mu_4} g^{\mu_1\mu_2}- \theta^{\mu_2\mu_4}(k_2+k_4)^{\mu_1}\Big)
\\&\phantom{XXXXXx}
-\theta^{\mu_1\mu_3}
\Big((\theta k_4)^{\mu_2} (k_1+k_3)^{\mu_4}
-(\theta k_2)^{\mu_4} (k_1+k_3)^{\mu_2}
-\frac{1}{2}\theta^{\mu_2\mu_4}(k_1+k_3)(k_2+k_4)\Big)\bigg]
\\&\phantom{XXXXXXXX.}
-\Big((\theta k_1)^{\mu_3} k_3^{\mu_1}+(\theta k_3)^{\mu_1} k_1^{\mu_3} \Big)
\Big((\theta k_2)^{\mu_4} k_4^{\mu_2}+(\theta k_4)^{\mu_2} k_2^{\mu_4} \Big)
\bigg\}\stackrel{\star}{U}. 
\end{split}
\label{Ustar}
\end{equation}
with self-evident shorthand notations:
\begin{equation}
\stackrel{\star}{S}=
\frac{\sin\frac{k_1\theta k_2}{2}}{\frac{k_1\theta k_2}{2}}
\frac{\sin\frac{k_3\theta k_4}{2}}{\frac{k_3\theta k_4}{2}},\;
\stackrel{\star}{T}=
\frac{\sin\frac{k_1\theta k_4}{2}}{\frac{k_1\theta k_4}{2}}
\frac{\sin\frac{k_2\theta k_3}{2}}{\frac{k_2\theta k_3}{2}},\;
\stackrel{\star}{U}=
\frac{\sin\frac{k_1\theta k_3}{2}}{\frac{k_1\theta k_3}{2}}
\frac{\sin\frac{k_2\theta k_4}{2}}{\frac{k_2\theta k_4}{2}},
\label{STAstar}
\end{equation}
The above $3\gamma$-vertex diagram $s$-, $t$-, $u$-channel amplitudes (\ref{Sstar}), (\ref{Tstar}), and (\ref{Ustar}), respectively, should be cancelled with the corresponding $s$-, $t$-, $u$-channel $4\gamma$-vertex diagram amplitudes we give next.

\subsubsection{$4\gamma$-vertices, $(s,t,u)$ channels: Non-SW($I$)$|$SW($II$) map-induced terms}

Starting with Feynman rules (\ref{A.1})--(\ref{A.4})  we perform simultaneous permutations $S_4$ of $\Gamma_A$ over the momentum-index pairs $\{k_i,\mu_i\},\;\;\forall i=1,....,4$. First, detailed permutation $S_4$ over relevant term $\Gamma_{A_1}^{\mu_1\mu_2\mu_3\mu_4}$ gives non-SW($I$)$|$SW($II$) induced four-photon terms: 
\begin{gather}
\begin{split}
S_4\Gamma_{A_1}^{\mu_1\mu_2\mu_3\mu_4}(k_1,k_2,-k_3,-k_4)\big|_{I}&=
4 \Big[(k_1\theta k_2)(k_3\theta k_4)\Big(g^{\mu_1\mu_3}g^{\mu_2\mu_4}-g^{\mu_1\mu_4}g^{\mu_2\mu_3}\Big)\stackrel{\star}{S}
\\&\phantom{X.}
+(k_1\theta k_4)(k_2\theta k_3)\Big(g^{\mu_1\mu_2}g^{\mu_3\mu_4}-g^{\mu_1\mu_3}g^{\mu_2\mu_4}\Big)\stackrel{\star}{T}
\\&\phantom{X.}
+(k_1\theta k_3)(k_2\theta k_4)\Big(g^{\mu_1\mu_2}g^{\mu_3\mu_4}-g^{\mu_1\mu_4}g^{\mu_2\mu_3}\Big)\stackrel{\star}{U}\Big],
\\\phantom{X.}
S_4\Gamma_{A_1}^{\mu_1\mu_2\mu_3\mu_4}(k_1,k_2,-k_3,-k_4)\big|_{II}&=0.
\end{split}
\label{S4A1}
\end{gather}
Note that the $S_4$ permutation over momenta $k_i,\forall i=1,2,3,4$ acting on 
the product of two functions $F_{\star_2}\left(k_1,k_2\right) F_{\star_2}\left(k_3,k_4\right)$, 
due to the momentum conservation and properties of the $\star_2$ product, gives only three different terms,  
$\stackrel{\star}{S}$, $\stackrel{\star}{T}$, and $\stackrel{\star}{U}$, respectively. Equation (\ref{S4A1}) describes  
the four-photon vertex arising from the pure $\star$-commutator, as already discussed in Appendix A after Eq. (\ref{A.3}),  and it corresponds exactly to the non-SW contribution  $\Gamma^{\mu_1\mu_2\mu_3\mu_4}_{I}(k_1,k_2,-k_3,-k_4)$ vertex. 
We anticipate in Eq. (\ref{S4A1}) that  there are no SW($II$) map-induced  contributions corresponding to 
the  $S_4\Gamma^{\mu_1\mu_2\mu_3\mu_4}_{A_1}(k_1,k_2,-k_3,-k_4)$ term. 
However, all other terms in Eqs.(\ref{A.2})--(\ref{A.5}) do arise due to the SW maps. 
Detailed $S_4$ permutation of remaining terms in Eq. (\ref{A.2}) from Appendix A starting with 
$\Gamma_{A_2}^{\mu_1\mu_2\mu_3\mu_4}$ gives 
\begin{gather}
\begin{split}
S_4\Gamma_{A_2}^{\mu_1\mu_2\mu_3\mu_4}(k_1,k_2,-k_3,-k_4)\big|_{I}&=0,
\\
S_4\Gamma_{A_2}^{\mu_1\mu_2\mu_3\mu_4}(k_1,k_2,-k_3,-k_4)\big|_{II}&=
4\Big[\theta^{\mu_1\mu_2}\theta^{\mu_3\mu_4}
\big((k_1k_3)(k_2k_4)-(k_1k_4)(k_2k_3)\big)\stackrel{\star}{S}
\\&\phantom{xx}
+\theta^{\mu_1\mu_4}\theta^{\mu_2\mu_3}
\big((k_1k_2)(k_3k_4)-(k_1k_3)(k_2k_4)\big)\stackrel{\star}{T}
\\&\phantom{xx}
+\theta^{\mu_1\mu_3}\theta^{\mu_2\mu_4}
\big((k_1k_2)(k_3k_4)-(k_1k_4)(k_2k_3)\big)\stackrel{\star}{U}\Big].
\end{split}
\label{S4A2}
\end{gather}
Here we anticipate that there are no non-SW($I$) map-induced contributions corresponding to the  $S_4\Gamma^{\mu\nu\rho\tau}_{A_2}(k_1,k_2,-k_3,-k_4)$ term. 
Permutations of the next terms give 
\begin{gather}
\begin{split}
&S_4\Gamma_{A_3}^{\mu_1\mu_2\mu_3\mu_4}(k_1,k_2,-k_3,-k_4)
\\&=4\Big\{\Big[(\theta k_1)^{\mu_2}\Big((\theta k_4)^{\mu_3}(k_1k_4)g^{\mu_1\mu_4}+(\theta k_3)^{\mu_4}(k_1k_3)g^{\mu_1\mu_3}\Big)
\\&\phantom{Xx.}
+(\theta k_2)^{\mu_1}\Big((\theta k_3)^{\mu_4}(k_1k_4)g^{\mu_2\mu_3}
+(\theta k_4)^{\mu_3}(k_1k_3)g^{\mu_2\mu_4}\Big)
\\&\phantom{XXX}
+\theta^{\mu_1\mu_2}\Big((k_1k_4)\big(k_1^{\mu_4}(\theta k_4)^{\mu_3}
-k_2^{\mu_3}(\theta k_3)^{\mu_4}\big)+(k_1k_3)\big(k_1^{\mu_3}(\theta k_3)^{\mu_4}-k_2^{\mu_4}(\theta k_4)^{\mu_3}\big)\Big)
\\&\phantom{XXx.}
+ \theta^{\mu_3\mu_4}\Big((k_1k_4)\big(k_3^{\mu_2}(\theta k_2)^{\mu_1}-k_4^{\mu_1}(\theta k_1)^{\mu_2}\big)+(k_1k_3)\big(k_3^{\mu_1}(\theta k_1)^{\mu_2}-k_4^{\mu_2}(\theta k_2)^{\mu_1}
\big)\Big)\Big]\stackrel{\star}{S}
\\&\phantom{Xx}
+ \Big[(\theta k_1)^{\mu_4}\Big((\theta k_2)^{\mu_3}(k_1k_2)g^{\mu_1\mu_2}
+(\theta k_3)^{\mu_2}(k_1k_3)g^{\mu_1\mu_3}\Big)
\\&\phantom{XX.}
+(\theta k_4)^{\mu_1}\Big((\theta k_2)^{\mu_3}(k_1k_3)g^{\mu_2\mu_4}
+(\theta k_3)^{\mu_2}(k_1k_2)g^{\mu_4\mu_3}\Big)
\\&\phantom{XXX.}
+ \theta^{\mu_1\mu_4}\Big((k_1k_2)\big(k_1^{\mu_2}(\theta k_2)^{\mu_3}
-k_4^{\mu_3}(\theta k_3)^{\mu_2}\big)+(k_1k_3)\big(k_1^{\mu_3}(\theta k_3)^{\mu_2}-k_4^{\mu_2}(\theta k_2)^{\mu_3}\big)\Big)
\\&\phantom{XXx..}
+\theta^{\mu_2\mu_3} \Big((k_1k_2)\big(k_2^{\mu_1}(\theta k_1)^{\mu_4}
-k_3^{\mu_4}(\theta k_4)^{\mu_1}\big)+(k_1k_3)\big(k_2^{\mu_4}(\theta k_4)^{\mu_1}-k_3^{\mu_1}(\theta k_1)^{\mu_4}\big)\Big)\Big]\stackrel{\star}{T}
\\&\phantom{Xx}
+ \Big[(\theta k_2)^{\mu_4}
\Big((\theta k_3)^{\mu_1}(k_1k_4)g^{\mu_2\mu_3}
+(\theta k_1)^{\mu_3}(k_1k_2)g^{\mu_1\mu_2}\Big)
\\&\phantom{XX.}
+(\theta k_4)^{\mu_2}
\Big((\theta k_3)^{\mu_1}(k_1k_2)g^{\mu_4\mu_3}
+(\theta k_1)^{\mu_3}(k_1k_4)g^{\mu_1\mu_4}\Big)
\\&\phantom{XXX.}
+\theta^{\mu_1\mu_3}\Big(
(k_1k_2)\big(k_1^{\mu_2}(\theta k_2)^{\mu_4}
-k_3^{\mu_4}(\theta k_4)^{\mu_2}\big)
+(k_1k_4)\big(k_1^{\mu_4}(\theta k_4)^{\mu_2}
-k_3^{\mu_2}(\theta k_2)^{\mu_4}\big)\Big)
\\&\phantom{XXx..}
+ \theta^{\mu_2\mu_4}\Big(
(k_1k_2)\big(k_2^{\mu_1}(\theta k_1)^{\mu_3}
-k_4^{\mu_3}(\theta k_3)^{\mu_1}\big)
+(k_1k_4)\big(k_2^{\mu_3}(\theta k_3)^{\mu_1}
-k_4^{\mu_1}(\theta k_1)^{\mu_3}\big)\Big)\Big]
\stackrel{\star}{U}\Big\}.
\end{split}
\label{S4A3}
\end{gather}
Also, we have
\begin{gather}
\begin{split}
&S_4\Gamma_{A_4}^{\mu_1\mu_2\mu_3\mu_4}(k_1,k_2,-k_3,-k_4)
\\&
=4 \Big\{\Big[\big(k_1\theta k_2\big)\Big((\theta k_4)^{\mu_3}\big(k_3^{\mu_2}g^{\mu_1\mu_4}-k_3^{\mu_1}g^{\mu_2\mu_4}\big)
-(\theta k_3)^{\mu_4}\big(k_4^{\mu_1}g^{\mu_2\mu_3}-
k_4^{\mu_2}g^{\mu_1\mu_3}\big)+\theta^{\mu_3\mu_4}
\big(k_3^{\mu_1}k_4^{\mu_2}-k_3^{\mu_2}k_4^{\mu_1}\big)\Big)
\\&\phantom{XX}
+ \big(k_3\theta k_4\big)\Big((\theta k_2)^{\mu_1}\big(k_1^{\mu_4}g^{\mu_2\mu_3}- k_1^{\mu_3}g^{\mu_2\mu_4}\big)
-(\theta k_1)^{\mu_2}\big(k_2^{\mu_3}g^{\mu_1\mu_4}-k_2^{\mu_4}g^{\mu_1\mu_3}\big)+\theta^{\mu_1\mu_2}
\big(k_1^{\mu_3}k_2^{\mu_4}-k_1^{\mu_4}k_2^{\mu_3}\big)
\Big)\Big]\stackrel{\star}{S}
\\&\phantom{Xx}
+ \Big[\big(k_1\theta k_4\big)\Big((\theta k_2)^{\mu_3}\big(k_3^{\mu_4}g^{\mu_1\mu_2}-k_3^{\mu_1}g^{\mu_2\mu_4}\big)
-(\theta k_3)^{\mu_2}\big(k_2^{\mu_1}g^{\mu_4\mu_3}-k_2^{\mu_4}
g^{\mu_1\mu_3}\big)+\theta^{\mu_2\mu_3}\big(k_2^{\mu_1}k_3^{\mu_4}-k_2^{\mu_4}k_3^{\mu_1}\big)\Big)
\\&\phantom{XX}
+ \big(k_2\theta k_3\big)\Big((\theta k_4)^{\mu_1}\big(k_1^{\mu_3}g^{\mu_2\mu_4}-k_1^{\mu_2}
g^{\mu_4\mu_3}\big)
-(\theta k_1)^{\mu_4}\big(k_4^{\mu_2}g^{\mu_1\mu_3}-k_4^{\mu_3}
g^{\mu_1\mu_2}\big)+\theta^{\mu_1\mu_4}
\big(k_1^{\mu_2}k_4^{\mu_3}-k_1^{\mu_3}k_4^{\mu_2}\big)
\Big)\Big]\stackrel{\star}{T}
\\&\phantom{Xx}
+\Big[\big(k_1\theta k_3\big) 
\Big((\theta k_4)^{\mu_2}\big(k_2^{\mu_3}g^{\mu_1\mu_4}
-k_2^{\mu_1}g^{\mu_4\mu_3}\big)
-(\theta k_2)^{\mu_4}\big(k_4^{\mu_1}g^{\mu_2\mu_3}-k_4^{\mu_3}
g^{\mu_1\mu_2}\big)+\theta^{\mu_2\mu_4}
\big(k_2^{\mu_1}k_4^{\mu_3}-k_2^{\mu_3}k_4^{\mu_1}\big)
\Big)
\\&\phantom{XX.}
+\big(k_2\theta k_4\big) \Big((\theta k_1)^{\mu_3}\big(k_3^{\mu_4}g^{\mu_1\mu_2}-k_3^{\mu_2}
g^{\mu_1\mu_4}\big)
-(\theta k_3)^{\mu_1}\big(k_1^{\mu_2}g^{\mu_4\mu_3}-k_1^{\mu_4}
g^{\mu_2\mu_3}\big)+\theta^{\mu_1\mu_3}
\big(k_1^{\mu_2}k_3^{\mu_4}-k_1^{\mu_4}k_3^{\mu_2}\big)
\Big)\Big]\stackrel{\star}{U}\Big\},
\end{split}
\label{S4A4}
\end{gather}
\begin{gather}
\begin{split}
S_4&\Gamma_{A_5}^{\mu_1\mu_2\mu_3\mu_4}(k_1,k_2,-k_3,-k_4)
\\&\phantom{x}
=-4 \Big\{\Big[(\theta k_1)^{\mu_2}\Big((\theta k_4)^{\mu_3}k_3^{\mu_1}k_2^{\mu_4}
+(\theta k_3)^{\mu_4}k_4^{\mu_1}k_2^{\mu_3}\Big)
+(\theta k_2)^{\mu_1}\Big((\theta k_4)^{\mu_3}k_1^{\mu_4}k_3^{\mu_2}
+(\theta k_3)^{\mu_4}k_1^{\mu_3}k_4^{\mu_2}\Big)\Big]
\stackrel{\star}{S}
\\&\phantom{XXX}
+ \Big[(\theta k_1)^{\mu_4}\Big((\theta k_3)^{\mu_2}k_2^{\mu_1}k_4^{\mu_3}
+(\theta k_2)^{\mu_3}k_3^{\mu_1}k_4^{\mu_2}\Big)
+(\theta k_4)^{\mu_1}\Big((\theta k_3)^{\mu_2}k_1^{\mu_3}k_2^{\mu_4}
+(\theta k_2)^{\mu_3}k_1^{\mu_2}k_3^{\mu_4}\Big)\Big]
\stackrel{\star}{T}
\\&\phantom{XXX}
+ \Big[(\theta k_1)^{\mu_3}\Big((\theta k_2)^{\mu_4}k_3^{\mu_2}k_4^{\mu_1}
+(\theta k_4)^{\mu_2}k_2^{\mu_1}k_3^{\mu_4}\Big)
+(\theta k_3)^{\mu_1}\Big((\theta k_2)^{\mu_4}k_1^{\mu_2}k_4^{\mu_3}
+(\theta k_4)^{\mu_2}k_1^{\mu_4}k_2^{\mu_3}\Big)\Big]
\stackrel{\star}{U}\Big\}.
\end{split}
\label{S4A5}
\end{gather}
Permuting additional relevant FR terms proportional to $W_{1_{1,2,3}}$ in Eq. (\ref{F.11}), we have: 
\begin{gather}
\begin{split}
&S_4\Big(W_{1_1}^{\mu_1\mu_2\mu_3\mu_4}(k_1,k_2,-k_3,-k_4)
F_{\star_2}(k_1,k_2)F_{\star_2}(k_3,k_4)\Big)
\\&
=4\Big\{\Big[\theta^{\mu_1\mu_2}\Big(
(\theta k_3)^{\mu_4}\big(k_1^{\mu_3}(k_2k_3)
-k_2^{\mu_3}(k_1k_3)\big)
+(\theta k_4)^{\mu_3}\big(k_1^{\mu_4}(k_2k_4)
-k_2^{\mu_4}(k_1k_4)\big)\Big)
\\&\phantom{XX}
+\theta^{\mu_3\mu_4}\Big(
(\theta k_1)^{\mu_2}\big(k_3^{\mu_1}(k_1k_4)
-k_4^{\mu_1}(k_1k_3)\big)
+(\theta k_2)^{\mu_1}\big(k_3^{\mu_2}(k_2k_4)
-k_4^{\mu_2}(k_2k_3)\big)
\Big)\Big]\stackrel{\star}{S}
\\&\phantom{Xx} - \Big[\theta^{\mu_1\mu_4}\Big(
(\theta k_2)^{\mu_3}\big(k_1^{\mu_2}(k_2k_4)
-k_4^{\mu_2}(k_1k_2)\big)
+(\theta k_3)^{\mu_2}\big(k_1^{\mu_3}(k_3k_4)
-k_4^{\mu_3}(k_1k_3)\big)\Big)
\\&\phantom{XX}+\theta^{\mu_2\mu_3} \Big(
(\theta k_1)^{\mu_4}\big(k_2^{\mu_1}(k_1k_3)
-k_3^{\mu_1}(k_1k_2)\big)
+(\theta k_4)^{\mu_1}\big(k_2^{\mu_4}(k_3k_4)
-k_3^{\mu_4}(k_2k_4)\big)
\Big)\Big]\stackrel{\star}{T}
\\&\phantom{Xx} - \Big[\theta^{\mu_1\mu_3}\Big(
(\theta k_2)^{\mu_4}\big(k_1^{\mu_2}(k_2k_3)
-k_3^{\mu_2}(k_1k_2)\big)
+(\theta k_4)^{\mu_2}\big(k_1^{\mu_4}(k_3k_4)
-k_3^{\mu_4}(k_1k_4)\big)\Big)
\\&\phantom{XX}+ \theta^{\mu_2\mu_4}\Big(
(\theta k_1)^{\mu_3}\big(k_2^{\mu_1}(k_1k_4)
-k_4^{\mu_1}(k_1k_2)\big)
+(\theta k_3)^{\mu_1}\big(k_2^{\mu_3}(k_3k_4)
-k_4^{\mu_3}(k_2k_3)\big)
\Big)\Big]\stackrel{\star}{U}
\Big\}
\end{split}
\label{S4W11}
\end{gather}
\begin{gather}
\begin{split}
&S_4\Big(W_{1_2}^{\mu_1\mu_2\mu_3\mu_4}(k_1,k_2,-k_3,-k_4)
F_{\star_2}(k_1,k_2)F_{\star_2}(k_3,k_4)\Big)
\\&
=-4\Big\{\Big[(\theta k_1)^{\mu_2}\Big((\theta k_4)^{\mu_3}
\big( k_3^{\mu_1} k_1^{\mu_4}+k_2^{\mu_4} k_4^{\mu_1}
-2g^{\mu_1\mu_4}(k_1k_3)\big)
+ (\theta k_3)^{\mu_4}\big(k_4^{\mu_1} k_1^{\mu_3}
+k_2^{\mu_3} k_3^{\mu_1}-2g^{\mu_1\mu_3}(k_1k_4)\big)\Big)
\\&\phantom{XXx.}
+(\theta k_2)^{\mu_1}\Big((\theta k_4)^{\mu_3}
\big( k_1^{\mu_4} k_4^{\mu_2}+ k_3^{\mu_2} k_2^{\mu_4}
-2g^{\mu_2\mu_4}(k_1k_4)\big)
+ (\theta k_3)^{\mu_4}\big(k_1^{\mu_3} k_3^{\mu_1}
+ k_4^{\mu_2} k_2^{\mu_3}-2g^{\mu_2\mu_3}(k_1k_3)\big)\Big)
\Big]\stackrel{\star}{S} 
\\&\phantom{XX.} 
- \Big[(\theta k_1)^{\mu_4}\Big((\theta k_3)^{\mu_2}
\big( k_2^{\mu_1} k_1^{\mu_3}+k_4^{\mu_3} k_3^{\mu_1}
-2g^{\mu_1\mu_3}(k_1k_2)\big)
+ (\theta k_2)^{\mu_3}\big(k_3^{\mu_1} k_1^{\mu_2}
+ k_4^{\mu_2} k_2^{\mu_1}
-2g^{\mu_1\mu_2}(k_1k_3)\big)
\\&\phantom{XXx.}
+(\theta k_4)^{\mu_1}\Big((\theta k_3)^{\mu_2}
\big( k_1^{\mu_3} k_3^{\mu_4}+ k_2^{\mu_4} k_4^{\mu_3}
-2g^{\mu_3\mu_4}(k_1k_3)\big)
+ (\theta k_2)^{\mu_3}\big(k_1^{\mu_2} k_2^{\mu_4}
+k_3^{\mu_4} k_4^{\mu_2}
-2g^{\mu_2\mu_4}(k_1k_2)\big)
\Big)\Big]\stackrel{\star}{T} 
\\&\phantom{XX.} 
-\Big[(\theta k_1)^{\mu_3}\Big((\theta k_4)^{\mu_2}
\big( k_2^{\mu_1} k_1^{\mu_4}+k_3^{\mu_4} k_4^{\mu_1}
-2g^{\mu_1\mu_4}(k_1k_2)\big)
+ (\theta k_2)^{\mu_4}\big(k_4^{\mu_1} k_1^{\mu_2}
+k_3^{\mu_2} k_2^{\mu_1}
-2g^{\mu_1\mu_2}(k_1k_4)\big)
\\&\phantom{XXX}
+(\theta k_3)^{\mu_1}\Big((\theta k_4)^{\mu_2}\big(k_2^{\mu_3} k_3^{\mu_4}
+ k_1^{\mu_4} k_4^{\mu_3}
-2g^{\mu_3\mu_4}(k_1k_4)\big)
+(\theta k_2)^{\mu_4}
\big( k_4^{\mu_3} k_3^{\mu_2}+ k_1^{\mu_2} k_2^{\mu_3}
-2g^{\mu_2\mu_3}(k_1k_2)\big)
\Big)\Big]\stackrel{\star}{U} \Big\}.
\end{split}
\label{S4W12}
\end{gather}
\begin{gather}
\begin{split}
&S_4\Big(W_{1_3}^{\mu_1\mu_2\mu_3\mu_4}(k_1,k_2,-k_3,-k_4)
F_{\star_2}(k_1,k_2)F_{\star_2}(k_3,k_4)\Big)
\\&=4\Big\{\Big[(k_1\theta k_2)
\Big((\theta k_3)^{\mu_4}\big(k_3^{\mu_2}g^{\mu_1\mu_3}-k_3^{\mu_1}g^{\mu_2\mu_3}\big)+(\theta k_4)^{\mu_3}\big(k_4^{\mu_2}g^{\mu_1\mu_4}-k_4^{\mu_1}g^{\mu_2\mu_4}\big)\Big)
\\&\phantom{XX}
+(k_3\theta k_4)\Big((\theta k_1)^{\mu_2}\big(k_1^{\mu_4}g^{\mu_1\mu_3}-k_1^{\mu_3}g^{\mu_1\mu_4}\big)+(\theta k_2)^{\mu_1}\big(k_2^{\mu_4}g^{\mu_2\mu_3}-k_2^{\mu_3}g^{\mu_2\mu_4}\big)\Big)
\Big]\stackrel{\star}{S}
\\&\phantom{Xx} 
- \Big[(k_1\theta k_4)\Big((\theta k_2)^{\mu_3}\big(k_2^{\mu_4}g^{\mu_1\mu_2}-k_2^{\mu_1}g^{\mu_2\mu_4}\big)
+(\theta k_3)^{\mu_2}\big(k_3^{\mu_4}g^{\mu_1\mu_3}-k_3^{\mu_1}g^{\mu_3\mu_4}\big)\Big)
\\&\phantom{XX}
+(k_2\theta k_3)\Big((\theta k_1)^{\mu_4}\big(k_1^{\mu_3}g^{\mu_1\mu_2}-k_1^{\mu_2}g^{\mu_1\mu_3}\big)+(\theta k_4)^{\mu_1}\big(k_4^{\mu_3}g^{\mu_2\mu_4}-k_4^{\mu_2}g^{\mu_3\mu_4}\big)\Big)
\Big]\stackrel{\star}{T}
\\&\phantom{Xx} 
- \Big[(k_1\theta k_3)\Big((\theta k_2)^{\mu_4}\big(k_2^{\mu_3}g^{\mu_1\mu_2}-k_2^{\mu_1}g^{\mu_2\mu_3}\big)
+(\theta k_4)^{\mu_2}\big(k_4^{\mu_3}g^{\mu_1\mu_4}-k_4^{\mu_1}g^{\mu_3\mu_4}\big)\Big)
\\&\phantom{XX}
+(k_2\theta k_4)\Big((\theta k_1)^{\mu_3}\big(k_1^{\mu_4}g^{\mu_1\mu_2}-k_1^{\mu_2}g^{\mu_1\mu_4}\big)+(\theta k_3)^{\mu_1}\big(k_3^{\mu_4}g^{\mu_2\mu_3}-k_3^{\mu_2}g^{\mu_3\mu_4}\big)\Big)
\Big]\stackrel{\star}{U}\Big\}.
\end{split}
\label{S4W13}
\end{gather}

Additional new terms proportional to $W_{2,3}$ in Appendix A (\ref{F.12}) and (\ref{F.W3}) are needed 
to be permuted to obtain all relevant four-photon vertex diagrams:
\begin{gather}
\begin{split}
&S_4\Big(W_2^{\mu_1\mu_2\mu_3\mu_4}(k_1,k_2,-k_3,-k_4)
F_{\star_2}(k_1,k_2)F_{\star_2}(k_3,k_4)\Big)
\\&=4\Big\{\Big[(\theta k_1)^{\mu_2}\Big((\theta k_3)^{\mu_4}\big( k_1^{\mu_3} k_3^{\mu_1}-g^{\mu_1\mu_3}(k_1k_3)\big)+(\theta k_4)^{\mu_3} \big(k_1^{\mu_4} k_4^{\mu_1}-g^{\mu_1\mu_4}(k_1k_4)\big)\Big)
\\&\phantom{XX}
+(\theta k_2)^{\mu_1}\Big((\theta k_3)^{\mu_4}\big( k_2^{\mu_3} k_3^{\mu_2}-g^{\mu_2\mu_3}(k_2k_3)\big)
+(\theta k_4)^{\mu_3} \big(k_2^{\mu_4} k_4^{\mu_2}-g^{\mu_2\mu_4}(k_2k_4)\big)\Big)\Big]\stackrel{\star}{S} 
\\&\phantom{Xx}
+\Big[(\theta k_1)^{\mu_4}\Big((\theta k_2)^{\mu_3}\big( k_2^{\mu_1} k_1^{\mu_2}
-g^{\mu_1\mu_2}(k_1k_2)\big)+(\theta k_3)^{\mu_2} \big(k_1^{\mu_3} k_3^{\mu_1}-g^{\mu_1\mu_3}(k_1k_3)\big)\Big)
\\&\phantom{XX}
+(\theta k_4)^{\mu_1}\Big((\theta k_2)^{\mu_3}\big( k_2^{\mu_4} k_4^{\mu_2}
-g^{\mu_2\mu_4}(k_2k_4)\big)+(\theta k_3)^{\mu_2} \big(k_3^{\mu_4} k_4^{\mu_3}-g^{\mu_3\mu_4}(k_3k_4)\big)\Big)\Big]\stackrel{\star}{T} 
\\&\phantom{Xx}
+\Big[(\theta k_1)^{\mu_3}\Big((\theta k_2)^{\mu_4}\big( k_1^{\mu_2} k_2^{\mu_1}
-g^{\mu_1\mu_2}(k_1k_2)\big)+(\theta k_4)^{\mu_2} \big(k_1^{\mu_4} k_4^{\mu_1}-g^{\mu_1\mu_4}(k_1k_4)\big)\Big)
\\&\phantom{XX}
+(\theta k_3)^{\mu_1}\Big((\theta k_2)^{\mu_4}\big( k_2^{\mu_3} k_3^{\mu_2}
-g^{\mu_2\mu_3}(k_2k_3)\big)+(\theta k_4)^{\mu_2} \big(k_3^{\mu_4} k_4^{\mu_3}-g^{\mu_3\mu_4}(k_3k_4)\big)\Big)\Big]
\stackrel{\star}{U}\Big\},
 \end{split}
\label{S4W2}
\end{gather}
\begin{gather}
\begin{split}
&S_4\Big(W_3^{\mu_1\mu_2\mu_3\mu_4}(k_1,k_2,-k_3,-k_4)
F_{\star_2}(k_1,k_2)F_{\star_2}(k_3,k_4)\Big)
\\&
=-2\bigg\{(k_1\theta k_2)
\Big[k_2^{\mu_1}\Big( 
2(\theta k_3)^{\mu_4}g^{\mu_2\mu_3}
+2(\theta k_4)^{\mu_3}g^{\mu_2\mu_4}
-\theta^{\mu_3\mu_4}(k_4 - k_3)^{\mu_2} \Big)
\\&\phantom{XXXXXXx}
-k_1^{\mu_2}\Big( 
2(\theta k_3)^{\mu_4}g^{\mu_1\mu_3}
+2(\theta k_4)^{\mu_3}g^{\mu_1\mu_4}
-\theta^{\mu_3\mu_4}(k_4 - k_3)^{\mu_1}\Big)
\\&\phantom{Xxx}
+g^{\mu_1\mu_2}\Big( 
2(\theta k_3)^{\mu_4}(k_1-k_2)^{\mu_3}
+2(\theta k_4)^{\mu_3}(k_1-k_2)^{\mu_4}
-\theta^{\mu_3\mu_4}(k_1 - k_2)(k_4 - k_3)\Big)\Big]
\stackrel{\star}{S} 
\\&\phantom{XXx}
+(k_3\theta k_4)
\Big[k_4^{\mu_3}\Big( 
2(\theta k_1)^{\mu_2}g^{\mu_1\mu_4}
+2(\theta k_2)^{\mu_1}g^{\mu_2\mu_4}
-\theta^{\mu_1\mu_2}(k_2 - k_1)^{\mu_4} \Big)
\\&\phantom{XXXXXXx}
-k_3^{\mu_4}\Big( 
2(\theta k_1)^{\mu_2}g^{\mu_1\mu_3}
+2(\theta k_2)^{\mu_1}g^{\mu_2\mu_3}
-\theta^{\mu_1\mu_2}(k_2 - k_1)^{\mu_3}\Big)
\\&\phantom{Xxx}+
g^{\mu_3\mu_4}\Big(2(\theta k_1)^{\mu_2}(k_3-k_4)^{\mu_1} 
+2(\theta k_2)^{\mu_1}(k_3-k_4)^{\mu_2}
-\theta^{\mu_1\mu_2}(k_3 - k_4)(k_2 - k_1)\Big)\Big]
\stackrel{\star}{S} 
\\&\phantom{XXx}
+(k_1\theta k_4)
\Big[k_1^{\mu_4}\Big( 
2(\theta k_2)^{\mu_3}g^{\mu_1\mu_2}
-2(\theta k_3)^{\mu_2}g^{\mu_1\mu_3}
+\theta^{\mu_2\mu_3}(k_2 + k_3)^{\mu_1}\Big)
\\&\phantom{XXXXXXx}
+k_4^{\mu_1}\Big( 
2(\theta k_2)^{\mu_3}g^{\mu_1\mu_2}
-2(\theta k_3)^{\mu_2}g^{\mu_1\mu_3}
+\theta^{\mu_2\mu_3}(k_2 + k_3)^{\mu_4} \Big)
\\&\phantom{Xxx}
-g^{\mu_1\mu_4}\Big(2(\theta k_2)^{\mu_3}(k_1+k_4)^{\mu_2} 
-2(\theta k_3)^{\mu_2}(k_1+k_4)^{\mu_3}
+\theta^{\mu_2\mu_3}(k_1 + k_4)(k_2 + k_3)\Big)\Big]
\stackrel{\star}{T} 
\\&\phantom{XXx}
+(k_2\theta k_3)
\Big[k_2^{\mu_3}\Big( 
2(\theta k_1)^{\mu_4}g^{\mu_1\mu_2}
-2(\theta k_4)^{\mu_1}g^{\mu_2\mu_4}
+\theta^{\mu_1\mu_4}(k_1 + k_4)^{\mu_2}\Big)
\\&\phantom{XXXXXXx}
+k_3^{\mu_2}\Big( 
2(\theta k_1)^{\mu_4}g^{\mu_1\mu_3}
-2(\theta k_4)^{\mu_1}g^{\mu_3\mu_4}
+\theta^{\mu_1\mu_4}(k_1 + k_4)^{\mu_3} \Big)
\\&\phantom{Xxx}
-g^{\mu_2\mu_3}\Big(2(\theta k_1)^{\mu_4}(k_2+k_3)^{\mu_1} 
-2(\theta k_4)^{\mu_1}(k_2+k_3)^{\mu_4}
+\theta^{\mu_1\mu_4}(k_2 + k_3)(k_1 + k_4)\Big)\Big]
\stackrel{\star}{T} 
\\&\phantom{XXx}
+(k_1\theta k_3)
\Big[k_1^{\mu_3}\Big( 
2(\theta k_4)^{\mu_2}g^{\mu_1\mu_4}
-2(\theta k_2)^{\mu_4}g^{\mu_1\mu_2}
-\theta^{\mu_2\mu_4}(k_2 + k_4)^{\mu_1}\Big)
\\&\phantom{XXXXXXx}
+k_3^{\mu_1}\Big( 
2(\theta k_4)^{\mu_2}g^{\mu_3\mu_4}
-2(\theta k_2)^{\mu_4}g^{\mu_2\mu_3}
-\theta^{\mu_2\mu_4}(k_2 + k_4)^{\mu_3} \Big)
\\&\phantom{Xxx}
-g^{\mu_1\mu_3}\Big(2(\theta k_4)^{\mu_2}(k_1+k_3)^{\mu_4} 
-2(\theta k_2)^{\mu_4}(k_1+k_3)^{\mu_2}
-\theta^{\mu_2\mu_4}(k_1 + k_3)(k_2 + k_4)\Big)\Big]
\stackrel{\star}{U} 
\\&\phantom{XXx}
+(k_2\theta k_4)
\Big[k_2^{\mu_4}\Big( 
2(\theta k_3)^{\mu_1}g^{\mu_2\mu_2}
-2(\theta k_1)^{\mu_3}g^{\mu_2\mu_1}
-\theta^{\mu_1\mu_3}(k_1 + k_3)^{\mu_2}\Big)
\\&\phantom{XXXXXXx}
+k_4^{\mu_2}\Big( 
2(\theta k_3)^{\mu_1}g^{\mu_3\mu_4}
-2(\theta k_1)^{\mu_3}g^{\mu_1\mu_4}
-\theta^{\mu_1\mu_3}(k_1 + k_3)^{\mu_4} \Big)
\\&\phantom{Xxx}
-g^{\mu_2\mu_4}\Big(2(\theta k_3)^{\mu_1}(k_2+k_4)^{\mu_3} 
-2(\theta k_1)^{\mu_3}(k_2+k_4)^{\mu_1}
-\theta^{\mu_1\mu_3}(k_1 + k_3)(k_2 + k_4)\Big)\Big]
\stackrel{\star}{U} \bigg\}
\end{split}
\label{S4W3}
\end{gather}

Finally, to prove vanishing of the SW$(II)$ map-induced contributions to the scattering amplitude 
in the pure gauge sector, i.e., that ${\cal M}^\gamma_{II}(\gamma\gamma\to\gamma\gamma)=0$  
(\ref{4gIIstu}), we split the total cancellation into the following four subsections to achieve a 
fully transparent presentation of our explicit proof. 
\subsubsection{Cancellation 1}

First, we extract third terms ``$(\frac{1}{2}\theta\theta)$'' from the ninth (next to last) 
lines in (\ref{Sstar}), (\ref{Tstar}), and (\ref{Ustar}), obtained from the computations in Fig.\ref{fig:FD6}, first three diagrams, and found the required sum of the $s$-, $t$-, and $u$-channel contributions as 
\begin{gather}
\begin{split}
&\sum\limits_{\frac{1}{2}\theta^{\mu_i\mu_j}\theta^{\mu_k\mu_l}}
\Big[{(\rm C4})+{(\rm C5})+{(\rm C6})\Big]_{II}\;\bigg|_{i\not=j\not=k\not=l=1,2,3,4}
\\&\phantom{.}=ie^2\frac{1}{4}\Big[(t^2-u^2) \theta^{\mu_1\mu_2}\theta^{\mu_3\mu_4}\stackrel{\star}{S}
+(u^2-s^2) \theta^{\mu_1\mu_4}\theta^{\mu_2\mu_3}\stackrel{\star}{T}
+(t^2-s^2) \theta^{\mu_1\mu_3}\theta^{\mu_2\mu_4}\stackrel{\star}{U}\Big].
\end{split}
\label{A.45-3}
\end{gather}
Now in terms of Mandelstam variables (\ref{Mandelstam}) we take the $\Gamma_{A_2}$ part of Eqs. (\ref{A.3}), 
i.e. (\ref{S4A2}), as a result of $S_4$ permutations of SW($II$) map-induced $4\gamma$-coupling 
term $(S_4\Gamma_{A_2})$ and obtain
\begin{gather}
\begin{split}
&
\frac{ie^2}{4}\Big[\Big(S_4\Gamma_{A_2}^{\mu_1\mu_2\mu_3\mu_4}(k_1,k_2,-k_3,-k_4)\Big|_{II}\Big)_
{(\rm C9)}\Big]\\&
=\frac{ie^2}{4}\Big[(u^2-t^2) \theta^{\mu_1\mu_2}\theta^{\mu_3\mu_4}\stackrel{\star}{S}
+(s^2-u^2) \theta^{\mu_1\mu_4}\theta^{\mu_2\mu_3}\stackrel{\star}{T}
+(s^2-t^2) \theta^{\mu_1\mu_3}\theta^{\mu_2\mu_4}\stackrel{\star}{U}\Big],
\end{split}
\label{A.45-4}
\end{gather}
which exactly cancels (\ref{A.45-3}).

$\phantom{Thus all XXxxxxxand given as (\ref{A.45-5}), and given as 
(\ref{A.45-5})and given as (\ref{A.45-5}) respectively respectivelyrespectivel}$Q.E.D.

\subsubsection{Cancellation 2}

The remaining first two terms in the ninth lines from (\ref{Sstar}), (\ref{Tstar}), and (\ref{Ustar}) are 
\begin{gather}
\begin{split}
&\sum\limits_{1^{\rm st}\;two\;terms\; in\;9^{\rm th}\;lines\;of}
\Big[{(\rm C4})+({\rm C5})+({\rm C6})\Big]_{II}
=-ie^2(k_3k_4)\theta^{\mu_1\mu_2}\Big[(\theta k_3)^{\mu_4} (k_1-k_2)^{\mu_3}
+(\theta k_4)^{\mu_3} (k_1-k_2)^{\mu_4}\Big]
\stackrel{\star}{S}
\\&\phantom{X\sum\limits_{1^{\rm st}\;two\;terms\;in\;9^{\rm th}\;lines\;of}
\Big[{(\ref{Sstar})}+{(\ref{Tstar})}
+{(\ref{Ustar})}\Big]_{II}}
+ie^2(k_2k_3)\theta^{\mu_1\mu_4}\Big[(\theta k_3)^{\mu_2} 
(k_1+k_4)^{\mu_3}-(\theta k_2)^{\mu_3} (k_1+k_4)^{\mu_2}\Big]
\stackrel{\star}{T}
\\&
\phantom{X\sum\limits_{1^{\rm st}\;two\;terms\;in\;9^{\rm th}\;lines\;of}
\Big[{(\ref{Sstar})}+{(\ref{Tstar})}
+{(\ref{Ustar})}\Big]_{II}}
+ie^2(k_2k_4)\theta^{\mu_1\mu_3}
\Big[(\theta k_4)^{\mu_2} (k_1+k_3)^{\mu_4}
-(\theta k_2)^{\mu_4} (k_1+k_3)^{\mu_2}\Big]
\stackrel{\star}{U},
\end{split}
\label{A.45-5}
\end{gather}
and we compare them with the sum of the corresponding contributions to the four-photon diagram in Fig.\ref{fig:FD6}  
denoted as a sum of all $S_4$ permutations of $\Gamma_{A_3}$(\ref{S4A3}) and $W_{1_1}$(\ref{S4W11}), 
respectively:
\begin{gather}
\begin{split}
&\sum\limits_{corresponding\;terms\;from}\frac{ie^2}{4}\Big[\Big(S_4\Gamma_{A_3}\Big)_{(\rm C10)}+\Big(S_4W_{1_1}\Big)_{(\rm C13)}\Big]_{II}
\\&
=+ie^2\theta^{\mu_1\mu_2}
\Big\{(\theta k_3)^{\mu_4}
\Big[k_1^{\mu_3}\big(k_1k_3\big)-k_2^{\mu_3}\big(k_1k_4\big)+k_1^{\mu_3}\big(k_2k_3\big)
-k_2^{\mu_3}\big(k_1k_3\big)\Big]
\\&
\phantom{XXXXXx}
+(\theta k_4)^{\mu_3}
\Big[k_1^{\mu_4}\big(k_1k_4\big)-k_2^{\mu_4}\big(k_1k_3\big)+k_1^{\mu_4}\big(k_2k_4\big)
-k_2^{\mu_4}\big(k_1k_4\big)\Big]+\cdot\cdot\cdot
\Big\}\stackrel{\star}{S}
\\&
\phantom{X}
-ie^2\theta^{\mu_1\mu_4}
\Big\{(\theta k_2)^{\mu_3}
\Big[k_1^{\mu_2}\big(k_1k_2\big)-k_4^{\mu_2}\big(k_1k_3\big)-k_1^{\mu_2}\big(k_2k_4\big)
+k_4^{\mu_2}\big(k_1k_2\big)\Big]
\\&
\phantom{XXXXXx}
+
(\theta k_3)^{\mu_2} 
\Big[k_1^{\mu_3}\big(k_1k_3\big)-k_4^{\mu_3}\big(k_1k_2\big)-k_1^{\mu_3}\big(k_3k_4\big)
+k_4^{\mu_3}\big(k_1k_3\big)\Big]+\cdot\cdot\cdot
\Big\}
\stackrel{\star}{T}
\\&
\phantom{X}
-ie^2\theta^{\mu_1\mu_3}
\Big\{(\theta k_2)^{\mu_4}
\Big[k_1^{\mu_2}\big(k_1k_2\big)-k_3^{\mu_2}\big(k_1k_4\big)-k_1^{\mu_2}\big(k_2k_3\big)
+k_3^{\mu_2}\big(k_1k_2\big)\Big]
\\&
\phantom{XXXXXx}
+(\theta k_4)^{\mu_2} 
\Big[k_1^{\mu_4}\big(k_1k_4\big)-k_3^{\mu_4}\big(k_1k_2\big)-k_1^{\mu_4}\big(k_3k_4\big)
+k_3^{\mu_4}\big(k_1k_4\big)\Big]+\cdot\cdot\cdot
\Big\}
\stackrel{\star}{U}.
\end{split}
\label{A.45-6}
\end{gather}
It is easy to see that Eq. (\ref{A.45-6}) exactly cancels the sum of the remaining first two terms arising 
from the ninth lines in Eqs. (\ref{Sstar}), (\ref{Tstar}), and (\ref{Ustar}) and given as Eq. (\ref{A.45-5}), respectively. 

$\phantom{Thus all XXxxxxxand given as (\ref{A.45-5}), and given as 
(\ref{A.45-5})and given as (\ref{A.45-5}) respectively respectivelyrespectivel}$Q.E.D.

\subsubsection{Cancellation 3}

Writing explicitly terms in the sum of $s$-, $t$-, and $u$-channel contributions to the four-photon 
diagram in Fig.\ref{fig:FD6}, we have:
\begin{gather}
\begin{split}
&\sum\limits_{1^{\rm st}\;terms\;from}\frac{ie^2}{4}\bigg[\Big(S_4\Gamma_{A_4}\Big)_{(\rm C11)}
+\Big(S_4W_{1_3}\Big)_{(\rm C15)}+\Big(S_4W_3\Big)_{(\rm C17)}\bigg]_{II}
\\&
=\frac{-ie^2}{4}\bigg\{2(k_1\theta k_2)\Big[2k_2^{\mu_1}\Big( 2(\theta k_3)^{\mu_4}g^{\mu_2\mu_3}
+2(\theta k_4)^{\mu_3}g^{\mu_2\mu_4}
+\theta^{\mu_3\mu_4}(k_3 - k_4)^{\mu_2} \Big)
\\&\phantom{XXXXXXXX}
-2k_1^{\mu_2}\Big( 
2(\theta k_3)^{\mu_4}g^{\mu_1\mu_3}
+2(\theta k_4)^{\mu_3}g^{\mu_1\mu_4}
+\theta^{\mu_3\mu_4}(k_3 - k_4)^{\mu_1}\Big)
\\&\phantom{XXXXXXXx}
+g^{\mu_1\mu_2}\Big( 
2(\theta k_3)^{\mu_4}(k_1-k_2)^{\mu_3}
+2(\theta k_4)^{\mu_3}(k_1-k_2)^{\mu_4}
+ \theta^{\mu_3\mu_4}(k_1 - k_2)(k_3 - k_4)\Big)\Big]
\stackrel{\star}{S} 
\\&\phantom{XXXx}
+ 2(k_3\theta k_4)
\Big[2k_4^{\mu_3}\Big( 
2(\theta k_1)^{\mu_2}g^{\mu_1\mu_4}
+2(\theta k_2)^{\mu_1}g^{\mu_2\mu_4}
+\theta^{\mu_1\mu_2}(k_1 - k_2)^{\mu_4} \Big)
\\&\phantom{XXXXXXXxx}
-2k_3^{\mu_4}\Big( 
2(\theta k_1)^{\mu_2}g^{\mu_1\mu_3}
+2(\theta k_2)^{\mu_1}g^{\mu_2\mu_3}
+\theta^{\mu_1\mu_2}(k_1 - k_2)^{\mu_3}\Big)
\\&\phantom{XXXXXXXX}
+g^{\mu_3\mu_4}\Big( 
2(\theta k_2)^{\mu_1}(k_3-k_4)^{\mu_2}
+2(\theta k_1)^{\mu_2}(k_3-k_4)^{\mu_1}
+ \theta^{\mu_1\mu_2}(k_1 - k_2)(k_3 - k_4)\Big)\Big]
\stackrel{\star}{S} 
\\&\phantom{XXXx}
+2(k_1\theta k_4)
\Big[2k_1^{\mu_4}\Big( 
2(\theta k_2)^{\mu_3}g^{\mu_1\mu_2}
-2(\theta k_3)^{\mu_2}g^{\mu_1\mu_3}
+\theta^{\mu_2\mu_3}(k_2 + k_3)^{\mu_1} \Big)
\\&\phantom{XXXXXXXxx}
+2k_4^{\mu_1}\Big( 
2(\theta k_2)^{\mu_3}g^{\mu_2\mu_4}
-2(\theta k_3)^{\mu_2}g^{\mu_3\mu_4}
+\theta^{\mu_2\mu_3}(k_2 + k_3)^{\mu_4}\Big)
\\&\phantom{XXXXXXXX}-
g^{\mu_1\mu_4}\Big(2(\theta k_2)^{\mu_3}(k_1+k_4)^{\mu_2} 
-2(\theta k_3)^{\mu_2}(k_1+k_4)^{\mu_3}
+\theta^{\mu_2\mu_3}(k_1 + k_4)(k_2 + k_3)\Big)\Big]
\stackrel{\star}{T} 
\\&\phantom{XXXX}
+2(k_2\theta k_3)
\Big[2k_2^{\mu_3}\Big( 
2(\theta k_1)^{\mu_4}g^{\mu_1\mu_2}
-2(\theta k_4)^{\mu_1}g^{\mu_2\mu_4}
+\theta^{\mu_1\mu_4}(k_1 + k_4)^{\mu_2} \Big)
\\&\phantom{XXXXXXXXX}
+2k_3^{\mu_2}\Big( 2(\theta k_1)^{\mu_4}g^{\mu_1\mu_3}
-2(\theta k_4)^{\mu_1}g^{\mu_3\mu_4}
+\theta^{\mu_1\mu_4}(k_1 + k_4)^{\mu_3}\Big)
\\&\phantom{XXXXXXXX}
-g^{\mu_2\mu_3}\Big(2(\theta k_1)^{\mu_4}(k_2+k_3)^{\mu_1} 
-2(\theta k_4)^{\mu_1}(k_2+k_3)^{\mu_4}
+\theta^{\mu_1\mu_4}(k_2 + k_3)(k_1 + k_4)\Big)\Big]
\stackrel{\star}{T} 
\\&\phantom{XXXX}
+2(k_1\theta k_3)
\Big[2k_1^{\mu_3}\Big( 2(\theta k_2)^{\mu_4}g^{\mu_1\mu_2}
-2(\theta k_4)^{\mu_2}g^{\mu_1\mu_4}+\theta^{\mu_2\mu_4}(k_2 + k_4)^{\mu_1} \Big)
\\&\phantom{XXXXXXXXx}
+2k_3^{\mu_1}\Big( 2(\theta k_2)^{\mu_4}g^{\mu_2\mu_3}
-2(\theta k_4)^{\mu_2}g^{\mu_3\mu_4}+\theta^{\mu_2\mu_4}(k_2 + k_4)^{\mu_3}\Big)
\\&\phantom{XXXXXXXX}-
g^{\mu_1\mu_3}\Big(2(\theta k_2)^{\mu_4}(k_1+k_3)^{\mu_2} -2(\theta k_4)^{\mu_2}(k_1+k_3)^{\mu_4}+\theta^{\mu_2\mu_3}(k_1 + k_3)(k_2 + k_4)\Big)\Big]
\stackrel{\star}{U} 
\\&\phantom{XXXX}
+2(k_2\theta k_4)\Big[2k_2^{\mu_4}\Big( 2(\theta k_1)^{\mu_3}g^{\mu_1\mu_2}
-2(\theta k_3)^{\mu_1}g^{\mu_2\mu_3}+\theta^{\mu_1\mu_3}(k_1 +k_3)^{\mu_2} \Big)
\\&\phantom{XXXXXXXXx}
+2k_4^{\mu_2}\Big( 2(\theta k_1)^{\mu_3}g^{\mu_1\mu_4}-2(\theta k_3)^{\mu_1}g^{\mu_3\mu_4}+\theta^{\mu_1\mu_3}(k_1 + k_3)^{\mu_4}\Big)
\\&\phantom{XXXXXXXX}-
g^{\mu_2\mu_4}\Big(2(\theta k_1)^{\mu_3}(k_2+k_4)^{\mu_1} 
-2(\theta k_3)^{\mu_1}(k_2+k_4)^{\mu_3}+\theta^{\mu_1\mu_3}(k_2 + k_4)(k_1 + k_3)\Big)\Big]
\stackrel{\star}{U} 
\bigg\}.
\end{split}
\label{A.45-9}
\end{gather}
By simple comparisons with Eq. (\ref{Sstar}), 
it is easy to see the cancellation of their first six lines with parts proportional 
to $\stackrel{\star}{S}$ in Eq. (\ref{A.45-9}). Since the same is working for $\stackrel{\star}{T}$ 
and $\stackrel{\star}{U}$ proportional terms, we finally have cancellations of 
corresponding contributions to the sum of diagrams from Fig.\ref{fig:FD6}.: 
\begin{equation}
\Big(\sum\limits_{1^{\rm st}\;6\;lines\;of}\Big[{(\rm C4})+({\rm C5})+({\rm C6)}\Big]_{II}
\Big)+{(\rm C22)}=0.
\label{A.45-10}
\end{equation}
$\phantom{Thus all XXxxxxxand given as (\ref{A.45-5}), and given as 
(\ref{A.45-5})and given as (\ref{A.45-5}) respectively respectivelyrespectivelyy}$Q.E.D.

\subsubsection{Cancellation 4}

First we extract the sum of the seventh, eighth, and tenth lines in Eqs. (\ref{Sstar}), (\ref{Tstar}), and
(\ref{Ustar}) and write explicitly just 
one of the terms in $s$-, $t$-, and $u$-channel contributions to the three-photon vertex diagrams from Fig.\ref{fig:FD6}:
\begin{gather}
\begin{split}
&\sum\limits_{(7+8+10)^{\rm th}\;lines\;of}\Big[{(\rm C4})+({\rm C5})+({\rm C6)}\Big]_{II}
\\&
=-ie^2(\theta k_1)^{\mu_2}\Big\{(k_3k_4)\Big[2(\theta k_3)^{\mu_4}g^{\mu_1\mu_3}+\theta^{\mu_3\mu_4}(k_3-k_4)^{\mu_1}\Big]
-(\theta k_3)^{\mu_4}k_2^{\mu_1}k_4^{\mu_3}\Big\}\stackrel{\star}{S}+\cdot\cdot\cdot
\\&\phantom{xx}
-ie^2(\theta k_1)^{\mu_4}\Big\{(k_2k_3)\Big[2(\theta k_2)^{\mu_3}g^{\mu_1\mu_2}+\theta^{\mu_2\mu_3}(k_2+k_3)^{\mu_1}\Big]
-(\theta k_2)^{\mu_3}k_3^{\mu_2}k_4^{\mu_1}\Big\}\stackrel{\star}{T}+\cdot\cdot\cdot
\\&\phantom{xx}
-ie^2(\theta k_1)^{\mu_3}\Big\{(k_2k_4)\Big[2(\theta k_2)^{\mu_4}g^{\mu_1\mu_2}
+\theta^{\mu_2\mu_4}(k_2+k_4)^{\mu_1}\Big]
-(\theta k_2)^{\mu_4}k_3^{\mu_1}k_4^{\mu_2}\Big\}\stackrel{\star}{U}+\cdot\cdot\cdot
\end{split}
\label{A.45-7}
\end{gather} 
Second, we write explicitly the first terms from (\ref{S4A3}), (\ref{S4A5}), (\ref{S4W11}), 
(\ref{S4W12}), and (\ref{S4W2}) in the $s$-, $t$-, and $u$-channels contributions to the four-photon diagram in Fig.\ref{fig:FD6}:
\begin{gather}
\begin{split}
\sum\limits_{1^{\rm st}\;terms\;in}&\frac{ie^2}{4}\bigg[\Big(S_4\Gamma_{A_3}\Big)_{(\rm C10)}
+\Big(S_4\Gamma_{A_5}\Big)_{(\rm C12)}
+\Big(S_4W_{1_1}\Big)_{(\rm C13)}+\Big(S_4W_{1_2}\Big)_{(\rm C14)}+\Big(S_4W_2\Big)_{(\rm C16)}\bigg]_{II}
\\
=&
+ie^2(\theta k_1)^{\mu_2}\Big\{\big((k_1k_3)+(k_1k_4)\big)\Big[2(\theta k_3)^{\mu_4}g^{\mu_1\mu_3}+\theta^{\mu_3\mu_4}(k_3-k_4)^{\mu_1}\Big]
\\&\phantom{XXXXXXXXXxx.}
-(\theta k_3)^{\mu_4}\Big[k_2^{\mu_1}k_4^{\mu_3}+k_4^{\mu_1}k_2^{\mu_3}
+k_3^{\mu_1}k_2^{\mu_3} + k_3^{\mu_1}k_1^{\mu_3}\Big]\Big\}
\stackrel{\star}{S}+\cdot\cdot\cdot
\\&
+ie^2(\theta k_1)^{\mu_4}\Big\{\big((k_1k_2)-(k_1k_3)\big)\Big[2(\theta k_2)^{\mu_3}g^{\mu_1\mu_2}+\theta^{\mu_2\mu_3}(k_2+k_3)^{\mu_1}\Big]
\\&\phantom{XXXXXXXXXXx}
-(\theta k_2)^{\mu_3}\Big[k_4^{\mu_1}k_3^{\mu_2}-k_3^{\mu_1}k_1^{\mu_2}
+k_2^{\mu_1}k_4^{\mu_2} -k_2^{\mu_1}k_1^{\mu_2}\Big]\Big\}
\stackrel{\star}{T}+\cdot\cdot\cdot
\\&
+ie^2(\theta k_1)^{\mu_3}\Big\{\big((k_1k_2)-(k_1k_4)\big)\Big[2(\theta k_2)^{\mu_4}g^{\mu_1\mu_2}+\theta^{\mu_2\mu_4}(k_2+k_4)^{\mu_1}\Big]
\\&\phantom{XXXXXXXXXXx}
-(\theta k_2)^{\mu_4}\Big[k_3^{\mu_1}k_4^{\mu_2}-k_2^{\mu_1}k_4^{\mu_2}
+k_2^{\mu_1}k_1^{\mu_2} -k_3^{\mu_1}k_1^{\mu_2}\Big]\Big\}
\stackrel{\star}{U}+\cdot\cdot\cdot,
\end{split}
\label{A.45-8}
\end{gather}
and see that they do cancel (\ref{A.45-7}) exactly. 
This way we have also showed that in the pure NCQED gauge sector all 
SW map-induced terms cancel out at tree level of 
the exclusive $\gamma\gamma\to\gamma\gamma$ scatterings. 

$\phantom{Thus all XXxxxxxand given as (\ref{A.45-5}), and given a
(\ref{A.45-5})and given as (\ref{A.45-5}) respectively respectivelyrespectively}$Q.E.D.  \\

The meaning of the above cancellations (\ref{A.45-3})$-$(\ref{A.45-8}) producing nontrivial zero for LbyL (\ref{4gIIstu}), 
together with nontrivial zero in Compton case (\ref{ComptSW(II)}) and with 
trivial zeros for  M\o ller (\ref{callMM}) and Bhabha (\ref{callMB}) cases, is that the SW map-induced contributions to all scattering amplitudes vanish, proving thus explicitly that in both the matter 
and the gauge sector of the on-shell U(1) NCQED scattering amplitudes 
with and without SW maps are equal or equivalent one to each other, as indicated generally in 
Refs. \cite{Martin:2016hji,Martin:2016saw}.

\end{document}